\documentclass[aps,pre,twocolumn,superscriptaddress]{revtex4x}
\usepackage{epsfig}
\usepackage{times}
\usepackage{float}
\usepackage{color}
\begin{document}

\title{Statistical physics of human cooperation}

\author{Matja{\v z} Perc}
\email{matjaz.perc@uni-mb.si}
\affiliation{Faculty of Natural Sciences and Mathematics, University of Maribor, Koro{\v s}ka cesta 160, SI-2000 Maribor, Slovenia}
\affiliation{CAMTP -- Center for Applied Mathematics and Theoretical Physics, University of Maribor, Mladinska 3, SI-2000 Maribor, Slovenia}

\author{Jillian J. Jordan}
\affiliation{Department of Psychology, Yale University, New Haven, Connecticut 06511, USA}

\author{David G. Rand}
\affiliation{Department of Psychology, Yale University, New Haven, Connecticut 06511, USA}
\affiliation{Department of Economics, Yale University, New Haven, Connecticut 06511, USA}
\affiliation{School of Management, Yale University, New Haven, Connecticut 06511, USA}

\author{Zhen Wang}
\affiliation{Center for Optical Imagery Analysis and Learning, Northwestern Polytechnical University, Xi'an 710072, China}

\author{Stefano Boccaletti}
\affiliation{CNR -- Institute of Complex Systems, Via Madonna del Piano, 10, 50019 Sesto Fiorentino, Florence, Italy}
\affiliation{The Italian Embassy in Israel, 25 Hamered st., 68125 Tel Aviv, Israel}

\author{Attila Szolnoki}
\affiliation{Faculty of Natural Sciences and Mathematics, University of Maribor, Koro{\v s}ka cesta 160, SI-2000 Maribor, Slovenia}
\affiliation{Institute of Technical Physics and Materials Science, Centre for Energy Research, Hungarian Academy of Sciences, P.O. Box 49, H-1525 Budapest, Hungary}

\begin{abstract}
Extensive cooperation among unrelated individuals is unique to humans, who often sacrifice personal benefits for the common good and work together to achieve what they are unable to execute alone. The evolutionary success of our species is indeed due, to a large degree, to our unparalleled other-regarding abilities. Yet, a comprehensive understanding of human cooperation remains a formidable challenge. Recent research in social science indicates that it is important to focus on the collective behavior that emerges as the result of the interactions among individuals, groups, and even societies. Non-equilibrium statistical physics, in particular Monte Carlo methods and the theory of collective behavior of interacting particles near phase transition points, has proven to be very valuable for understanding counterintuitive evolutionary outcomes. By studying models of human cooperation as classical spin models, a physicist can draw on familiar settings from statistical physics. However, unlike pairwise interactions among particles that typically govern solid-state physics systems, interactions among humans often involve group interactions, and they also involve a larger number of possible states even for the most simplified description of reality. The complexity of solutions therefore often surpasses that observed in physical systems. Here we review experimental and theoretical research that advances our understanding of human cooperation, focusing on spatial pattern formation, on the spatiotemporal dynamics of observed solutions, and on self-organization that may either promote or hinder socially favorable states.
\end{abstract}

\maketitle

\tableofcontents

\section{Introduction}

\subsection{Human cooperation}
Human cooperation is the result of our evolutionary struggles for survival. Approximately two million years ago the jaw-closing muscle of some hominids mutated, thus giving space for larger brains, which in turn needed larger body size to carry. As a result, our ancestors begun to mature more slowly than other apes, which likely led to serious challenges in rearing offspring that survived \cite{peters_83, calder_84}. Alloparental care and provisioning for the young of others have therefore been put forward as the impetus for the evolution of remarkable other-regarding abilities of the genus \textit{Homo} that we witness today \cite{hrdy_11}. There also exist evidence that the conflicts between groups have been instrumental for strengthening our cooperative drive and for enhancing our in-group solidarity \cite{bowles_11}. Since all of this took place in the very distant past, evidence supporting one or the other thesis is scarce and circumstantial. But regardless of whether it was the slow development of our offspring towards self-sustained existence, or the fear of being wiped-out by our neighbors, extensive and comprehensive cooperation, also among unrelated individuals, became the cornerstone of our evolutionary success story \cite{axelrod_84}.

Fast forward to the present time, it is clear that many of the challenges that pressured our ancestors into cooperation are gone. Nevertheless, we are still cooperating, and on ever larger scales, to the point that we may deserve being called ``SuperCooperators'' \cite{nowak_11}. A more critical look, however, reveals several ups and downs to our more mature existence. Undeniably, there is an abundance of technological breakthroughs and innovations that make our lives better. The 20th century is often referred to as the century of physics. From x-rays to the semiconductor industry, the human society today would be very different were it not for the progress made in physics laboratories around the world \cite{perc2013self, sinatra2015century}. Moreover, we have basically conquered our planet, to the point that the only real threat to us is ourselves. We are also compassionate, we care for one another, and we are civilized and social. But at the same time, and in stark contrast to these ups, our societies are also home to millions that live on the edge of existence. We deny people shelter, we deny people food, and we deny people their survival. Many human societies are seriously failing to meet the most basic needs of millions around the world \cite{humans_14}. Interestingly, the book \textit{The Better Angels of our Nature} \cite{pinker_11} argues that our world is more peaceful today than it ever was in the past. While statistically this may be true, it is probably easy to agree that peace is more than just the absence of war and death by armed forces.

\begin{figure*}
\begin{center}
\centerline{\epsfig{file=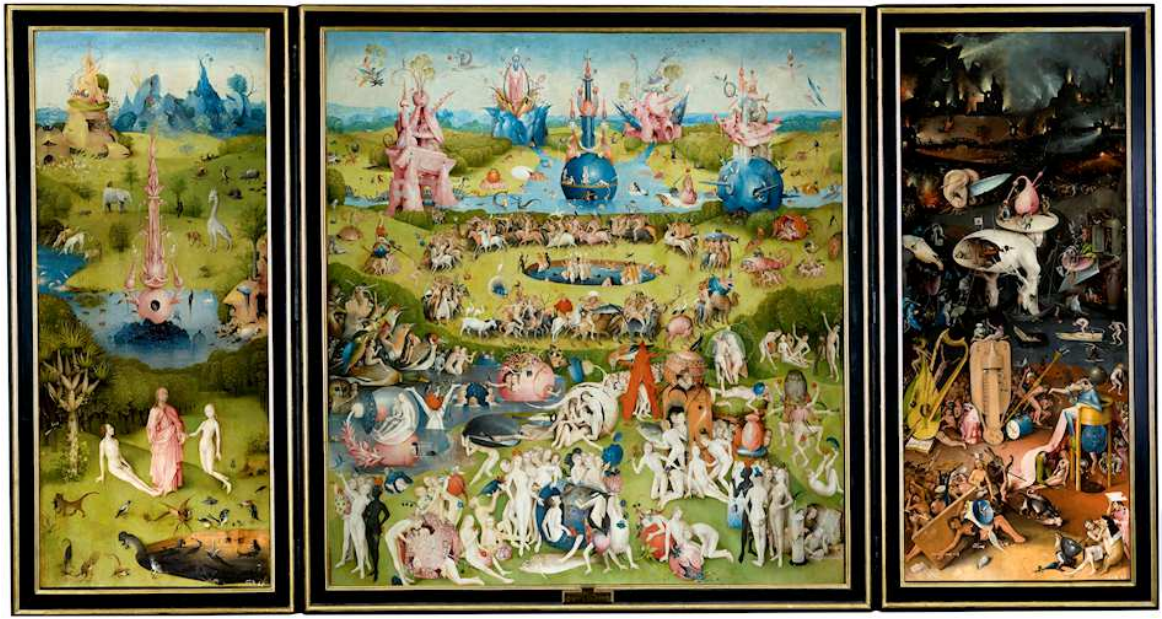,width=15.4cm}}
\caption{\textit{The Garden of Earthly Delights} is a triptych painted by Hieronymus Bosch around the turn of the 16th century. The painting is housed in the Museo del Prado in Madrid, serving as a didactic warning on the perils of life's temptations. From left to right, we first have the meeting of Adam and Eve, followed by a socially balanced state of wellbeing in the center, and ending with a hellscape that portrays the torments of damnation. The picture was taken by Alonso de Mendoza, and reproduced here from the public domain of Wikimedia Commons.}
\label{garden}
\end{center}
\end{figure*}

We should be able to do better. In fact, we must do better, because the current state seems largely unsustainable even in the short term, let alone in the long term, with the tragedy of the commons and doomsday scenarios as the one depicted in Fig.~\ref{garden} looming in the not so distant future. We thus need to learn how to cooperate better with one another, and we have to understand that our actions and the choice that we make everyday have consequences that go far beyond our local communities. The problem, however, is that to cooperate more or better, or even to cooperate at all, is in many ways unnatural, and this is true for us as well as for all living organisms. According to Darwin's \textit{The Origin of Species}, natural selection favors the fittest and the most successful individuals, which in turn implies an innate selfishness that greatly challenges the concept of cooperation. In short, cooperation is costly. As such, exercising it can weigh heavily on individual wellbeing and prosperity. If only the fittest survive, why should one perform an altruistic act that is costly to perform but benefits another? Why should we care for and contribute to the public good if freeriders can enjoy the same benefits for free? Since intact cooperation forms the bedrock of our efforts for a sustainable and better future, understanding cooperative behavior in human societies has been declared as one of the grand scientific challenges of the 21st century \cite{pennisi_s05, kennedy_s05}.

\subsection{The role of statistical physics}
At this point, one may question the relevance of physics in all of this. Methods of statistical physics have recently been applied to subjects that, in the traditional sense, could be considered as out of scope. Statistical physics of social dynamics \cite{castellano_rmp09}, of evolutionary games in structured populations \cite{szabo_pr07, perc_bs10, wang_z_epjb15, szabo_pr16}, of crime \cite{orsogna_plr15}, and of epidemic processes and vaccination \cite{pastor_rmp15, wang_z_pr16}, are all recent examples of this exciting development. The latter comes with a slight delay but strong support from network science \cite{barabasi_16}, which has been going from strength to strength during the past decade and a half \cite{albert_rmp02, newman_siamr03, boccaletti_pr06, fortunato_pr10, holme_sr12, kivela_jcn14, boccaletti_pr14, barabasi_16}, delivering inspirational results, models, and methods, that have revived not just statistical physics, but many other fields of natural and social sciences.

In this regard, the evolution of cooperation is no exception. While research in the realm of biology has delivered kin selection theory \cite{hamilton_wd_jtb64}, which rests on the fact that by helping a close relative to reproduce still allows indirect passing of the genes to the next generation, and while other key mechanisms have been identified that promote cooperation (including direct and indirect reciprocity as well as group selection), network reciprocity \cite{nowak_s06}, in particular, has been the main motivator for the involvement of statistical physics in this line of research.

The manifestation of network reciprocity relies on pattern formation in a structured population, which provides a more realistic description of reality than exactly solvable well-mixed models. In the simplest case, a structured population is described by a square lattice \cite{nowak_n92b}, where cooperators form compact clusters and can thus avoid, at least those in the interior of such clusters, being exploited by defectors. In short, cooperators do better if they are surrounded by other cooperators. However, the emergence of cooperation and the phase transitions leading to other counterintuitive evolutionary outcomes depend sensitively on the structure of the interaction network and the type of interactions, as well as on the number and type of competing strategies \cite{santos_prl05, pacheco_prl06, gomez-gardenes_prl07, ohtsuki_prl07, roca_plr09, lee_s_prl11,  mathiesen_prl11, szolnoki_prl12, assaf_prl12, gomez_prl13, knebel_prl13, pinheiro_prl14}. Studies that are unique to physicists have led to significant advances in our understanding of the evolution of cooperation, for example by expanding our understanding of the role of heterogeneity of interaction networks \cite{santos_prl05} or competing agents \cite{szolnoki_epl07, perc_pre08}, the dynamical organization of cooperation \cite{gomez-gardenes_prl07} and population growth \cite{poncela_njp09}, the spontaneous emergence of hierarchy \cite{szolnoki_njp08, lee_s_prl11, szolnoki_srep16, gomez-gardenes_jtb08}, as well as the intriguing role of strategic complexity \cite{szolnoki_prl12, szolnoki_prx13}, to name only some examples.

Human cooperation is special in that we are intelligent enough to enforce it when it is failing. As such, human cooperation is subject to both positive and negative incentives \cite{andreoni_aer03, rand_tcs13, okada_ploscb15, rand_pone15, kraft_cobs15}. Positive incentives typically entail rewards for behaving prosocially \cite{dreber_n08, rand_s09, hilbe_prsb10, hauert_jtb10, szolnoki_epl10, szolnoki_njp12, szolnoki_prsb15}, while negative incentives typically entail punishing free-riding \cite{fehr_aer00, boyd_pnas03, gardner_a_an04, henrich_s06b, sigmund_tee07, raihani_tee12, szolnoki_jtb13, hauser_jtb14, jordan_n16}. However, just like cooperation incurs a cost for the benefit of the common good, so does the provisioning of rewards or sanctions incur a cost for the benefit or harm of the recipients. Individuals that abstain from dispensing such incentives therefore become second-order freeriders \cite{fehr_n04}, and they are widely believed to be amongst the biggest impediments to the evolutionary stability of rewarding and punishing \cite{panchanathan_n04, hauert_s07, helbing_ploscb10, hilbe_srep12, chen_xj_njp14}.

Another rather unique human ability is tolerance, which is the willingness to steadfastly endure something, in particular a trying circumstance such as the existence of opinions or behavior, with a fair and objective attitude. Although natural selection favors the fittest and thus challenges cooperation, tolerance and social norms in human societies may just be the missing ingredient for cooperative behavior to prevail \cite{henrich2006culture, boyd2009culture, szolnoki_rsif15, chen_xj_pre09b, szolnoki_pre15, chen_xj_pre09, capraro_prsb15, szolnoki_njp16}.

The complexity of the mathematical models that result out of taking into account the above considerations requires the usage of methods of non-equilibrium statistical physics. In particular, Monte Carlo methods and the theory of collective behavior of interacting particles near phase transition points have proven to be very valuable for understanding counterintuitive evolutionary outcomes that allow cooperation to prevail. In what follows, we review statistical physics research done to advance our understanding of human cooperation.

In the first place, however, we present an overview of human experiments, where we describe the goals and the methodology \cite{rand_jtb12}, as well as review experiments measuring prosociality \cite{herrmann_s08, pillutla_jes03, barclay_06, fehrler_ehb13}, punishment \cite{guth_jebo82, falkinger_aer00, fehr_n02, henrich_s06b, balafoutas_el4,  charness_jebo08, jordan_n16} and rewarding \cite{sefton_ei07, rand_s09, hauser_srep16} (or both \cite{rand_s09, pedersen_prsb13, almenberg_pp17}), and network effects \cite{gracia-lazaro_pnas12, grujic_pone10, grujic_srep12, grujic_pone12, traulsen_pnas10, rand_pnas11, wang_j_pnas12, shirado_nc13, jordan_pone13}.

We then present an overview of mathematical models, where we focus on the public goods game as the null model for human cooperation \cite{perc_jrsi13}, with extensions towards incorporating punishment \cite{helbing_njp10, szolnoki_pre11, perc_njp12}, rewarding \cite{szolnoki_epl10, szolnoki_njp12, szolnoki_prsb15}, correlated positive and negative reciprocity \cite{szolnoki_prx13}, as well as tolerant players \cite{szolnoki_pre15}. Next, we briefly review the methodology, in particular the Monte Carlo simulation technique \cite{binder_88, newman_99}, the theory of phase transitions \cite{stanley_71, liggett_85, marro_99, hinrichsen_ap00}, and the important concept of the stability of subsystem solutions in structured populations. We then proceed with the overview of results, where we separately consider peer- \cite{helbing_njp10, szolnoki_epl10, helbing_ploscb10, helbing_pre10c} and pool-based strategies \cite{szolnoki_pre11, szolnoki_pre11b, chen_xj_fbs14}, the self-organization of incentives for cooperation \cite{perc_njp12, szolnoki_njp12}, antisocial strategies \cite{rand_jtb10, rand_nc11, szolnoki_prsb15}, as well as tolerance \cite{szolnoki_pre15, szolnoki_njp16}. We conclude with an outline of possible directions for future research in the realm of statistical physics of human cooperation.

\section{Human experiments}
\label{humex}
Human experiments are a critical tool for testing predictions of theoretical models and investigating human cooperation. As the following subsections show, experiments provide clear evidence that people behave prosocially, even in anonymous, one-shot interactions when real money is at stake. Moreover, people are willing to enforce prosociality, punishing selfishness and rewarding cooperation. It also matters whether the structure of the interaction network is taken into account, and whether the latter is fixed or changing over time.

\subsection{Goals of human experiments}
\label{exgoals}
Models investigate what is theoretically possible, and the focus of this review is models of the evolution and maintenance of human cooperation. Experiments with human subjects compliment theoretical work by describing human behavior, and testing predictions generated by models. In other words, experiments test not what is theoretically possible, but what empirically occurs.

Experiments investigating human cooperation typically employ economic games, in which subjects make decisions about how to divide real money between themselves and other subjects (who are typically anonymous strangers). Within the game theory community, the use of incentivized economic games started gaining traction in the early 80s (e.g., \cite{guth_jebo82}), and has steadily grown in popularity ever since (for an overview, see \cite{camerer_03}).

The most basic use of experiments is to measure subjects' behavioral tendencies in simple games \cite{camerer_04}. To this end, researchers ask subjects to make decisions about how to allocate their money, often in contexts where they are anonymous and thus there is no overt social pressure to behave a certain way. For example, a simple game that can be used to measure generosity towards strangers is the dictator game \cite{forsythe_geb94}. This game involves a dictator and a recipient. The dictator starts with an endowment of money (e.g., \$10), and decides how much to share with the recipient. The canonical dictator game is one-shot (i.e., the dictator makes one decision and then the dictator and recipient never interact again) and anonymous (i.e., the dictator and recipient do not know each other's identities). Behavior in this game can thus be used to measure generosity towards strangers, in the absence of any strategic or self-interested incentive to give.

\begin{figure}
\centerline{\epsfig{file=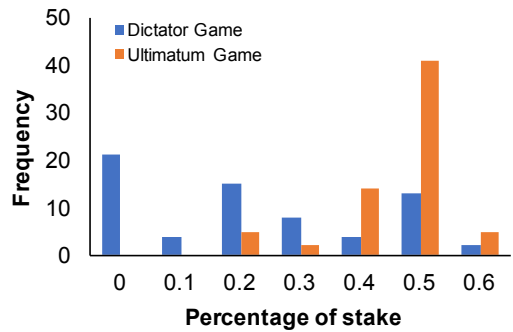,width=6.8cm}}
\caption{Distribution of offers in a dictator game (blue) versus an ultimatum game (orange). Data come from collapsing across all incentivized games in \cite{forsythe_geb94}. Subjects offer more in the ultimatum game (where the modal offer is 50\%) than in the dictator game (where the modal offer is nothing), demonstrating that the threat of rejection motivates increased giving.}
\label{dgvsug}
\end{figure}

Experiments can also be used to investigate how play varies across different versions of a game. For example, the interaction structure of the dictator game can be altered to give the recipient power, as is done, for example, in the ultimatum game \cite{guth_jebo82}. In the ultimatum game, a proposer receives an endowment and proposes an amount to allocate to a responder, as in the dictator game. Unlike the dictator game, however, in the ultimatum game the responder can either accept the proposal, or reject it -- in which case both players earn nothing. Thus, while in the dictator game the dictator can unilaterally decide how to allocate the endowment, in the ultimatum game the proposer only receives her proposed allocation if the responder is willing to accept it. By giving power to the responder, the ultimatum game measures not simple generosity but strategic generosity: how much does the proposer choose to share, given the threat of rejection? Offers are generally higher in the ultimatum game than the dictator game, demonstrating that the threat of rejection motivates increased giving, as illustrated in Fig.~\ref{dgvsug} \cite{forsythe_geb94}.

Experiments can also manipulate whether a game is one-shot or repeated, allowing researchers to investigate the power of reciprocal play as a tool to promote cooperation. This commonly occurs in the context of a prisoner's dilemma, which models cooperation between pairs of individuals \cite{rapoport_gs66}. In this symmetric game, there are two players. They simultaneously decide whether to cooperate by reducing their own payoff to increase their partner's, or to defect by maximizing their own payoff at the expense of their partner. If both players cooperate, they each earn more, i.e., they each get the reward $R$, than if they both defect, in which case they both get the punishment $P$. However, if one player cooperates and the other defects, the cooperator earns less than he would have if both players had defected, i.e., he gets the sucker's payoff $S<P$, while the defector earns more than she would have if both players had cooperated, i.e., she gets the temptation $T>R$. Thus, cooperation is positive-sum, increasing the total payoff, but it is always in an individual's self-interest to defect.

\begin{figure}
\centerline{\epsfig{file=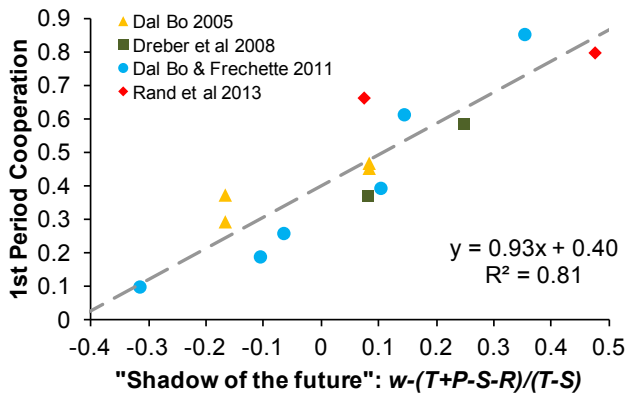,width=8.5cm}}
\caption{Cooperative behavior in repeated prisoner's dilemma experiments as a function of the ``shadow of the future'', or the extent to which future consequences exist for actions in the current period. Data are reproduced from a review of prisoner's dilemma experiments in \cite{rand_tcs13}. Specifically, the $x-$axis shows the amount by which the continuation probability $w$ (probability that two subjects play another prisoner's dilemma round together) exceeds the critical payoff threshold $(T + P - S - R)/(T - S)$ necessary for tit-for-tat to risk-dominate always defect. In a population that is $1/2$ tit-for-tat and $1/2$ always defect, $w < (T + P - S - R)/(T - S)$ means that always defect earns more than tit-for-tat; $w = (T + P - S - R)/(T - S)$ means that tit-for-tat and always defect do equally well; and the more $w$ exceeds $(T + P - S - R)/(T - S)$, the more tit-for-tat earns compared to always defect. The $y-$axis indicates the probability of cooperation in the first round of each repeated prisoner's dilemma game. When the shadow of the future is stronger, subjects are more likely to open their game play with cooperation, demonstrating the power of repeated interactions to promote cooperation.}
\label{pdex}
\end{figure}

If the prisoner's dilemma is one-shot, then theory predicts universal defection, as defection earns a strictly higher payoff than cooperation. However, if the game is repeated, and it is possible for people to condition their cooperation on the previous play of their partner. In this case, cooperation can be sustained through reciprocal strategies (\cite{axelrod_84,  fudenberg_e86,  trivers_qrb71}): if I know that you will only cooperate with me next round if I cooperate with you now, then cooperation can be self-interested -- provided that the probability that we play again for another round (i.e., the ``continuation probability'') is sufficiently high. This prediction is supported by experimental evidence: people cooperate more in repeated prisoner's dilemma games than in one-shot prisoner's dilemma games, and cooperation levels are increasing in the continuation probability, as illustrated in Fig.~\ref{pdex} \cite{bo_jel17, rand_tcs13}. Repeated play even promotes cooperation among $10$ to $12$ year old children \cite{blake_srep15}.

Relatedly, reputation effects can be investigated by manipulating the observability of behavior. A large theoretical literature shows that reputation systems can promote the evolution of cooperation \cite{nowak_n05}, and experiments confirm that people are more likely to cooperate when their decisions are observable to others \cite{milinski_n02, pfeiffer_jrsi12, wedekind_s00}.

\subsection{Experimental methodology}
\label{exmethods}
Economic game experiments have traditionally been conducted in laboratories, in which experimental subjects (typically undergraduate students) are recruited to play games. Typically, groups of subjects are recruited to participate at the same time, so that they can interact with each other and be paid accordingly. However, individual subjects usually do not know who they are paired with, and decisions are kept anonymous. Subjects are usually fully informed of the rules and payoff structure of the game, and are paid a fixed ``show-up fee'' for participating, as well as a ``bonus'' payment based on their game choices, and the choices of other players.

In recent years, conducting economic game experiments on the Internet has become increasingly popular, particularly through Amazon Mechanical Turk \cite{rand_jtb12}. Amazon Mechanical Turk is an online labor market in which employers pay workers to complete ``human intelligence tasks'', which are short tasks for relatively low pay -- and can include participating in an experiment. Subjects (workers) can be paid a show-up fee for completing the human intelligence task, and then a bonus payment based on their behavior, and the behavior of the other subjects (also Amazon Mechanical Turk workers) who they are paired with. Amazon Mechanical Turk makes it possible to quickly recruit large samples, and to obtain a relatively diverse sample that is closer to nationally representative than a university undergraduate sample. Moreover, evidence demonstrates that, across a wide range of economic games, game play is very similar on Amazon Mechanical Turk and in the physical laboratory -- even though stakes are generally an order of magnitude higher in the lab \cite{amir_pone12, horton_ee11}.

\subsection{Measuring prosociality}
\label{exprosoc}

\subsubsection{Cooperation in groups}
\label{excogroups}
The most common experimental game for examining cooperation among groups of players is the public goods game \cite{fehr_aer00}. In the public goods game, each member of a group (typically consisting of four subjects) starts with an endowment, and decides how much to contribute to the public good. Contributions are multiplied by the experimenter (e.g., doubled), and then divided equally among all players. Thus, as in the prisoner's dilemma game, contributing to the public good is positive sum, but strictly costly to the individual: no matter how much other group members contribute, it is payoff-maximizing to contribute nothing.

In one-shot public goods games, substantial cooperation is observed: the average contribution is typically around fifty percent (although this varies based on factors such as the contribution multiplier and the particular subject pool used). Furthermore, it is common to observe a bimodal distribution, where the most common responses are to contribute nothing or to contribute everything. Thus, many people are willing to make substantial contributions to the public good, but a sizeable proportion contribute nothing. As a result of these defectors, it is uncommon for public goods game cooperation to be sustained over time when the game is repeated \cite{fehr_aer00, rand_s09}. The key problem is that there is no way for players to cooperate only with group members who are willing to contribute, but not with defectors. Thus, ``conditional cooperators'', or those who wish to cooperate with other cooperators but not with defectors, typically switch to defection as the game progresses -- destroying cooperation.

\begin{figure}
\centerline{\epsfig{file=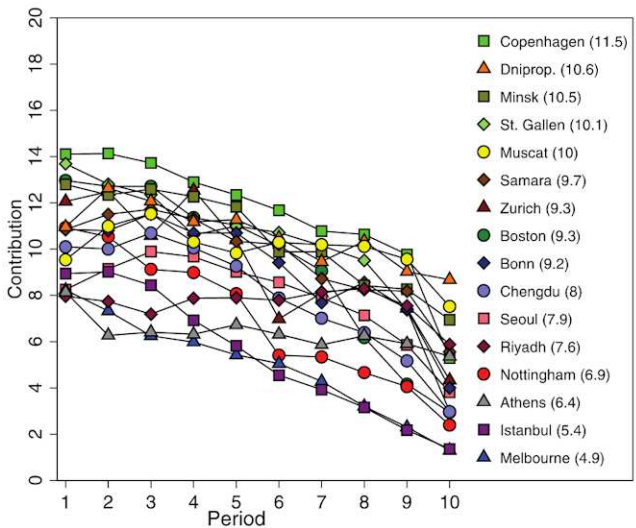,width=8.5cm}}
\caption{Contributions in a repeated public goods game in $15$ countries. Data come from the no punishment treatment of \cite{herrmann_s08}. While there is substantial cross-cultural variation in contributions, contributions decline over time (period) in almost every population. Figure reproduced with permission from \cite{herrmann_s08}.}
\label{countries}
\end{figure}

To truly understand human behavior, however, experiments must investigate play in economic games across diverse cultures. The vast majority of experiments focus on societies that are western, educated, industrialized, rich, and democratic (WEIRD) \cite{henrich_n10}, and thus leave open questions of generalizability and variation. Research on play in the public goods game demonstrates that there is substantial cross-cultural variation in cooperation. In one study of 16 complex and developed societies across the world \cite{herrmann_s08}, rates of cooperation in the first period varied between an average of approximately 70\% contributed (e.g., Copenhagen) to an average of approximately 40\% contributed (e.g., Athens). However, the finding that rates of cooperation declined over time in a repeated game was remarkably consistent across cultures (Fig.~\ref{countries}).

\subsubsection{Cooperation in dyads}
\label{excodyads}
While the public goods game measures contributions to groups, the dictator game and the prisoner's dilemma game, described above, model prosocial behavior between dyads. Play in these games also varies across cultures. For example, the canonical result from the dictator game that play is bimodal: most participants either share half of their endowment, or nothing \cite{engel_ee11}. In other words, equality and complete selfishness are the most common behaviors. However, in a study of small-scale societies, there was sizeable variation both in average sharing, and in the distribution of responses. The Hadza hunter gatherers, for example, show a single mode of sharing 10 percent \cite{henrich_aer01}.

Another popular two-player game in the domain of prosociality is the trust game \cite{berg_geb95}. The trust game models a situation in which one person can trust another, and trust can be met with either trustworthiness or exploitation. The trust game has two players: a trustor and a trustee. The trustor starts with an endowment, and decides how much, if anything, to send to the trustee; anything that gets sent is tripled by the experimenter. Then, the trustee decides how much, if anything, to return. Thus, the trustor stands to gain money if the trustee is trustworthy and will return more than $1/3$, but the trustee always faces an incentive to return nothing. The amount the trustor sends is, therefore, a measure of trust, and the amount the trustee returns is a measure of trustworthiness. In the trust game, trustors show substantial trust of trustees, and substantial trustworthiness is also observed. Furthermore -- implicating reciprocity effects -- larger percentages are returned when the trustor sends to the trustee \cite{pillutla_jes03}. Moreover, evidence suggests that trustors show more trust of trustees who have previously cooperated with others, implicating reputation effects \cite{barclay_06, fehrler_ehb13, jordan_n16}.

\subsection{Measuring the enforcement of prosociality}
\label{exenforce}
In addition to measuring prosocial behavior, experiments are also used to measure the enforcement of prosociality. In other words, are people willing to punish defectors? In economic game experiments, punishment is usually operationalized as the opportunity to pay a cost to impose a larger cost on somebody who has violated a norm, or transgressed in some way. Punishment is thus different than ``spite'' because it is targeted at a defector (as a way to enforce prosociality), rather than a competitor (as a way to increase one's relative payoff).

\subsubsection{Punishment in public goods games}
\label{expunishment}

\begin{figure}
\begin{center}
\centerline{\epsfig{file=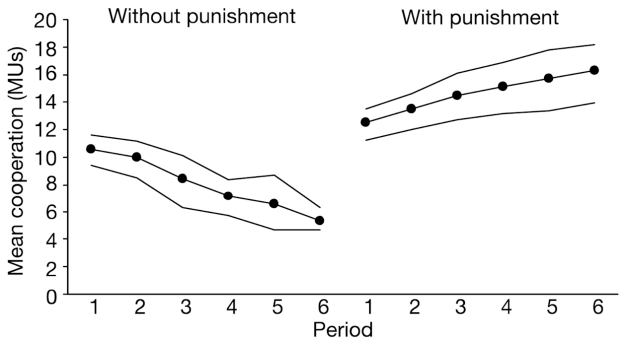,width=8.3cm}}
\caption{Public goods game contributions without and with punishment. When punishment is not possible, contributions decline over time. In contrast, when punishment is possible, contributions increase over time. Figure reproduced with permission from \cite{fehr_n02}.}
\label{wwpun}
\end{center}
\end{figure}

A canonical paradigm involves modifying the public goods game to include the opportunity for punishment \cite{falkinger_aer00, fehr_n02}. After each round, subjects learn how much each member of their group contributed, and then have the opportunity to punish other players. Typically, a substantial proportion of subjects are willing to pay to punish, and punishment is systematically targeted at low contributors. Furthermore, as illustrated in Fig.~\ref{wwpun}, including the opportunity for punishment increases public goods game contributions -- even in the first round of a repeated public goods game. This result demonstrates the power of punishment to promote cooperation, and shows that deterrence starts operating even before defectors have the chance to personally experience punishment.

However, the positive effect of punishment on cooperation is not universal across cultures. In some societies, people engage in anti-social punishment, or punishment targeted at cooperators \cite{herrmann_s08}. Theoretical models show that natural selection can actually favor anti-social punishment \cite{rand_jtb10, rand_nc11}, and empirical work shows that anti-social punishment is more common in societies with relatively weak norms of civic cooperation and rule of law \cite{gachter_prsb10, herrmann_s08}. Furthermore, anti-social punishment can actually prevent punishment from functioning to deter defectors (because it is still possible to be punished after cooperating) \cite{herrmann_s08}.

Nonetheless, many people -- especially in societies with relatively strong cooperative norms -- are willing to pay to punish defectors in the public goods game with punishment. This finding demonstrates a drive to reduce the payoffs of non-contributors. But what is the basis for this desire to punish? In the public goods game, an individual who punishes a defector is retaliating against somebody who has harmed both himself and other group members. Thus, it is unclear whether punishment in the public goods game reflects a drive to engage in retaliation (``second-party punishment'', or punishment by the individual who has been harmed) or a more impartial desire to enforce norms of good behavior (``third-party punishment'', or punishment by an unaffected observer).

\subsubsection{Second-party punishment}
\label{ex2pun}
Other punishment games allow researchers to tease apart motivations for retaliation versus norm enforcement. The ultimatum game is the canonical game used to measure second-party punishment \cite{guth_jebo82}. In the ultimatum game, the payoff-maximizing action for the responder is to accept any non-zero offer: accepting means getting something, rather than nothing. However, if the proposer makes a small, unfair offer, rejecting is a way to punish the proposer. It is costly to the responder, but more costly to the proposer (because when the proposer makes an unfair offer, she has more to gain from it being accepted than the responder does).

As noted above, proposers in the ultimatum game offer more money than dictators in the dictator game, reflecting the power of punishment to motivate prosociality \cite{guth_jebo82}. Indeed, this proposer behavior is rational: the payoff-maximizing offer in the ultimatum game is typically substantially higher than zero (and can be as high as 50\% in some studies \cite{stagnaro_sp17}), because responders are willing to pay to punish selfishness by rejecting low offers. Generally, about half of responders reject offers below 30\% \cite{camerer_03}. However, there is substantial cross cultural variation in ultimatum game play, with some populations almost universally rejecting very unfair offers (e.g., the Gusii) and others almost always accepting them (e.g., the Shuar) \cite{henrich_aer01, henrich_s06b}. Nevertheless, in almost all societies, there is some rejection of unfairness, and rejection rates decline as offer size increases.

\subsubsection{Third-party punishment}
\label{ex3pun}

\begin{figure}[b]
\centerline{\epsfig{file=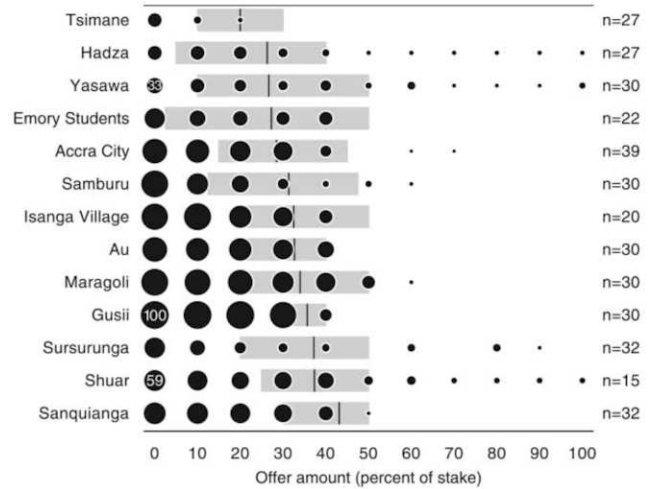,width=8.5cm}}
\caption{Punishment in a third-party punishment game as a function of offer amount across $15$ small-scale societies, with areas of the black bubbles showing the fraction of player 2's who were willing to punish that offer. Also shown are offers, with solid vertical lines marking the mean offer, and gray shaded rectangle highlighting the interquartile of offers. While there is substantial cross-cultural variation, all societies show some third-party punishment of unfairness, and declining punishment as offers move from zero to half of the stake. Punishment of hyper-generous offers (i.e., greater than half) is almost never observed. Figure reproduced with permission from \cite{henrich_s06b}.}
\label{3rdparty}
\end{figure}

To measure third-party punishment, the canonical approach is to modify the dictator game by adding a third-party punishment stage \cite{fehr_ehb04}. Specifically, a third player is given an endowment, and can sacrifice some of this endowment to take money away from the dictator. This third player can condition his punishment on how much the dictator shared with the recipient. Experiments show that providing the option for third-party punishment increases prosociality \cite{balafoutas_el4,  charness_jebo08, jordan_n16}. Moreover, as in the ultimatum game, some third-party punishment of selfishness is near-universal across cultures, and punishment declines as dictators transfer more to recipients (Fig.~\ref{3rdparty}) \cite{henrich_s06b}. Thus, people are willing to pay personal costs to punish selfishness, even when they haven't been directly harmed or affected in any way. However, comparing rates of second- and third-party punishment reveals that the drive to retaliate when harmed directly is generally stronger than the drive to punish the mistreatment of others (e.g., \cite{fehr_ehb04}).

Recent research provides evidence that third-party punishment may be motivated by reputational benefits \cite{barclay_06}. Just as trustors in the trust game send more to trustees who have previously cooperated, they also send more to trustees who have previously engaged in third-party punishment \cite{barclay_06, horita_ebs10, jordan_n16}. These reputation benefits also appear to motivate punishment: third parties are more likely to punish when their behavior is observable \cite{kurzban_ehb07}. Moreover, rates of third-party punishment are reduced when potential punishers have the opportunity to signal their trustworthiness more directly by sharing with others \cite{jordan_n16}, as predicted by theoretical models of third-party punishment as a costly signal of trustworthiness \cite{jordan_n16, jordan_jtb17}.

\subsubsection{Rewarding}
\label{exrewarding}
Lab experiments can also be used to study reward, rather than punishment. For example, public goods games can be modified by allowing subjects to pay to reward members of their group (i.e., by interspersing rounds of the public goods game with the prisoner's dilemma game) \cite{rand_s09, sefton_ei07}. These experiments show that people preferentially reward those who contribute to the group, and that this rewarding can sustain cooperation when members of the same group interact repeatedly \cite{rand_s09} -- even when groups are very large and each player can only be rewarded by one or two others \cite{hauser_srep16}. Games involving punishment can also be modified to include both reward and punishment options, in which case both options are typically used \cite{almenberg_pp17, pedersen_prsb13, rand_s09}. Evidence suggests that in some circumstances, people prefer compensating victims to punishing transgressors \cite{feldmanhall_nc14}.

\subsection{Experiments on network effects}
\label{exnetworks}
Economic game experiments have also been used to explore the impact of non-random interaction on cooperation. A great deal of theoretical work has suggested that cooperation in the prisoner's dilemma can evolve when populations are structured into interaction networks - certain population structures can support cooperation in contexts where well-mixed populations cannot (for a review, see \cite{nowak_pt10}). In apparent contrast to this theory, however, numerous experiments failed to stable cooperation when players were arranged over network structures \cite{gracia-lazaro_pnas12, grujic_pone10, grujic_srep12, grujic_pone12, traulsen_pnas10}. A closer inspection of the theory reveals a potential explanation for this mismatch: one particularly influential line of theoretical work found that cooperation is only evolutionary stable on networks when the benefit-to-cost ratio of cooperation is larger than the average number of network neighbors \cite{ohtsuki_n06} (see also \cite{lieberman_n05, allen2017evolutionary}), and this condition was not satisfied by most experiments. Indeed, a more recent set of experiments show that fixed networks can stabilize cooperative behavior, but only when this condition is satisfied \cite{rand_pnas14}.

\begin{figure}
\centerline{\epsfig{file=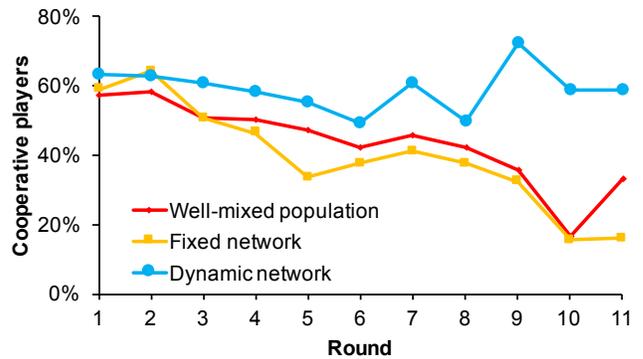,width=8.5cm}}
\caption{Cooperation as a function of network structure. Data come from \cite{rand_pnas11}. Subjects play a repeated cooperation game -- in which they choose whether to pay to cooperate with all of their neighbors -- within either a well-mixed population, a fixed network, or a dynamic network where 30\% of pairs can update their network connection status each round. While cooperation declines over time in fixed networks and well-mixed populations, it is stably maintained in dynamic networks where people have control over whom they are connected to.}
\label{exnets}
\end{figure}

Another body of experimental work on networks considers the act that network structure is often dynamic: individuals have control over who they are connected to, and can make or break connections. These choices can thus be used as a way to reward cooperative behavior (by forming or maintaining connections with cooperators) or punish selfishness (by avoiding or breaking connections with defectors). Theoretical models predict that dynamic networks should be highly effective at promoting cooperation (for a review, see \cite{perc_bs10}), and indeed, this prediction is validated by experimental work. When networks are dynamic, people preferentially connect to cooperators, resulting in stable cooperation (Fig.~\ref{exnets}) \cite{rand_pnas11, shirado_nc13, wang_j_pnas12}. Interestingly, the positive effect of dynamic networks on cooperation appears to be driven largely by the fact that defectors are willing to try switching to cooperation, even when their neighbors are defecting, in order to attract new partners -- rather than by cooperators being willing to maintain cooperation when their neighbors are defecting \cite{jordan_pone13}.

\subsection{Pitfalls and the experiment-model divide}
\label{expitfall}
We conclude our review of human experiments by highlighting a common pitfall associated with using experiments to test whether particular mechanisms for the evolution of cooperation are actually effective at promoting human prosociality in the real world. It is common for experiments to test a model by exactly recreating the incentive structure present in that model in the lab, and then measuring human behavior. However, in such experiments, people can use domain-general strategic reasoning to determine what is in their self-interest -- which will, by design, be to behave as predicted by the model. Thus, positive results in such experiments do not necessarily mean that the mechanism in question actually operates to promote cooperation outside of the laboratory (but merely that the mechanism can promote cooperation when implemented).

For example, the fact that people engage in reciprocal cooperation in repeated prisoner's dilemma games does not necessarily mean that direct reciprocity operates to promote cooperation in daily life. While this is certainly possible, it is also possible that subjects behave as predicted by reciprocity models because they use reason to determine that in the experiment, reciprocal cooperation is payoff-maximizing. Thus, experimenters need to test predictions that can only be true if the mechanism operates in daily life. For example, researchers should search for evidence that people have preferences, or show behavioral patterns, that would be beneficial if the mechanism operated in daily life, but are not payoff-maximizing within the laboratory experiment -- such as responding reciprocally even in non-repeated games where there is no financial incentive to do so \cite{fischbacher_el01}, and doing so in an intuitive way which is undermined by careful deliberation and consideration of the details of the laboratory setup \cite{rand_n12, rand2016cooperation}.

As another example, consider models suggesting that third-party punishment can serve as a costly signal of trustworthiness, because the costs of punishing are lower for individuals who face incentives to be trustworthy than for individuals who face incentives to exploit others \cite{jordan_n16, jordan_jtb17}. Consistent with the hypothesis that this mechanism operates in daily life, economic game experiments show that people who engage in third-party punishment are trusted more than non-punishers \cite{barclay_06, horita_ebs10} -- and actually are more trustworthy \cite{jordan_n16} -- in a one-shot trust game. Because the trust game is not repeated, it is always payoff-maximizing for the trustee to return nothing, and thus for the truster to send nothing. Furthermore, nothing is built into the incentive structure of these experiments to create a link between the cost of punishment (which is always the same for all subjects) and the incentives to be trustworthy (of which there are never any for any subjects).

Nonetheless, people expect individuals who punish to return more money, and they actually do -- even though doing so is strictly costly within the game. These experiments thus provide evidence that punishment operates as a signal of trustworthiness in daily life, and is unlikely to be explained by subjects reasoning about what is payoff-maximizing within the experiment. In contrast, domain-general reasoning could lead subjects to the predicted results if the experiments were designed to exactly recreate the model (e.g., by giving some subjects an incentive to return money in the trust game and a reduced cost of punishment). In sum, then, designing experiments to test theoretical models in ways that shed interesting light on human behavior is an important challenge.

\section{Mathematical models}
\label{mathmodels}

\begin{figure*}
\begin{center}
\centerline{\epsfig{file=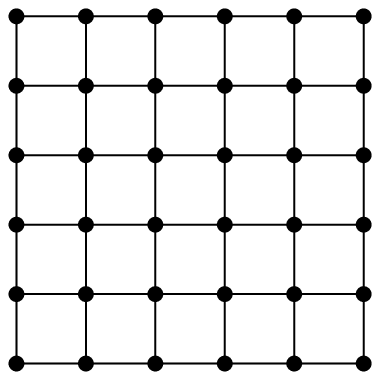,width=3.9cm}\epsfig{file=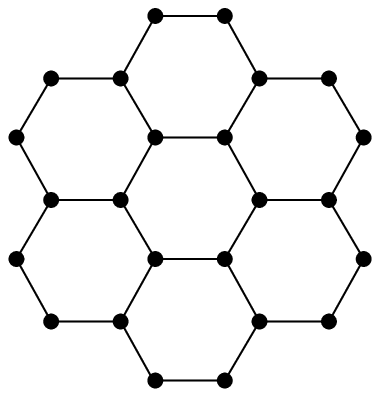,width=3.9cm}\epsfig{file=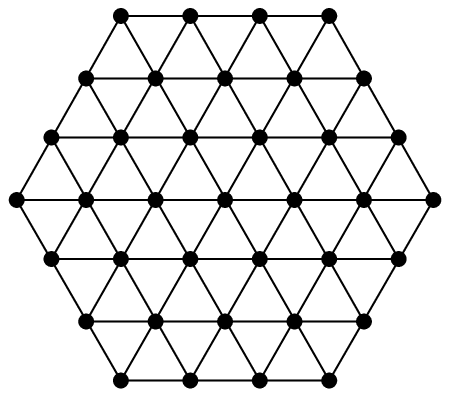,width=3.9cm}\epsfig{file=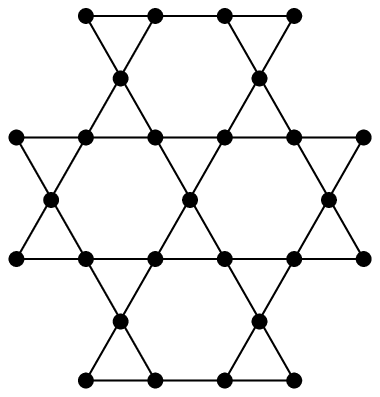,width=3.9cm}}
\caption{Different types of lattices. On the square lattice (left) each site has four nearest neighbors, while on the honeycomb lattice (second from left) each site has three. In both cases the clustering coefficient ${\cal C}$ is zero. The triangular (second from right) and the kagom{\'e} lattice (right) both feature percolating overlapping triangles, having clustering coefficients ${\cal C}=2/5$ and ${\cal C}=1/3$, respectively.}
\label{lattices}
\end{center}
\end{figure*}

While the use of statistical physics for the better understanding of human cooperation can be considered a relatively recent development, evolutionary game theory \cite{sigmund_93, weibull_95, hofbauer_98, nowak_06, sigmund_10} is long established as the theory of choice for studying the evolution of cooperation among selfish individuals, including humans \cite{rand_n12, gracia-lazaro_pnas12, rand_tcs13, rand_pnas14, capraro2015social}. Competing strategies vie for survival and reproduction through the maximization of their utilities, which are traditionally assumed to be payoffs that are determined by the definition of the contested game. The most common assumption underlying the evolution in structured populations has been that the more successful strategies are imitated and thus spread based on their success in accruing the highest payoffs. Evolutionary dynamics based on these basic principles is considered as the main driving force of evolution, reflecting the individual struggle for success and the pressure of natural selection.

Before presenting specific games to model human cooperation, a note is in order regarding the networks that ought to describe our interactions. In statistical physics, lattices are used to describe structured interactions among particles in the simplest terms \cite{binder_88, newman_99}. Examples of the most frequently employed lattices are presented in Fig.~\ref{lattices}. Despite their dissimilarity with social networks that describe how we are actually connected with one another \cite{wasserman_94, christakis_09}, lattices enjoy remarkable popularity in human cooperation research \cite{gracia-lazaro_pnas12, rand_pnas14}, with the square lattice also being used for the seminal discovery of network reciprocity \cite{nowak_n92b}.

There are compelling reasons for this being the case. A lattice is the simplest of networks that allows us to go beyond the well-mixed population assumption, and as such it allows us to take into account the fact that the interactions among humans are limited to a finite number of partners, and that these interactions are also inherently structured rather than random. Moreover, as is well known, the structure of a network can significantly affect evolutionary outcomes \cite{szabo_pr07, roca_plr09, perc_bs10}. Frequently, we want to eschew these effects in order to focus on the mechanisms that may stem from sources other than the properties of the interaction network. In general, lattices can be regarded as an even field for all competing strategies, and especially for games that are governed by group interactions, such as the public goods game \cite{perc_jrsi13}, using the square lattice suffices to reveal all feasible evolutionary outcomes. Unless stated otherwise, all the results reviewed in what follows were obtained on one of the regular lattices depicted in Fig.~\ref{lattices}.

\subsection{Public goods game as the null model}
\label{nullmodel}
The public goods game is simple and intuitive, as already described in Section~\ref{humex}, and it is in fact a generalization of the pairwise prisoner's dilemma game to group interactions \cite{perc_jrsi13}. In a group of players, each one can decide whether to cooperate or defect. Cooperators ($C$) contribute $c=1$ to the common pool, while defectors ($D$) contribute nothing. The sum of all contributions is multiplied by a multiplication factor $r>1$, which takes into account synergistic effects of cooperation. In particular, there is an added value to a joint effort that is often more than just the sum of individual contributions. After the multiplication, the resulting amount of public goods is divided equally amongst all group members, irrespective of their strategy. In a group $g$ containing $G$ players the resulting payoffs are thus
\begin{eqnarray}
\Pi_C^{g} &=& r(N_C+1)/G-1,\\
\Pi_D^{g} &=& rN_C/G,
\end{eqnarray}
where $N_C$ is the number of cooperators around the player for which the payoff is calculated. Evidently, the payoff of a defector is always larger than the payoff of a cooperator, if only $r<G$. With a single parameter, the public goods game hence captures the essence of a social dilemma in that defection yields highest short-term individual payoffs, while cooperation is optimal for the group, and in fact for the society as a whole. If nobody cooperates public goods vanish and we have the tragedy of the commons \cite{hardin_g_s68}.

In a well-mixed population, where groups are formed by selecting players uniformly at random, $r=G$ is a threshold that marks the transition between defection and cooperation. If players imitate strategies of their neighbors with a higher payoff, then for $r<G$ everybody defects, while for $r>G$ everybody in the population cooperates.

Interactions among humans, however, are seldom random, and it is therefore important for the null model to take this into account. The square lattice, as already argued above, is among the simplest of networks that one can consider. Thus, let the public goods game be staged on a square lattice with periodic boundary conditions where $L^2$ players are arranged into overlapping groups of size $G=5$ such that everyone is connected to its $G-1$ nearest neighbors. In this case a player $x$ obtains its payoffs by playing the public goods game with its $G-1$ partners as a member of all the $g=1, \ldots, G$ groups where it is member. Its overall payoff $\Pi_{s_x}$ is thus the sum of all the payoffs $\Pi_{s_x}^{g}$ acquired in each individual group.

This null model -- the spatial public goods game -- has been studied in detail in \cite{szolnoki_pre09c}, where it was shown that, for an intermediate selection intensity, cooperators survive only if $r>3.74$, and they are able to defeat defectors completely for $r>5.49$. Both the $D \to (C+D)$ and the $(C+D) \to D$ phase transition are continuous. Subsequently, the impact of critical mass \cite{szolnoki_pre10, archetti_jtb12}, i.e., the evolution of cooperation under the assumption that the collective benefits of group membership can only be harvested if the fraction of cooperators within the group exceeds a threshold value, and the effects of different group sizes \cite{isaac_qje88, janssen_jtb06, szolnoki_pre11c}, have also been studied in detail.

In general, it is important that in structured populations, due to network reciprocity, cooperators are able to survive at multiplication factors that are well below the $r=G$ limit that applies to well-mixed populations. The $r>3.74$ threshold for cooperators to survive on the square lattice can be considered as a benchmark value, below and above which we have harsh and lenient conditions for the evolution of human cooperation, respectively.

\subsection{Punishment}
Punishment is a form of retaliation or negative reciprocity that, in whichever form, entails paying a cost for somebody else to incur a cost \cite{sigmund_tee07}. In peer punishment, individual players take it upon themselves to punish defectors. In pool punishment, those willing to punish invest into a common pool, from where resources are taken when it is time to punish a defector. Antisocial punishment is a corrupt form of sanctioning, where punishers exploit their status by punishing those that behave prosocially \cite{herrmann_s08, helbing_ploscb10, rand_nc11}. In the latter case, punishment simply becomes a self-interested tool for protecting oneself against potential competitors. In what follows, we review commonly used variants of the spatial public goods game with punishment.

\subsubsection{Peer punishment}
\label{peerpunmodel}
The null model introduced above can be easily upgraded to account for peer punishment \cite{fehr_n02, brandt_prsb03, gardner_a_an04}. Cooperators that punish defectors $(s_x=P)$ can be introduced as the third competing strategy. In this case, both cooperative strategies ($C$ and $P$) contribute $c=1$ to the common pool, while defectors contribute nothing. Moreover, a defector is fined with $\beta/(G-1)$ from each punishing cooperator within the group, which in turn requires each punisher to bear the cost $\gamma/(G-1)$ for each defector that is punished. A defector thus suffers the maximal fine $\beta$ if it is surrounded solely by punishers ($N_P=G-1$), while a lonely punisher bears the largest cost $\gamma$ if it is surrounded solely by defectors ($N_D=G-1$). We note that $\beta$ and $\gamma$ are introduced normalized with the number of other players in each group $(G-1)$, simply to facilitate comparisons with results obtained on other interaction networks or by using differently sized groups (this is the case also in all subsequently reviewed models in the Sections that follow). In agreement with these rules, the payoff values of the three competing strategies obtained from each group $g$ are
\begin{eqnarray}
\Pi_C^{g} &=& R(N_C+N_P+1)/G-1,\\
\Pi_P^{g} &=& R(N_C+N_P+1)/G-1-\gamma N_D /(G-1),\\
\Pi_D^{g} &=& R(N_C+N_P)/G-\beta N_P /(G-1),
\end{eqnarray}
where $N_{s_x}$ denotes the number of players with strategy $s_x$ around the player for which the payoff is calculated. For further details on the public goods game with peer punishment we refer to \cite{helbing_njp10, brandt_prsb03, helbing_pre10c, helbing_ploscb10}.

\subsubsection{Pool punishment}
\label{poolpunmodel}
Pool punishment is synonymous to institutionalized punishment, where the contributions of punishers are meant to cover the costs of institutions like the police or other elements of the justice system independently of their necessity or efficiency \cite{sigmund_n10}. The situation thus changes in comparison to peer punishment, where the punishers pay the cost of punishment only if it is necessary, i.e., when the defectors are identified in the group and sanctioned. In the absence of defectors the income of peer-punishers is thus identical to that of traditional cooperators. On the other hand, because of their permanent contributions to the punishment pool the income of pool punishers is always smaller than that of cooperators.

For the public goods game to accommodate pool (institutionalized) punishment \cite{szolnoki_pre11}, we introduce cooperators that punish defectors, whereby taking resources from the common pool. The pool-punishers ($s_x = O$), like traditional cooperators ($s_x = C$), contribute a fixed amount $c=1$ to the common pool, while defectors contribute nothing. As always, the sum of all contributions in each group is multiplied by the multiplication factor $r>1$ and equally divided among group members. In addition, pool punishment requires precursive allocation of resources, and therefore each pool punisher contributes an amount $\gamma$ to the punishment pool irrespective of the strategies in its neighborhood. Defectors, on the other hand, must bear the punishment fine $\beta$, but only if there is at least one pool punisher present in the group. Denoting the total number of cooperators ($C$), pool-punishers ($O$) and defectors ($D$) in a given group $g$ by $N_{C}$, $N_{O}$ and $N_{D}$, respectively, the payoffs
\begin{eqnarray}
\Pi_{C}^g &=& r(N_{C}+N_{O})/G - 1, \\
\Pi_{O}^g &=& \Pi_{C}^g - \gamma , \\
\Pi_{D}^g &=& r(N_{C}+N_{O})/G-\beta f(N_{O}),
\end{eqnarray}
are obtained by each player $x$ depending on its strategy $s_x$, where the step-like function $f(Z)$ is $1$ if $Z>0$ and $0$ otherwise. For further details on the public goods game with pool punishment we refer to \cite{szolnoki_pre11, szolnoki_pre11b, sasaki_srep15}.

\subsubsection{Self-organized punishment}
\label{adaptivepunmodel}
Models presented in above Sections~\ref{peerpunmodel} and \ref{poolpunmodel} assume that, once set, the fine and cost of punishment do not change over time. By allowing players to adapt their sanctioning efforts in dependence on the success of cooperation, we arrive at a model with self-organized punishment. More precisely, on top of the traditional two-strategy public goods game presented in Section~\ref{nullmodel}, here each player is assigned an additional parameter $\mu_x$ keeping score of its punishing activity. Initially $\mu_x=0$ for all players. Subsequently, whenever a defector succeeds in passing its strategy, all the remaining cooperators in all the groups containing the defeated cooperator increase their punishing activity by one, i.e., $\mu_x=\mu_x+1$.

We emphasize that the presence of defectors alone never triggers an increase in $\mu_x$. Only when defectors spread do the cooperators resort to sanctioning. Conversely, if defectors fail to spread, then at every second round all cooperators decrease their punishing activity by one, as long as $\mu_x \geq 0$. We emphasize that if the population contains cooperators with $\mu_x>0$, then all the cooperators having $\mu_x=0$ become second-order freeriders \cite{fehr_n04}. Due to the presence of punishers, i.e., cooperators having $\mu_x>0$, the accumulation of payoffs changes as well. In particular, each defector is fined with an amount $\pi_x \Delta/(G-1)$ from every punishing cooperator that is a member of the group, while at the same time the punishing cooperators that execute the punishment bear the cost $\pi_x \alpha \Delta/(G-1)$ for every defector punished. Here $\Delta$ determines the incremental step used for the punishing activity and $\alpha$ is a free parameter determining whether the sanctions are costly ($\alpha>1$) or not ($\alpha<1$). Taking adaptive punishment into account, the payoff of player $x$ in a given group $g$ is thus
\begin{eqnarray}
\Pi_C^g&=&r \frac{N_C}{G} - 1 - \frac{1}{(G-1)}N_D \mu_x \alpha \Delta \,\,\,{\rm if} \,\,s_x=C,\\
\Pi_D^g&=&r \frac{N_C}{G} - \frac{1}{(G-1)} \sum_{y \in g} \mu_y \Delta \,\,\,\,\,\,\,\,\,\,\,\,\,\,\,\,{\rm if} \,\,s_x=D.
\end{eqnarray}
In above two equations $N_C$ and $N_D$ are the total numbers of cooperators and defectors in the group $g$, respectively. For further details on the public goods game with self-organized punishment we refer to \cite{perc_njp12}.

\subsection{Rewarding}
Rewarding is common in human societies as a sign of positive reciprocity towards well-behaved, prosocial, or otherwise kind behavior \cite{sigmund_pnas01, hilbe_prsb10}. Contrary to punishment, rewarding entails paying a cost for somebody else to incur a benefit. In peer rewarding, individual players take it upon themselves to reward other cooperators, whereas the institutionalized variant of this behavior foresees investments into a common pool that are then used for rewarding cooperators. Similarly to antisocial punishment, antisocial rewarding seeks to sway unworthy recipients into actions that benefit those that reward \cite{dos-santos_m_prsb15, szolnoki_prsb15}. In what follows, we review commonly used variants of the spatial public goods game with rewarding.

\subsubsection{Peer rewarding}
\label{peerewmodel}
To accommodate peer rewarding, the null model presented in Section~\ref{nullmodel} obtains rewarding cooperators ($s_x =R$) as the third competing strategy. Using standard parametrization used thus far, the two cooperating strategies ($C$ and $R$) contribute $c=1$ to the public good while defectors contribute nothing, and the sum of all contributions is multiplied by the multiplication factor $r>1$ and then shared equally among all $G$ group members. In addition, here each cooperator ($C$ or $R$) receives the reward $\beta/(G-1)$ from every rewarding cooperator that is a member of the group, and every rewarding cooperator from this group therefore bears an additional cost $\gamma/(G-1)$. Denoting the number of cooperators, defectors, and rewarding cooperators around the player for which the payoff is calculated by $N_C$, $N_D$, and $N_R$, respectively, the payoffs for each group $g$ are
\begin{eqnarray}
\Pi_C^g &=& r(N_C+N_R+1)/G - 1 + \beta N_R/(G-1),\\
\Pi_R^g &=& \Pi_C - \gamma (N_C+N_R)/(G-1)\,,\\
\Pi_D^g &=& r(N_C+N_R)/G\,.
\end{eqnarray}
For further details on the public goods game with rewarding we refer to \cite{szolnoki_epl10}.

\subsubsection{Pool rewarding}
\label{poolrewmodel}
In comparison to pool punishment, pool rewarding is a significantly more scarcely phenomenon in human societies. There are plenty of institutions that enforce the law, but none officially that would reward citizens for upholding it. As such, pool rewarding exists mainly in certain subcultures of the population, where it is also often mixed with the antisocial variant of the same phenomenon (for example in criminal organizations, where successful members are rewarded if they have done their crime right) \cite{szolnoki_prsb15}. We therefore present a model that incorporates both prosocial and antisocial pool rewarding. In parallel to the traditional version of the public goods game entailing cooperators ($C$) and defectors ($D$), two additional strategies run an independent pool rewarding scheme. These are rewarding cooperators ($R_C$) and rewarding defectors ($R_D$), who essentially establish a union-like support to aid akin players.

Accordingly, rewarding cooperators contribute $c=1$ to the prosocial rewarding pool. The sum of all contributions in this pool is subsequently multiplied by the synergy factor $r_2>1$, and the resulting amount is distributed equally amongst all $R_C$ players in the group. Likewise, at each instance of the public goods game all rewarding defectors contribute $c=1$ to the antisocial rewarding pool. The sum of all contributions in this pool is subsequently multiplied by the same synergy factor $r_2>1$ that applies to the prosocial rewarding pool, and the resulting amount is distributed equally amongst all $R_D$ players in the group. Here the focus is thus on the consequences of union-like support to akin players, without considering second-order free-riding. It is therefore important that we consider strategy-neutral pool rewarding in that individual contributions to the prosocial and the antisocial rewarding pool are the same ($c=1$), as is the multiplication factor $r_2$ that is subsequently applied. Otherwise, if an obvious disadvantage would be given to either the prosocial or the antisocial rewarding pool, the outcome of the game would become predictable. We also emphasize that, in order to consider the synergistic consequence of mutual efforts and to avoid self-rewarding of a lonely player \cite{brandt_pnas06}, $r_2=1$ is always applied if only a single individual contributed to the rewarding pool. For further details on the public goods game with pool rewarding we refer to \cite{szolnoki_prsb15, sasaki_jtb11}.

\subsubsection{Self-organized rewarding}
\label{adaptiverewmodel}
Models presented in above Sections~\ref{peerewmodel} and \ref{poolrewmodel} assume that, once set, the fine and cost of rewarding do not change over time. By allowing players to adapt their rewarding efforts in dependence on the success of cooperation, we arrive at a model with self-organized rewarding. More precisely, in addition to cooperators ($s_x = C$) and defectors ($s_x = D$), the game is contested also by rewarding cooperators ($s_x = R$). Each rewarding cooperator received an additional parameter $\mu_x$, which keeps score of its rewarding activity. While this parameter is initially zero, subsequently, whenever a defector succeeds in passing its strategy, all the remaining rewarding cooperators in all the groups containing the defeated player increase their rewarding activity by one, i.e., $\mu_x=\mu_x+1$. The related costs increase accordingly. However, to maintain the latter is unwanted, and hence at every second round all rewarding cooperators decrease their rewarding activity by one, as long as $\mu_x \geq 0$. The payoff of player $x$ adopting $s_x = C$ in a given group $g$ of size $G$ is thus
\begin{equation}
\Pi_C^g=r \frac{N_C+N_R+1}{G} - 1 + \frac{\Delta}{(G-1)} \sum_{i \in g}\mu_i,
\end{equation}
where $N_C$, $N_D$ and $N_R$ are the number of other cooperators, defectors and rewarding cooperators in the group $g$, respectively. The sum runs across all the neighbors in the group, while $\pi_i$ is the actual rewarding activity of player $i$. The corresponding payoff of a rewarding cooperator at site $x$ is
\begin{equation}
\Pi_R^g=P_C^g - \frac{\alpha \Delta}{(G-1)} \pi_x (N_C+N_R),
\end{equation}
while a defector, who's payoff is derived exclusively from the contributions of others, gets
\begin{equation}
\Pi_D^g=r \frac{N_C+N_R}{G}.
\end{equation}
It follows that each player adopting $s_x = C$ or $s_x = R$ is rewarded with an amount $\mu_i \Delta/(G-1)$ from every rewarding cooperator, having rewarding activity $\mu_i$, that is a member of the same group. At the same time, each rewarding cooperator bears the cost $\mu_i \alpha \Delta/(G-1)$ for every cooperator that was rewarded. Self-rewarding is excluded. Here $\Delta$ and $\alpha$ are key free parameters, determining the incremental step used for the rewarding activity and the cost of rewards, respectively. Note that $\alpha$ is actually the ratio between the cost of rewarding and the reward that is allotted to cooperators. For further details on the public goods game with self-organized rewarding we refer to \cite{szolnoki_njp12}.

\subsection{Correlated positive and negative reciprocity}
\label{correlatedmodel}
As already emphasized, reciprocity is long considered an important piece of the puzzle of human cooperation. If someone is kind to us, we are kind in return. We reward cooperation. On the other hand, if someone is unfair or exploitative, we tend to retaliate. We punish defection. According to the strong reciprocity hypothesis \cite{gintis_jtb00, fehr_hn02, boyd_pnas03, bowles_tpb04}, positive and negative reciprocity are correlated to give us optimal evolutionary predispositions for the successful evolution of cooperation. But is this really true? Should we reward and punish, or should we do just one of the two, or maybe neither? Recent economic experiments tend to reject the strong reciprocity hypothesis \cite{yamagishi_pnas12, egloff_pnas13}, and everyday experience also leaves us with the impression that people will either reward cooperation or punish defection, but seldom will they do both. Methods of statistical physics can contribute relevantly to the resolution of this so-called ``stick versus carrot'' dilemma \cite{andreoni_aer03, hilbe_prsb10, gaechter_n12, chen_xj_jrsi14}, as demonstrated in \cite{szolnoki_prx13}.

In the public goods game with correlated positive and negative reciprocity, we have defectors ($s_x = D$), cooperators that punish defectors ($s_x = P$), cooperators that reward other cooperators ($s_x = R$), and cooperators that both punish defectors as well as reward other cooperators ($s_x = B$) contesting the game. As always, all three cooperative strategies ($P$, $R$ and $B$) contribute $c=1$ to the public good, while defectors contribute nothing. Moreover, a defector is fined with $\beta/(G-1)$ from each punisher ($P$ or $B$) within the group, which in turn requires each punisher to bear the cost $\gamma/(G-1)$ for each defector that is punished. Similarly, every cooperator is given the reward $\beta/(G-1)$ from every $R$ and $B$ player within the group, while each of them has to bear the cost of rewarding $\gamma/(G-1)$ for every cooperator that is rewarded. In agreement with these rules, the payoff values of the four competing strategies obtained from each group $g$ are
\begin{widetext}
\begin{eqnarray}
\Pi_D^{g} &=& r(N_P+N_R+N_B)/G - \beta (N_P+N_B) /(G-1),\\
\Pi_P^{g} &=& r(N_P+N_R+N_B+1)/G - \gamma N_D /(G-1) + \beta (N_R+N_B) / (G-1) - 1,\\
\Pi_R^{g} &=& r(N_P+N_R+N_B+1)/G - \gamma (N_P+N_R+N_B) /(G-1)+ \beta (N_R+N_B)/(G-1) - 1,\\
\Pi_B^{g} &=& r(N_P+N_R+N_B+1)/G - \gamma + \beta (N_R+N_B) / (G-1) - 1,
\end{eqnarray}
\end{widetext}
where $N_{s_x}$ denotes the number of players with strategy $s_x$ around the player for which the payoff is calculated. For further details on the public goods game with correlated positive and negative reciprocity we refer to \cite{szolnoki_prx13}.

\subsection{Tolerance}
\label{tolerancemodel}
Tolerance implies enduring trying circumstances with a fair and objective attitude. To determine whether evolutionary advantages might be stemming from diverse levels of tolerance in a population, a variant of the spatial public goods game can be devised \cite{szolnoki_pre15, szolnoki_njp16}, where in addition to cooperators ($s_x = C$), defectors ($s_x = D$), and loners ($s_x = L$), tolerant players ($s_x = M_i$) are also present in the population. Here loners are players that simply abstain from the game and settle for a small, but secure payoff instead \cite{hauert_s02}. Previous research has shown, however, that defectors, cooperators, and loners become entailed in a closed loop of dominance \cite{kerr_n02, kirkup_n04, reichenbach_pre06, arenas_jtb11, wang_wx_pre11, szolnoki_jrsif14, groselj_pre15}, which maintains a Red Queen existence of cooperative behavior that on average is no better than if everybody would abstain \cite{hauert_jtb02, semmann_n03}. Tolerant players might yield a more favorable evolutionary outcome, whereby depending on the number of defectors $i$ within a group, a tolerant player can either cooperate in or abstain from a particular instance of the game.

Evidently, there are as many levels of tolerance as there are possible defectors in the group, so that $i=0, \ldots, G-1$. If the number of defectors in a group is smaller than $i$ the player $M_i$ acts as a cooperator, while otherwise it acts as a loner. As such, the value of $i$ determines the level of tolerance a particular $M_i$ player has. The higher the value of $i$, the higher the number of defectors that are tolerated by an $M_i$ player within a group without it refusing cooperation. As the two extreme cases, $i=0$ indicates that an $M_i$ player will always remain in the non-participatory loner state, while $i=G-1$ indicates that an $M_i$ player will switch to a loner state only if all the other neighbors are defectors. Importantly, regardless of the choice an $M_i$ player makes, it always bears the cost $\gamma$ as a compensation for knowing the number of defectors in a group.

As is standard practice, all cooperative strategies contribute a fixed amount $c=1$ to the public good while defectors and loners contribute nothing. The sum of all contributions in each group is multiplied by the synergy factor $r$ and the resulting public goods are distributed equally amongst all the group members that are not loners. Importantly, the $r>1$ factor is applied only if there are at least two contributions made to the common pool from within the group. Otherwise, a lonely contributor is unable to utilize on the synergistic effect of a group effort, and hence $r=1$ applies. It is also a widely accepted protocol that loners, who do not participate in the game, obtain a moderate but secure payoff $\sigma=1$. Moreover, let $N_D$ be the number of defectors within a group. Diverse tolerance thresholds can be introduced by using different $\delta_i$ prefactors, which are $\delta_i=0$ if $N_D \ge i$ and $\delta_i=1$ if $N_D < i$. Hence, the total number of contributors to the common pool is
\begin{equation}
N_{TC} = N_C + \sum_{i=0}^{G-1} \delta_i N_{M_i},
\end{equation}
where $N_s$ denotes the number of players in the group who follow strategy $s$. By using this notation, the payoff values of the competing strategies obtained from each group $g$ are:
\begin{eqnarray}
\Pi_D\,\,\, &=& r\frac{N_{TC}}{N_D+N_{TC}},\\
\Pi_C\,\,\, &=& \Pi_D - 1,\\
\Pi_L\,\,\, &=& \sigma,\\
\Pi_{M_{i}} &=& \delta_i \Pi_C + (1-\delta_i) \sigma - \gamma.
\end{eqnarray}
For further details on the public goods game with diverse tolerance levels we refer to \cite{szolnoki_njp16}. Notably, a simplified version of this model, with only a single type of tolerant players present in the population at any given time, with the number of defectors that are tolerated designated with the threshold $H$, has also been proposed and studied in \cite{szolnoki_pre15}.

\section{Monte Carlo methods}
\label{mcmethods}
The use of computers to solve problems in statistical physics has a long and fruitful history, dating as far back as the Manhattan Project, where analog computers were used so frequently they often broke down. Digital computers, such as the ENIAC (Electronic Numerical Integrator and Computer), were intertwined with nuclear science from the beginning. In fact, one of the first real uses of ENIAC was by Edward Teller, who used the machine in his early work on nuclear fusion reactions \cite{manhattan}. Today, computers are used in practically all areas of physics, and it is indeed difficult to imagine scientific progress without them. Monte Carlo methods form the largest and most important class of numerical methods used for solving statistical physics problems \cite{binder_88, newman_99}.

\subsection{Random sequential strategy updating}
\label{simulatemethod}
When studying evolutionary games in structured populations presented in Section~\ref{mathmodels}, the Monte Carlo simulation procedure is used for random sequential strategy updating. The usage of this established method ensures that the treatment is aligned with fundamental principles of statistical physics, and it enables a comparison of obtained results with generalized mean-field approximations \cite{dickman_pre01, szolnoki_pre02, dickman_pre02, szolnoki_pre05, szabo_pr07} as well as a proper determination of phase transitions between different stable strategy configurations. In what follows, we describe the simulation procedure if the public goods game is staged on a square lattice, although the elementary steps are of course easily adapted to any other interaction network.

\begin{figure*}
\centerline{\epsfig{file=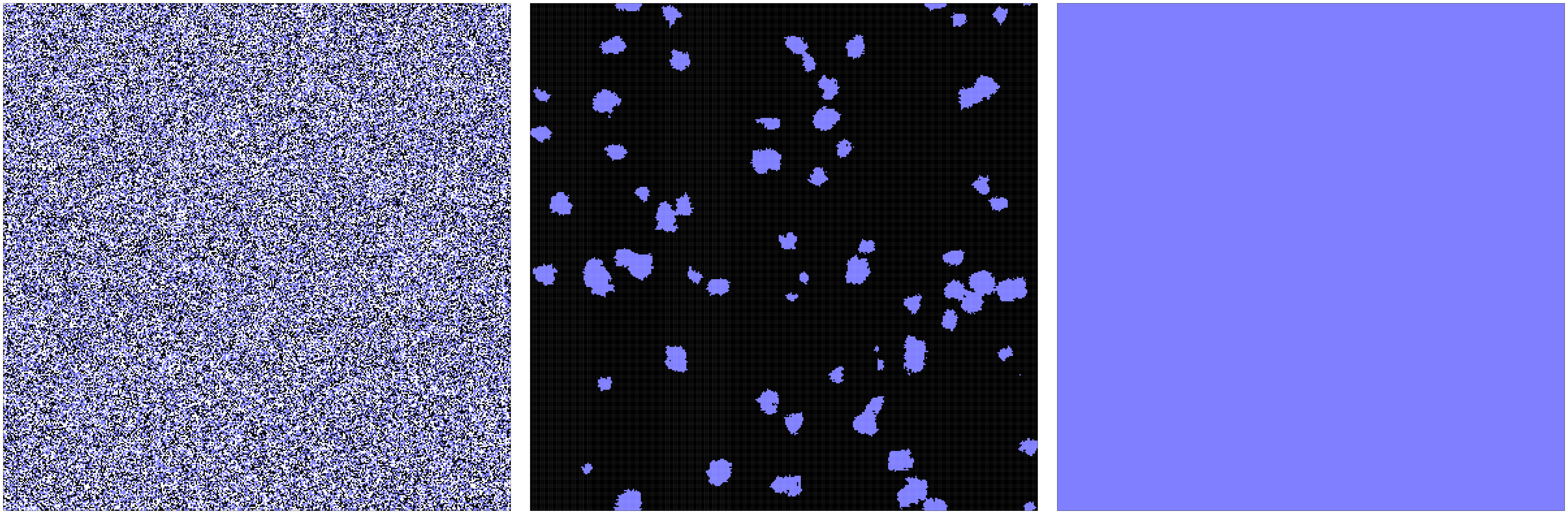,width=16cm}}
\vspace{0.1cm}
\centerline{\epsfig{file=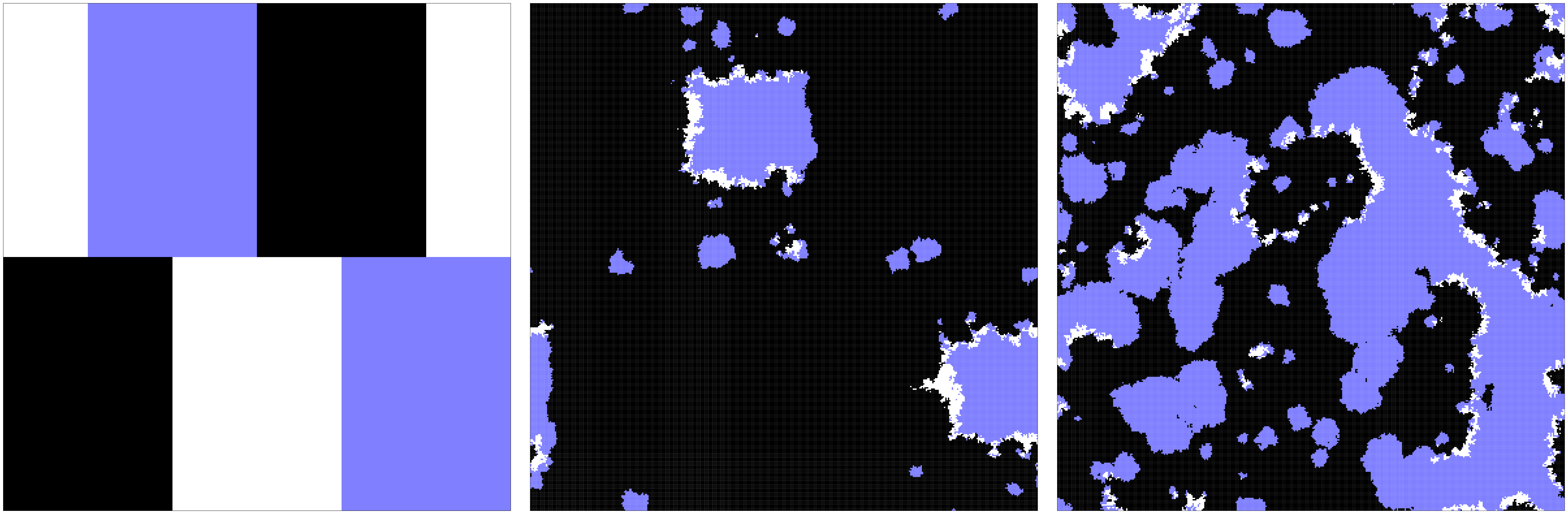,width=16cm}}
\caption{Failure of a random initial state to yield a relaxation to the most stable solution of the public goods game with pool punishment (see Section~\ref{poolpunmodel} for the definition of the model). Presented is the evolution of strategy distribution for two different initial states. Upper row shows the evolution from a random initial state, while the bottom row shows the evolution from a prepared initial state. Black, white, and blue denotes players with defector ($D$), cooperator ($C$), and pool punisher ($O$) strategy, respectively. The correct stationary state is the $D+C+O$ phase in the rightmost bottom panel, not the $O$ phase in the rightmost upper panel. In both cases identical parameters were used, namely $L=390$, $r=2.0$, $\gamma=0.1$, $\beta=0.79$, and $K=0.5$. Figure reproduced with permission from \cite{szolnoki_pre11}.}
\label{preparedinitfig}
\end{figure*}

Initially, all competing strategies are distributed uniformly at random (see also Section~\ref{randominitimethod} below) on a $L \times L$ square lattice with periodic boundary conditions. The microscopic dynamics involves the following elementary steps. First, a randomly selected player $x$ with strategy $s_x$ plays the public goods game with its $G-1$ partners as a member of all the $g=1,\ldots,G$ groups where it is member, whereby its overall payoff $\Pi_{s_x}$ is thus the sum of all the payoffs $\Pi_{s_x}^{g}$ acquired in each individual group. Next, player $x$ chooses one of its nearest neighbors at random, and the chosen co-player $y$ also acquires its payoff $\Pi_{s_y}$ in the same way as previously player $x$. Finally, player $y$ imitates the strategy of player $x$ with a probability given by the Fermi function
\begin{equation}
W(s_x \to s_y)=\frac{1}{1+\exp[(\Pi_{s_y}-\Pi_{s_x}) /K]}\,,
\label{fermi}
\end{equation}
where $K$ quantifies the uncertainty by strategy adoptions \cite{szabo_pre98, szolnoki_pre09c}. In the $K \to 0$ limit, player $y$ copies the strategy of player $x$ if and only if $\Pi_{s_x} > \Pi_{s_y}$. Conversely, in the $K \to \infty$ limit, payoffs seize to matter and strategies change as per flip of a coin. Between these two extremes players with a higher payoff will be readily imitated, although under-performing strategies may also be adopted, for example due to errors in the decision making, imperfect information, and external influences that may adversely affect the evaluation of an opponent. Repeating these elementary steps $L^2$ times constitutes one full Monte Carlo step (MCS), which thus gives a chance to every player to change its strategy once on average.

As an alternative to the above-described imitation dynamics we also mention the logit rule, which is mathematically equivalent to the statistics used in physics to describe the dynamics of spins in a Fermi-Dirac distribution \cite{glauber_jmp63, binder_prb80}. The logit rule is also known as the myopic best response rule in evolutionary game theory \cite{szabo_jtb12, amaral2017role}. According to the logit rule, a player will change its strategy $s_x$ to another randomly selected strategy $s_x^{\prime}$ with probability
\begin{equation}
W(s_x^{\prime} \to s_x) =\frac{1}{1+\exp[(\Pi_{s_x}-\Pi_{s_x^{\prime}}) /K]}\,,
\label{logit}
\end{equation}
where $\Pi_{s_x^{\prime}}$ is the player's new payoff if it changed to the other strategy while the strategies of other players in the groups where he is member remain unchanged. Unlike with imitation, here a player is basically asking himself what would be the benefits of changing his strategy, even if there is nobody with a different strategy around. The logit rule thus yields an innovative dynamic since new strategies can appear spontaneously \cite{traulsen_pnas09}, and recently many have considered this particular updating dynamics as it can leads to different results if compared to imitation dynamics \cite{sysiaho_epjb05, roca_pre09, szabo_pre10, szabo_jtb12b, wang_z_srep12, vilone2014social, szolnoki_pre14, szolnoki_srep14, amaral2016stochastic}. Instead of the reproduction of the fittest returned by imitation dynamics, the logit rule is more akin to a rational analysis of a particular situation, and as such it may be particularly applicable to model human behavior. Nevertheless, the logit rule and its variants have been seldom considered in the realm of the public goods game, which is why we focus our review on results obtained with the imitation dynamics.

Regardless of which microscopic dynamics is used, the average density or fraction of each particular strategy in the population ($\rho_{s_x}$) is determined in the stationary state, after a sufficiently long relaxation time, i.e., when the average fraction of the strategies becomes time independent. Depending on the actual conditions, such as the proximity to phase transition points and the typical size of emerging spatial patterns, the linear system size has to be varied from $L=400$ to $7200$, and the relaxation times range from $10^3$ to $10^6$ MCS. That is, if one wants to ensure that the statistical error of results is small, for example comparable with the line thickness in the figures, although precise values depend on each individual case. It is also important to note that random initial conditions may not necessarily yield a relaxation to the most stable solution of the game, even at a large system size. Therefore, to verify the stability of different solutions, the usage of specially prepared initial conditions is often necessary, or at least recommended, as shown in Fig.~\ref{preparedinitfig}. More information on the later issue is provided in Section~\ref{subsystemethod} below.

\subsection{Random initial conditions}
\label{randominitimethod}
This subsection is devoted to clarifying an important misconception that has to do with the use of random initial conditions in Monte Carlo simulations of evolutionary games in structured populations. It is often written that strategies are initially distributed uniformly at random over a lattice or a network to give each the same chance of evolutionary success. Evidently, this has to do with the fact that random initial conditions make certain that each strategy occupies about the same amount of space in a population. But this alone does not confer equal chances of survival to all strategies, in particularly not if the competing strategies are three or more.

In fact, it is quite impossible to engineer initial conditions that would achieve this, because stable solutions are not made up solely of single strategies, but also of two-strategy alliances, three-strategy alliances that are perhaps entailed in a closed loop of dominance, even four-strategy phases, and so on. These are called subsystem solutions (see also Section~\ref{subsystemethod} below)), and they can be stable individually, if they are kept separate from other subsystem solutions. Of course, when we begin studying an evolutionary game for the first time, we are quite clueless as to which subsystem solutions are stable and which are not at certain parameter values. Importantly, some strategies can survive only within two- or three- or four-strategy subsystem solutions, and these subsystem solutions first have to form in the population before they can compete against each other. But different time scales characterize the formation of different subsystem solutions. Accordingly, no matter how hard we try, equal chances for survival in spatial evolutionary games with three or more competing strategies are rarely achievable, especially not with random initial conditions. What random initial conditions, paired with a very large system size, do accomplish, is they give a chance to each subsystem solution to emerge somewhere locally in the population, and the most stable one can subsequently invade the whole population. At small system sizes, however, only those subsystem solutions can evolve whose characteristic formation times are sufficiently short.

Since we have no way of knowing which initial configuration of strategies will yield a stable subsystem solution, our best option is to use random initial conditions with a very large system  size, and hope for all of them to emerge at some point in time. After we identify them, however, it is much more efficient and fair in terms of equal survival chances to use prepared initial states, and to do a proper stability analysis of subsystem solutions as described in Section~\ref{subsystemethod}.

When considering emergent phenomena in human societies, it is also practically always the case that a movement or a rebelion or an initiative starts locally, from a select initial state that is comprised of like-minded individuals. In this sense, random initial conditions are perhaps the most unnatural way to study human cooperation, but unavoidable due to the reasons stated above.

\subsection{Phase transitions}
\label{phasetranmethod}
Phase transitions are at the heart of statistical physics \cite{stanley_71, liggett_85, marro_99, hinrichsen_ap00}. The statistical physics of human cooperation is no exception, as indeed most of the research revolves around determining phase diagrams of the studied evolutionary games and determining the properties of the phase transitions that separate different stable strategy configurations. Traditionally, of course, phase transitions describe transitions between solid, liquid and gaseous states of matter. In general, near a phase transition the thermodynamic features of a system depend only on a small number of variables, but are insensitive to the details of the underlying microscopic dynamics. Thus, many macroscopic phenomena may be grouped into a small set of universality classes, specified by the shared sets of relevant observables. The universality classes are defined by critical exponents, which can be identical for very different physical systems. This coincidence of critical exponents is explained by the renormalization group theory \cite{kadanoff_p66, kadanoff_rmp67, wilson_rmp75, hohenberg_rmp77}, which shows that the differences are traceable to irrelevant observables while the relevant observables are shared in common.

For the purpose of this review, we first recall that phase transitions can be continuous and discontinuous, or equivalently, second-order or first-order, respectively. Continuous phase transitions can be characterized by critical exponents, whereby in human cooperation, due to the spatiotemporal dynamics of the competing strategies, the critical exponents that characterize spreading processes play the most prominent role. In such processes, phase transitions may exist to absorbing states, where the density of the strategy that is receding drops to zero. The order parameter is usually the density of the strategy
\begin{equation}
\rho_{s_x}(t)= \frac{1}{L^2} \langle \sum_x s_x(t) \rangle  \,,
\end{equation}
which in the supercritical phase vanishes as
\begin{equation}
\rho^{\infty} \propto |p-p_c|^\zeta
\end{equation}
as the control parameter $p$ approaches the critical value $p_c$. Thus far, as we will review from Section~\ref{peeresults} onwards, the predominant universality class characterizing models of human cooperation has been found to be that of directed percolation \cite{marro_99, hinrichsen_ap00}, where $\zeta=0.584(4)$ \cite{odor_rmp04}.

In addition to continuous phase transitions, discontinuous phase transitions are also common in models of human cooperation, occurring for example as a consequence of indirect territorial competition \cite{helbing_njp10}, or the spontaneous emergence of cyclic dominance \cite{szolnoki_epl10, szolnoki_jrsif14}. A particularly exotic solution was observed in the public goods game with correlated positive and negative reciprocity \cite{szolnoki_prx13} (see Section~\ref{correlatedmodel} for the definition of the model), where the amplitude of oscillations that characterize the cyclic dominance phase $D+P+B$ was found to be divergent and ultimately terminating in an absorbing phase. To demonstrate this phenomenon, we have to measure the fluctuations of a strategy in the stationary state according to
\begin{equation}
\chi= {L^2 \over M} \sum_{t_i = 1} ^M \left \langle \left( \rho_{s_x}(t_i)- \overline{\rho_{s_x}} \right)^2 \right \rangle,
\label{divergeamp}
\end{equation}
where $\overline{\rho_{s_x}}$ is the average value of the fraction of strategy $s_x$ in the population. It turns out that in this case the scaled quantity $\chi$ is size-independent, thus indicating a divergent fluctuation as the control parameter approaches the critical value. The cyclic dominance phase is therefore unable to exist beyond this value despite the fact that the average fractions of all three strategies are far from zero. Instead, the phase terminates via a discontinuous phase transition.

\subsection{Stability of subsystem solutions}
\label{subsystemethod}

\begin{figure*}
\centerline{\epsfig{file=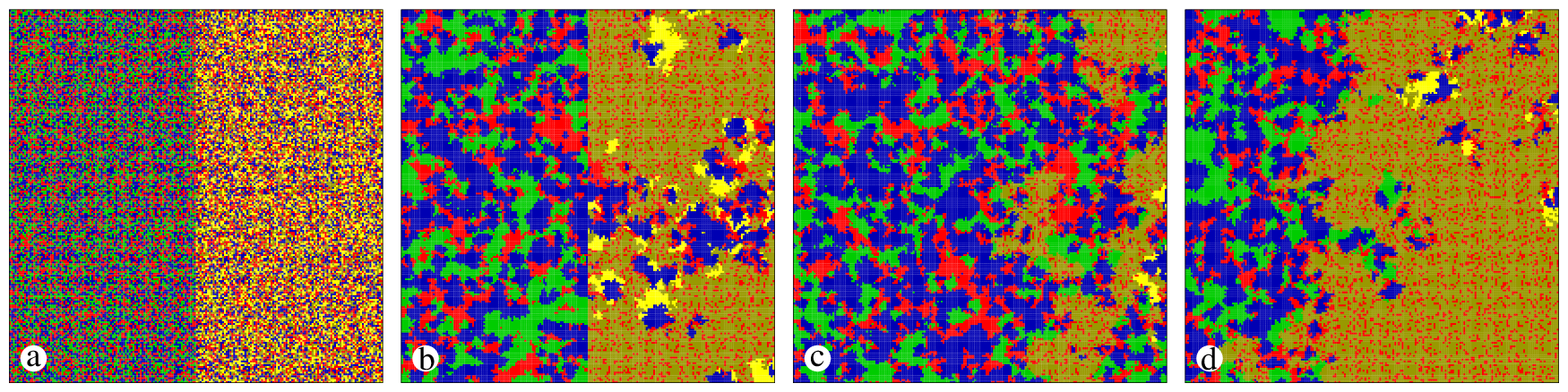,width=16.7cm}}
\centerline{\epsfig{file=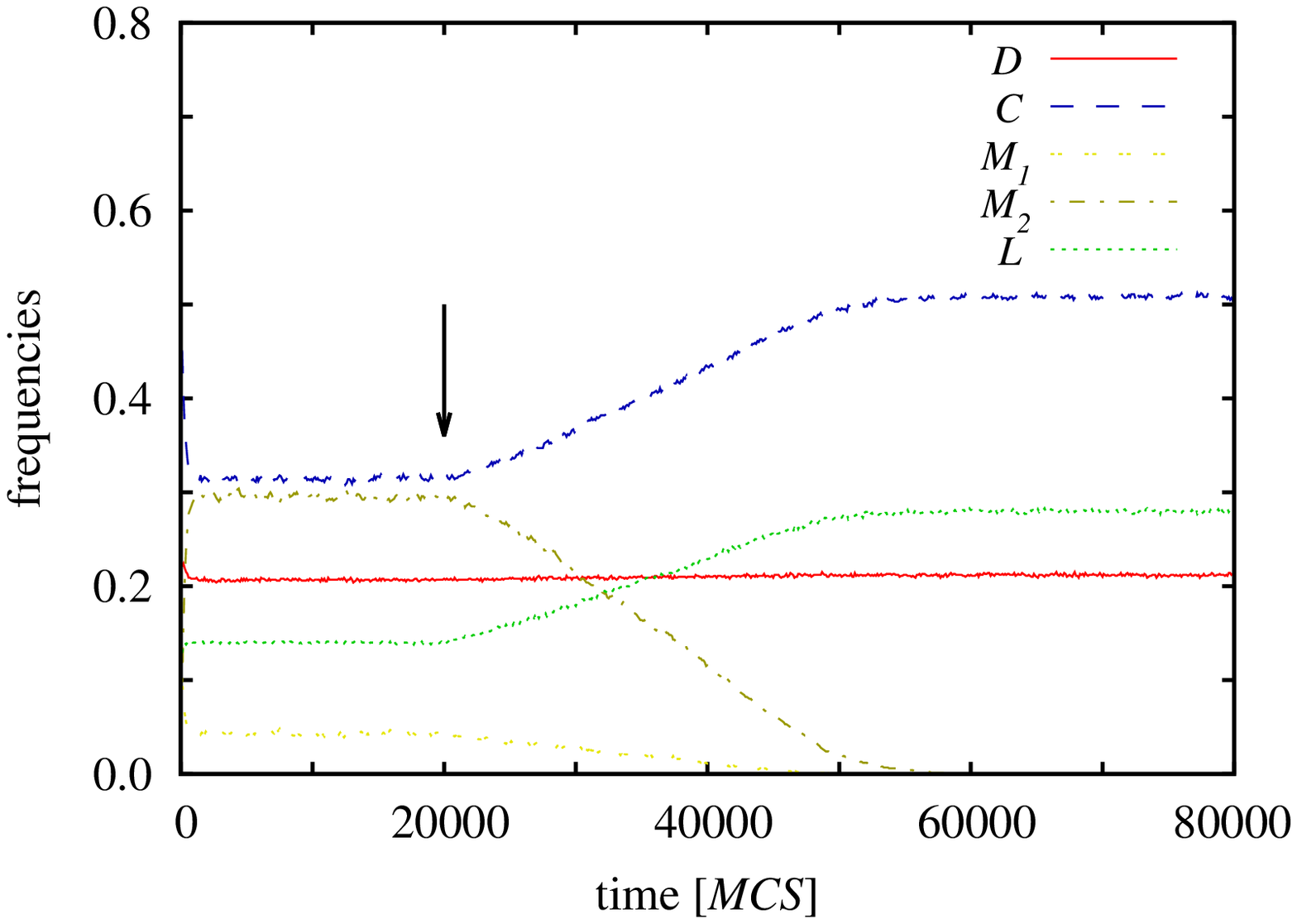,width=8.2cm}\epsfig{file=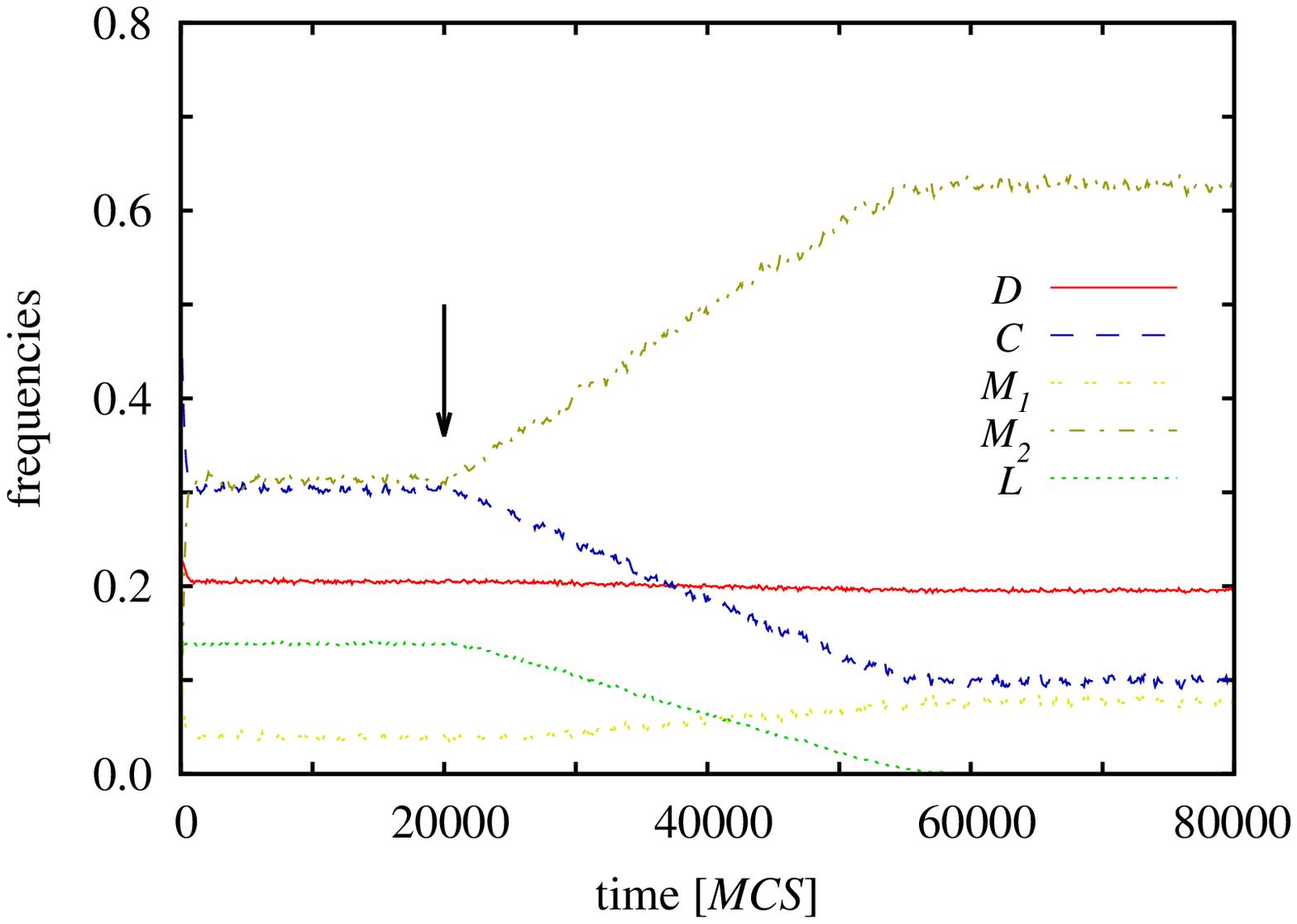,width=8.2cm}}
\caption{An example of a proper stability analysis of two subsystem solutions, namely the three-strategy $D+C+L$ and the four-strategy $D+C+M_1+M_2$ phase in the public goods game with tolerant players (see Section~\ref{tolerancemodel} for the definition of the model and Section~\ref{diversetoleranceresults} for more results), which are separated by a discontinuous phase transition. The series of snapshots in the upper row shows the separation of the lattice in two parts, each of which is initially randomly populated with the strategies that will form one of the two competing subsystem solutions [panel~(a)]. In panel (b), the subsystem solutions are formed in both halves of the square lattice, and accordingly, their competition can start by removing the border between them (thus allowing strategy transfer across the border). Panel (c) shows an intermediate state during the competition in which the $D+C+L$ phase will ultimately turn out to be the winner. Here $r=2.80$ and $\gamma=0.35$. Panel (d), on the other hand, shows an intermediate state during the competition in which the $D+C+M_1+M_2$ phase will ultimately turn out to be the winner. Here $r=2.81$ while the value of $\gamma$ is unchanged. Importantly, for both values of $r$ the state depicted in panel (b) is qualitatively exactly the same (both phases are individually stable regardless of which value of $r$ is used). The two graphs in the bottom row depict the corresponding ($r=2.80$ left and $r=2.81$ right) time evolution of the strategy densities. After a relaxation of 20~000 MCS (marked by an arrow), the two subsystem solutions start competing for space. On the left side the $D+C+L$ solution wins, while on the right side the $D+C+M_1+M_2$ solution wins. The linear system size used for this example was $L=2400$, but the snapshots in the upper row contain just a $200 \times 200$ cutoff of the whole population for clarity. Figure reproduced with permission from \cite{szolnoki_njp16}.}
\label{subsystemstability}
\end{figure*}

It is important to emphasize difficulties and pitfalls that are frequently associated with Monte Carlo simulations of evolutionary games with three or more competing strategies in structured populations. Foremost, it is crucial to choose a sufficiently large system size and to use long enough relaxation times. If these conditions are not met, Monte Carlo simulations can yield incorrect one- and/or two-strategy solutions that are unstable against the introduction of a group of mutants. For example, a homogeneous phase of cooperators or pool punishers in the public goods game with pool punishment can be invaded completely by the offspring of a single defector inserted into the population if only the value of $r$ is sufficiently low \cite{szolnoki_pre11}. At the same time, defectors can be invaded by a single group of pool punishers or cooperators if initially they form a sufficiently large compact cluster, such as a large-enough rectangular box.

As a short detour, we note that even if the competing strategies are only two, such as for example in the null model of human cooperation (see Section~\ref{nullmodel}), quenched heterogeneities in the population, for example due to differences in the distribution of public goods within a group \cite{perc_njp11}, may also significantly complicate Monte Carlo simulations. More precisely, quenched heterogeneities may evoke the existence of the Griffiths phase \cite{griffiths_prl69}, which has recently attracted considerable attention \cite{munoz_prl10, vazquez_prl11}, also in studies concerning the evolution of cooperation \cite{droz_epjb09}.

The essence of the problem of quenched heterogeneities for the extinction processes has been well described in \cite{noest_prl86, noest_prb88}, where it was shown that such systems are frequently characterized by patches of different sizes, providing better conditions for one of the strategies (or species) to survive. Due to the localization, the subordinate strategy can die out very slowly on the separated (or weakly interacting) patches, with an average lifetime increasing with the patch size.

Noest \cite{noest_prl86, noest_prb88} demonstrated that for suitable conditions (determined by the distribution of patch sizes) the extinction of the subordinate strategy follows a power law, whereby the exponent depends on the parameters. The latter fact can cause serious technical difficulties in the classification of the final stationary state, especially related to the $(C+D) \to C$ transition in game theoretical models, as demonstrated for example in Fig.~3 of \cite{droz_epjb09}, where it can be inferred that even very long simulation times might not be enough to reach the final stationary state, although the trend (power law behavior) clearly indicates the disappearance of the subordinate strategy in the limit when the time goes to infinity. We note that in such cases, an additional time-dependence in the background can significantly shorten the relaxation time. For example, if the quenched heterogeneities are varied on an extremely slow time scale (much slower than is characteristic for the main evolutionary process), the final conclusions remain the same, yet the occasional variations can accelerate the extinction significantly.

Turning back to the stability of subsystem solutions, indeed, an evolutionary game with three or more competing strategies has a large number of possible solutions because all the solutions of each subsystem are also solutions of the whole system \cite{szolnoki_pre11, szolnoki_pre11b, szolnoki_prl12, szolnoki_prx13, szolnoki_njp16}. The accurate location of phase transition points in phase diagrams, as well as the nature of these phase transitions as described in Section~\ref{phasetranmethod}, can therefore be determined accurately only by means of a stability analysis of competing subsystem solutions. A subsystem solution can be formed by any subset of all the competing strategies, and on their own (if separated from other strategies) these subsystems solutions are stable. This is trivially true if the subsystem solution is formed by a single strategy, but is likewise true if more than one strategy forms such a solution. Evidently then, for any specific set of parameter values, more than one subsystem solution exists. The dominant subsystem solution, and hence the phase that is ultimately depicted in the phase diagram as the stable solution of the whole system, can only be determined by letting all the subsystem solutions compete against each other.

The winner between two subsystem solutions can be determined by the average velocity of the invasion front that separates them. It is of course crucial that the competing subsystem solutions, if they are formed by two or more strategies, are characterized by a proper composition and spatiotemporal structure before the competition starts, i.e., before the front/interface between them is opened up to strategy invasions. In general, one must perform a systematic stability check between all possible pairs of subsystem solutions. Fortunately, this analysis can often be performed simultaneously if we choose a suitable patchy structure of subsystem solutions where all possible interfaces are present in the lattice. The lattice is in this case divided into several large rectangular boxes with different initial strategy distributions (containing one, two, or three strategies), and the strategy adoptions across the interfaces are initially forbidden for a sufficiently long initialization period. By using this approach, one can avoid the difficulties associated either with the fast transients from a random initial state or with the different time scales that characterize the formation of different subsystem solutions. One is then able to observe quite extravagant yet stable solutions, like for example the cyclic dominance between defectors, cooperators, and an alliance between pool punishers and defectors, in the public goods game with pool punishment (see Fig.~9 in \cite{szolnoki_pre11} or Section~\ref{poolpunresults} below).

As an example, we show in Fig.~\ref{subsystemstability} the stability analysis between two subsystem solutions, namely the three-strategy $D+C+L$ phase and the four-strategy $D+C+M_1+M_2$ phase in the public goods game with tolerant players (see Section~\ref{tolerancemodel} for the definition of the model). In the phase diagram (see Fig.~\ref{diversetolerancephase} in Section~\ref{diversetoleranceresults} below), the two phases are separated by a discontinuous phase transition point, although on both sides the two phases are individually stable, i.e., are proper subsystem solutions. To monitor the competition between them, we launch the evolution from a prepared initial state, where one half of the lattice initially contains only strategies $D, C$, and $L$ distributed uniformly at random, while the other half of the lattice contains only the strategies $D, C, M_1$, and $M_2$ distributed uniformly at random. As the next step, we let the two subsystem solutions evolve to their representative state in terms of the strategy frequencies and the typical spatial pattern. Only when both reach their stationary state we open the border by allowing strategy invasion through the separating interface. Lastly, we monitor how the competition between the two solutions evolves, i.e., which subsystem solution will turn out as the winner. The example depicted in Fig.~\ref{subsystemstability} demonstrates clearly that the final outcome depends sensitively on the value of the synergy factor $r$. At the smaller value of $r$ the $D+C+L$ solution wins, while at the slightly larger value of $r$ the $D+C+M_1+M_2$ solution turns out to be the dominant one.

\begin{figure*}
\centerline{\epsfig{file=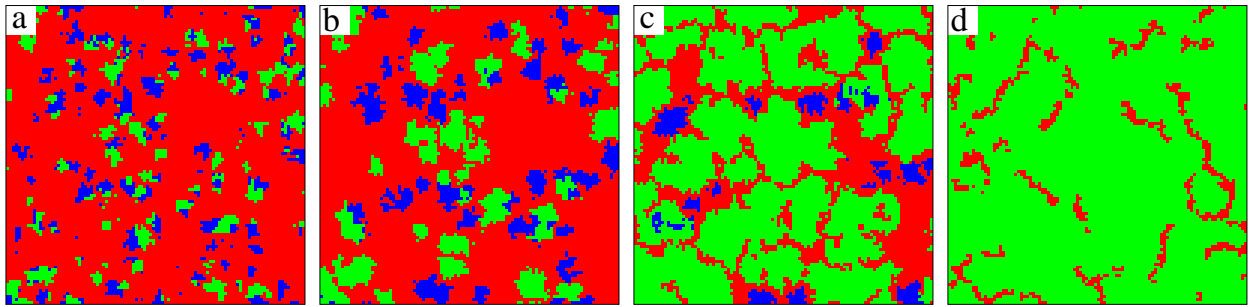,width=16.7cm}}
\caption{Spatiotemporal distribution of the three strategies in the public goods game with peer punishment at $t=10$, $40$, $150$, and $1000$ Monte Carlo steps (MCS) from panel (a) to (d). Cooperators ($C$) are represented by blue, defectors ($D$) by red, and peer punishers ($P$) by green color. The snapshots clearly demonstrate that the homogeneous domains of the $C$ and $P$ strategies compete separately against $D$, and the more successful $P$ strategy ultimately wins this indirect territorial competition. The cost of punishment is $\gamma=0.2$, the fine is $\beta=0.3$, and the multiplication factor is $r=3.8$. Figure reproduced with permission from \cite{helbing_njp10}.}
\label{indirectsnaps}
\end{figure*}

We emphasize that finite-size effects can easily play an obstructive role in the stability analysis of subsystem solutions illustrated in Fig.~\ref{subsystemstability}. If we start the evolution from a random initial state using a small system size, it can easily happen that we observe a misleading evolutionary outcome, simply because the phase that would be a stable solution at a large system size has no chance to emerge -- for example, one of the strategies that would be necessary to form it dies out beforehand due to the small system size. But that is not the only caveat. Even if we use prepared initial states for the stability analysis, we should be careful because the space (part of the lattice) allocated to each potential subsystem solution should be large enough for the latter to emerge. For example, the fluctuations of strategies in the cyclically dominant $D+C+L$ phase could be extremely large, and therefore this subsystem solution alone requires a large population to avoid fixation before the characteristic stationary pattern emerges. In this regard, we note that the upper panels of Fig.~\ref{subsystemstability} show just a small patch around the border where the two solutions meet, which is cut out of a large $2400 \times 2400$ lattice (not shown). This is also why the periodic boundary conditions cannot be detected in the four depicted snapshots.

Taken together, the stability analysis of subsystem solutions is a key procedure that must be used to properly determine stationary states in evolutionary games in structured populations if the competing strategies are three or more. Prepared initial conditions, a large enough system size, the partitioning of the lattice with proper interfaces, and a sufficiently long relaxation time for all subsystem solutions to obtain their stationary spatiotemporal dynamics before the interfaces are opened up for strategy invasions, are thereby the essentials.

\section{Peer-based strategies and the evolution of cooperation}
\label{peeresults}
With this section, we turn to the review of the most interesting results in the realm of statistical physics of human cooperation. In peer-based strategies, individual players take it upon themselves to either punish defectors \cite{helbing_ploscb10, helbing_njp10, helbing_pre10}, or to reward cooperators \cite{szolnoki_epl10}, or, as in the case of correlated negative and positive reciprocity \cite{szolnoki_prx13}, to do both. In the following subsections, we review results obtained with upgrading the null model with peer-based strategies.

\subsection{Peer punishment and indirect territorial competition}
\label{peerpunresults}

One of the more fascinating evolutionary outcomes of peer punishment in the spatial public goods game (see Section~\ref{peerpunmodel} for the definition of the model) is the emergence of indirect territorial competition \cite{helbing_ploscb10, helbing_njp10}, which allows the counterintuitive survival of peer punishers even though their payoffs in the presence of defectors are by definition always lower than those of traditional cooperators. Accordingly, in well-mixed populations peer punishers can not survive \cite{hauert_s07, ohtsuki_n09}. In structured populations, however, peer punishers and cooperators segregate spontaneously into homogeneous compact clusters, and then compete independently against defectors. Whoever is more effective in accruing space from the defectors lays the groundwork for the extinction of the other cooperative strategy. One example of this phenomenon is illustrated in Fig.~\ref{indirectsnaps}, where the blue cooperators and green peer punishers segregate and compete separately against the red defectors. Ultimately, cooperators die out because punishers are more effective in this competition.

\begin{figure}[b]
\centerline{\epsfig{file=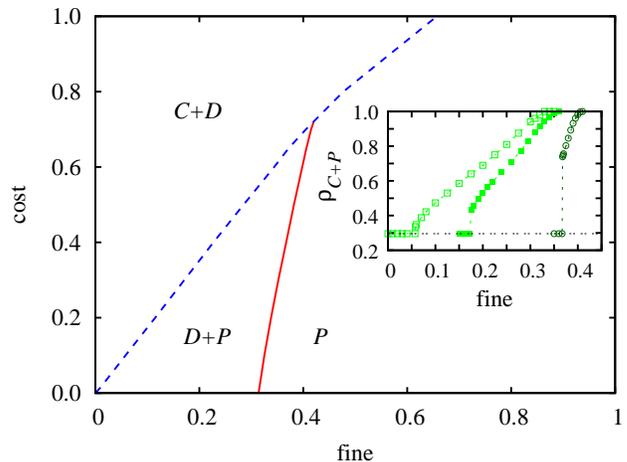,width=8.5cm}}
\caption{Full fine-cost ($\beta-\gamma$) phase diagram for the public goods game with peer punishment, as obtained for $r=3.8$ and $K=0.5$. Solid red line denotes continuous phase transitions, while dashed blue line denotes discontinuous phase transitions. Different phases are denoted by the symbols of the strategies that survive in the stationary state. Inset shows the overall fraction of both cooperative strategies ($\rho_{C+P}$) in dependence on the fine $\beta$, as obtained for punishment costs $\gamma=0.1$, $0.3$, and $0.65$ from left to right. Similar phase diagrams can be obtained for smaller values of $r$, where, however, only strategy $D$ survives at small values of the punishment fine. We recall that $r=3.74$ is the benchmark value beyond which a mixed $C+D$ phase is stable in the null model without punishment. Figure reproduced with permission from \cite{helbing_njp10}.}
\label{peerpunphased}
\end{figure}

Importantly, indirect territorial competition is commonly the source of discontinuous phase transitions in game-theoretical models. The control parameter in this case is the ration between the cost of punishment $\gamma$ and the punishment fine is $\beta$. If punishment is sufficiently cheap, then peer punishers do better than cooperators. In the opposite case, if punishment is too costly, cooperators do better. Evidently, there is precisely defined switch in this case, which gives rise to a discontinuous phase transition between the $C+D$ and the $C+P$ phase. The phase diagram depicted in Fig.~\ref{peerpunphased} confirms this exactly. The inset, on the other hand, shows how the overall cooperation level increases monotonously with the fine for three different values of the punishment cost $\gamma$. It can be observed that punishing cooperators always prevail for a sufficiently large fine, independently of the punishment cost $\gamma$. If the cost is lower than a critical value ($\gamma \approx 0.65$ at $r=3.8$), the application of a sufficiently large fine will lead to a discontinuous $(C+D) \to (D+P)$ phase transition, where punishing cooperators replace pure cooperators in the two-strategy phase. It is important to note that, in the beginning of the evolutionary process, $C$ and $P$ players may form mixed cooperative islands. However, when defectors are not in the neighborhood, the two strategies have identical payoffs and thus become equivalent, and the strategy update dynamics defined by Eq.~(\ref{fermi}) results in logarithmic coarsening that is otherwise characteristic for the voter model \cite{dornic_prl01}. Although the coarsening is logarithmically slow, $C$ and $P$ players in these islands segregate quickly, given that their size is typically very small. After this segregation, homogeneous clusters of pure cooperators and punishing cooperators compete separately against the defectors. When the punishment fine is sufficiently large, punishing cooperators suddenly become more effective against defectors than pure cooperators, so that eventually the later are crowded out and replaced by the former. It is worth noting that discontinuous phase transitions due to indirect territorial competition appear to be common in evolutionary games in structured populations, as they have been observed also in the public goods game with pool punishment \cite{szolnoki_pre11} and in the public goods game with correlated positive and negative reciprocity \cite{szolnoki_prx13}, the results for both of which will be reviewed below.

\begin{figure}
\centerline{\epsfig{file=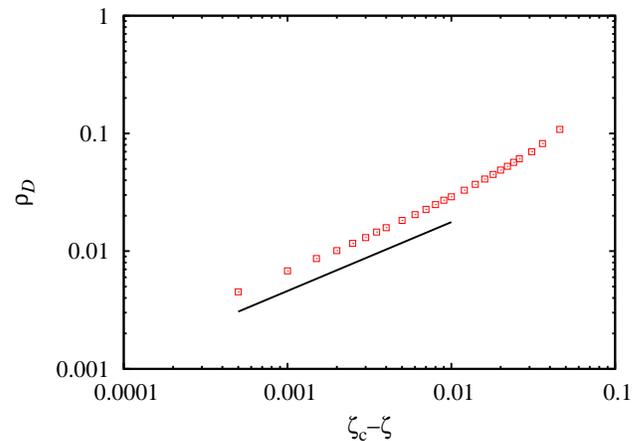,width=8.5cm}}
\caption{Critical scaling behavior of the order parameter $\rho_D$ for the spatial public goods game with peer punishment, as obtained for the punishment cost $\gamma= 0.1$ when the fine $\beta$ approaches the critical value $\zeta_c=0.3262(1)$. The solid line indicates the slope $0.584$, characterizing directed percolation. Figure reproduced with permission from \cite{helbing_njp10}.}
\label{peerpunscale}
\end{figure}

Further with regards to the nature of the phase transitions in Fig.~\ref{peerpunphased}, we note that while the population always leaves the $C+D$ phase via a discontinuous phase transition, the transition between the $D+P$ phase and the $P$ phase is always continuous. The $(D+P) \to P$ continuous phase transition agrees with the directed percolation universality class conjecture \cite{janssen_zpb81, grassberger_zpb82}. Namely, the interactions amongst players are short-ranged, and the order parameter, which is the fraction of defectors $\rho_D$, becomes zero at the critical value of the fine $\beta_c$, where the population arrives at the absorbing $P$ phase. Accordingly, the exponent of the phase transition is expected to belong to the universality class of directed percolation, for which $\rho_D \propto (\beta_c - \beta)^\zeta$ with $\zeta = 0.584(4)$ in two spatial dimensions \cite{marro_99} (see also Section~\ref{phasetranmethod}). Figure~\ref{peerpunscale} shows the decay of the fraction of defectors at a fixed punishment cost $\gamma= 0.1$ when the fine $\beta$ approaches the critical value $\beta_c=0.3262(1)$. The numerically determined critical exponent is in good agreement with the mentioned exponent $0.584$ for directed percolation, as indicated by the black solid line.

In addition to the results reviewed above, the consideration of peer punishment in the public goods game offers additional insights into human cooperation that deserve mentioning. In the first place, there is the so-called ``who laughs last laughs best effect'' \cite{helbing_ploscb10}, where peer punishers defeat cooperators even when the defectors are eventually eliminated, but this process is very slow. That is, the system behavior can change its character significantly even after very long times. The finally winning strategy can be in a miserable situation in the beginning, and its victory may take very long. Secondly, there is also a phenomenon dubbed the ``Lucifer's positive side effect'', where a permanent but modest generation of defectors through small mutation rates can considerably accelerate the spreading of peer punishers \cite{helbing_pre10}, effectively counteracting the aforementioned ``who laughs last laughs best effect''. Lastly, it has been shown that peer punishment can lead to hysteresis in the population that is otherwise characteristic for systems that exhibit super-heating and super-cooling, ultimately leading to punishment being a double-edged sword in that it can stabilize cooperation below the critical point, but it can also stabilize defectors above the critical point \cite{hintze_pb15}. We conclude this subsection by noting that more biologically-oriented research on punishment in structured populations can be found in \cite{brandt_prsb03, nakamaru_jtb06}, while research on peer punishment in some form relying or being connected to statistical physics is in \cite{short_pre10, amor_pre11, xu_c_pa11, gao_j_pa12, chan_pa13, vukov_pcbi13, luo_js_csf13, wang_z_srep13c, chen_xj_srep15, chen_xj_pre15}.

\subsection{Peer rewarding and the emergence of cyclic dominance}
\label{peerewresults}
While peer punishment has traditionally been considered more successful than rewarding \cite{sigmund_pnas01, sigmund_tee07}, the fact that the cost of punishment frequently fails to offset gains from enhanced cooperation, as well as the fact that the act of punishment itself can be considered hostile and thus subject to revenge, puts rewards back in contention as a viable catalyst for human cooperation \cite{dreber_n08, rand_s09}. The problem of whether to use punishment or rewarding for promoting prosocial behavior is known as the ``stick versus carrot'' dilemma \cite{andreoni_aer03, hilbe_prsb10, milinski_jtb12}. By studying the evolution of cooperation in the spatial public goods game, where besides the traditional cooperators ($C$) and defectors ($D$), rewarding cooperators ($R$) supplement the array of possible strategies (see Section~\ref{peerewmodel} for the definition of the model), we can understand better the benefits and risk of arguably the much more gentle form of cooperation enforcement \cite{szolnoki_epl10}.

\begin{figure}
\centerline{\epsfig{file=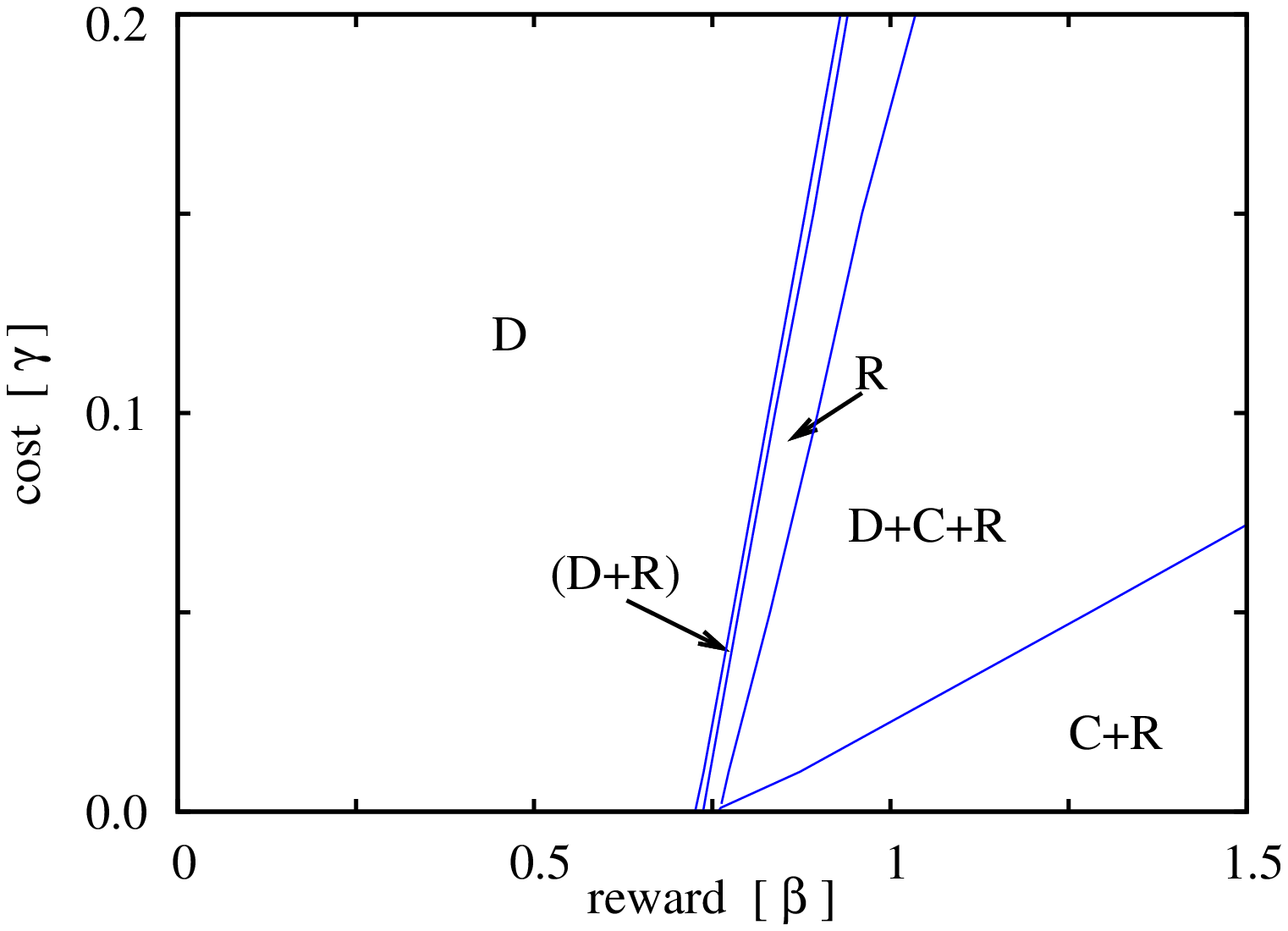,width=8cm}}
\centerline{\epsfig{file=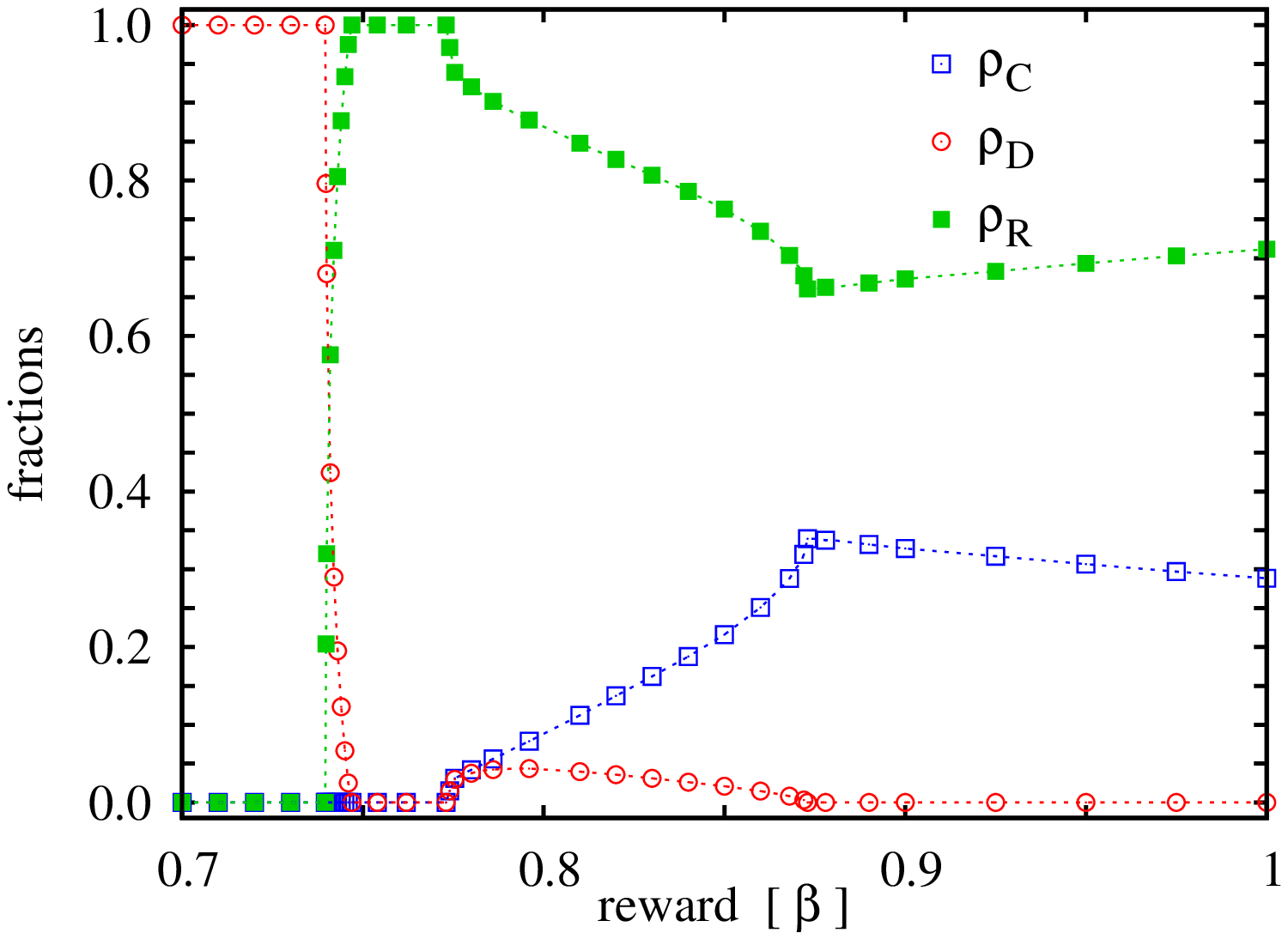,width=8cm}}
\caption{Full reward-cost ($\beta-\gamma$) phase diagram for the public goods game with peer rewarding, as obtained for $r=2.0$ and $K=0.5$. Different phases are denoted by the symbols of the strategies that survive in the final strategy distribution. Solid blue lines indicate continuous phase transitions. Bottom panel shows a typical cross-section of the phase diagram at the cost $\gamma=0.01$, depicting the fraction of cooperators $\rho_{C}$, defectors $\rho_{D}$ and rewarding cooperators $\rho_{R}$ in dependence on the reward $\beta$. Figure reproduced with permission from \cite{szolnoki_epl10}.}
\label{peerewphased}
\end{figure}

\begin{figure*}
\centerline{\epsfig{file=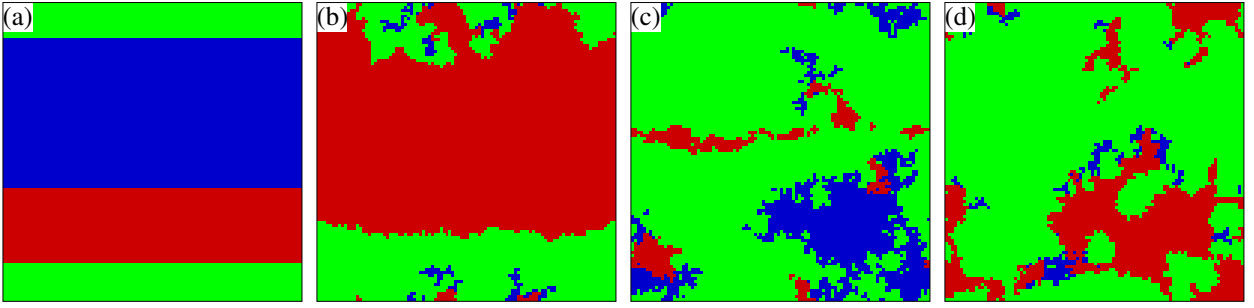,width=16.7cm}}
\caption{Spatiotemporal distribution of the three strategies in the public goods game with peer rewarding on a $100 \times 100$ square lattice with specially prepared initial conditions. Colors red, green and blue depict the location of defectors ($D$), rewarding cooperators ($R$) and cooperators ($C$), respectively. It can be observed that defectors invade cooperators, cooperators invade rewarding cooperators, and rewarding cooperators in turn invade defectors, thus forming a closed loop of dominance that give rise to cyclical interactions and a spatiotemporal dynamics akin to the rock-paper-scissors game \cite{szolnoki_jrsif14}. The snapshots were taken at 0 (a), 140 (b), 560 (c) and 600 (d) full MCS, and the parameter values were $r=2.0$, $\gamma=0.05$ and $\beta=0.9$. Figure reproduced with permission from \cite{szolnoki_epl10}.}
\label{peerewsnap}
\end{figure*}

Figure~\ref{peerewphased} shows the full reward-cost phase diagram in the low-$r$ limit, where it can be observed that the pure $D$ phase first gives way to a very narrow region of coexistence of $D+R$ and shortly thereafter reaches the pure $R$ phase as the reward increases. The blue transition lines, indicating continuous second-order phase transitions, lean towards higher rewards for larger costs, yet this effect is expected and validates the behavior of the examined model. Most remarkable is the reappearance of defectors in a stable $D+C+R$ phase if the reward is increased further, thus giving rise to a stable coexistence of all three strategies. Finally, if the reward is higher still and the costs remain moderate (note that the slope of the rightmost transition line is considerably larger), defectors again die out and leave $C$ and $R$ as the only remaining strategies. Notably, here $C$ and $R$ in the absence of defectors are not equivalent strategies as was the case above for peer punishment (see Section~\ref{peerpunresults} and \cite{helbing_njp10}), and thus their stable coexistence is possible, which is a pure consequence of spatiality.

Turning to the reappearance of defectors for intermediate rewards, we show in the bottom panel of Fig.~\ref{peerewphased} a characteristic cross-section of the phase diagram obtained for $\gamma=0.01$. In agreement with the four blue lines depicted in the phase diagram, we can observe four continuous phase transitions. From left to right we have, first, the emergence of rewarding cooperators ($\rho_{R}>0$), which is quickly followed by the extinction of defectors ($\rho_{D} = 0$). Subsequently, defectors reaper with traditional cooperators to form the coexistence of all three strategies, and finally, at $\beta \approx 0.873$ defectors die out again. Interpreting these observations, for sufficiently large $\beta$ the rewarding cooperators can support each other and protect themselves against the invasion of defectors. In accordance with the well-known network reciprocity mechanism, rewarding cooperators aggregate into compact clusters with a smooth interface. At still higher $\beta$, the efficiency of rewarding cooperators is so strong that defectors cannot survive. Remarkably, for $\beta>0.775$ the support of cooperative actions becomes powerful enough to enable not just the proliferation of rewarding cooperators, but also the survivability of traditional cooperators. But since the synergy factor is low ($r=2.0$), the traditional cooperators are susceptible to exploitation by defectors and can therefore survive only in the vicinity of rewarding cooperators. Nevertheless, the emergence of traditional cooperators simultaneously enables also the survivability of defectors via a stable $D+C+R$ phase that is governed by cyclic dominance.

The workings of this cyclic dominance can be demonstrated by examining the snapshots of strategy distributions. Figure~\ref{peerewsnap}(a) depicts a prepared initial state, whereafter the movements of the boundaries that separate the three strategies give vital insight into the dominance between them. Due to the small synergy factor $r$, the defectors (red) can easily invade the blue region of traditional cooperators. Simultaneously, since the reward is large, rewarding cooperators (green) can outperform defectors. In the midst of rewarding cooperators, however, traditional cooperators (blue) can spread as well because they enjoy the significant benefits of reward but do not bear any costs. But as soon as some of the traditional cooperators depart from the safe haven of rewarding cooperators, the whole circle of invasion starts anew, leading to an uprise of defectors (red), who are then again conquered by rewarding cooperators, who then again foster the spreading of traditional cooperators, and so on. Clearly thus, the three strategies form a closed loop of dominance, which can be observed nicely if following the snapshots presented in Fig.~\ref{peerewsnap} from left to right. It is important to point out that qualitatively identical spatial patterns emerge from random initial conditions if the system size is sufficiently large (see Section~\ref{randominitimethod} for a discussion).

The stability of cyclic dominance phases in structured populations is always rather precarious. Namely, if one of the three strategies dies out by chance due to a small system size, the balance within the closed loop of dominance is broken, and accordingly, one of the remaining two strategies spreads across the whole population. To avoid this, it is therefore paramount to use sufficiently large system sizes. Interestingly, the stationary density of defectors is considerable, but the increase of the $\rho_{D}(\beta)$ function is the more dramatic the larger the cost of reward $\gamma$. This is in agreement with the behavior of predator-prey systems where the direct support of prey will ultimately be beneficial for the predators. Naturally, if the reward $\beta$ is even larger, defectors cannot survive and the system arrives to the mixed $C+R$ phase, as depicted in Fig.~\ref{peerewphased}. Thereafter, the qualitative behavior does not change and the fraction of cooperators and rewarding cooperators converges to a nonzero value. This, however, is a unique consequence of the spatial structure since in well-mixed populations cooperators, i.e., second-order freeriders due to their unwillingness to bear the additional costs of rewarding, clearly perform better than rewarding cooperators and are thus dominant. In fact, the mechanism that allows rewarding cooperators to survive in the sea of second-order freeriders is identical to the one which allows cooperators to survive in the sea of defectors in the traditional public goods game. In both cases the subordinate strategy forms compact clusters, and from there on classical network reciprocity works its magic \cite{nowak_n92b}.

As the above results show, rewarding is certainly viable for promoting human cooperation, but it is also quite a bit more tricky than punishment. This has to do with the spontaneous emergence of cycling dominance between the three competing strategies, which leads to counterintuitive effects, such as for example that moderate rewards may promote cooperation better than high rewards. Needless to say that these results do not settle the ``stick versus carrot'' dilemma \cite{andreoni_aer03, hilbe_prsb10}, but they do offer valuable insights that are unique to the statistical physics approach to the problem.

\subsection{Correlated strategies and an exotic first-order phase transition}
\label{peercorrresults}
In addition to the ``stick versus carrot'' dilemma addressed in the preceding subsection, there is also the strong reciprocity hypothesis \cite{gintis_jtb00, fehr_hn02, bowles_tpb04}, which asserts that optimal enforcement of cooperation entails both punishing and rewarding. This hypothesis, however, is not aligned with recent economic experiments \cite{yamagishi_pnas12, egloff_pnas13}. By considering the public goods game with correlated positive and negative reciprocity (see Section~\ref{correlatedmodel} for the definition of the model) and methods of statistical physics, we can relevantly add to the preceding research on this subject and help bring clarity as to the differing conclusions \cite{szolnoki_prx13}.

\begin{figure}[b]
\centerline{\epsfig{file=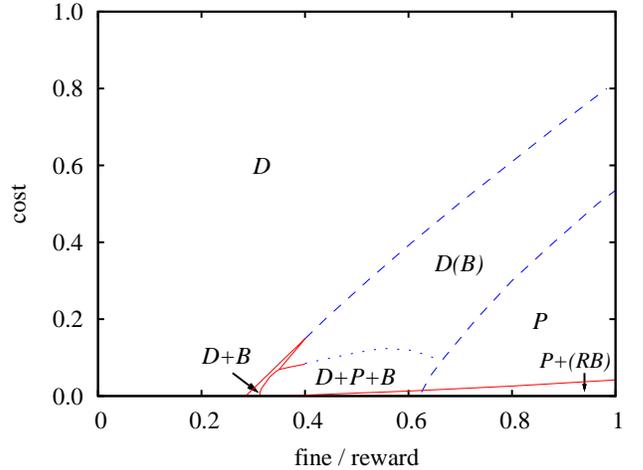,width=8.4cm}}
\caption{Full fine-cost ($\beta-\gamma$) phase diagram for the public goods game with correlated positive and negative reciprocity, as obtained for $r=2.5$ and $K=0.5$. Solid red lines denote continuous phase transitions, while dashed blue lines denote discontinuous phase transitions. More precisely, the dashed blue line between phases $D$ and $D(B)$ indicates that there would be a discontinuous phase transition if only $D$ and $B$ strategies were initially present in the system. The same holds for the dashed blue line separating phases $D(B)$ and $P$. The three-strategy $D+P+B$ phase is separated with a dotted blue line to emphasize that there are two different ways in which this solution can give way to the $D(B)$ phase. In particular, at smaller fines the transition is continuous because the average fraction of strategy $P$ gradually decays to zero. At larger fines, the averages of all three strategies remain finite, but the amplitude of oscillations diverges regardless of the system size (see Fig.~\ref{corrscaling}), which ultimately results in an abrupt termination of cyclic dominance between the three strategies and an exotic first-order phase transition (see Fig.~\ref{r2_5_F0_550}). Figure reproduced with permission from \cite{szolnoki_prx13}.}
\label{phd_r2_5}
\end{figure}

The phase diagram presented in Fig.~\ref{phd_r2_5} reveals that discontinuous phase transitions dominate, which has to do with the spontaneous emergence of cyclic dominance \cite{reichenbach_n07, reichenbach_prl07, mobilia_jtb10, mobilia_epl11} between strategies $D$, $P$ and $B$. In particular, within the three-strategy $D+P+B$ phase strategy $D$ outperforms strategy $P$, strategy $P$ outperforms strategy $B$, while strategy $B$ again outperforms strategy $D$. As was frequently the case before \cite{wu_zx_pre05, szolnoki_pre11, szolnoki_pre11b, szolnoki_prl12} (see \cite{szolnoki_jrsif14} for a review), here too the spontaneous emergence of cyclic dominance brings with it fascinating dynamical processes that are driven by pattern formation, by means of which the phase may terminate. There are two qualitatively very different ways for the $D+P+B$ cyclic dominance phase to give way to the $D(B)$ phase [here $D(B)$ indicates that either a pure $D$ or a pure $B$ phase can be the final state if starting from random initial conditions]. The first, around $\beta=0.37$, is relatively straightforward, in that the average fractions of strategies $P$ and $B$ decay due to the increasing cost $\gamma$, which ultimately results in the vanishing average value of the fraction of strategy $P$. The closed cycle of dominance is therefore interrupted and the $D+P+B$ phase terminates.

\begin{figure}
\centerline{\epsfig{file=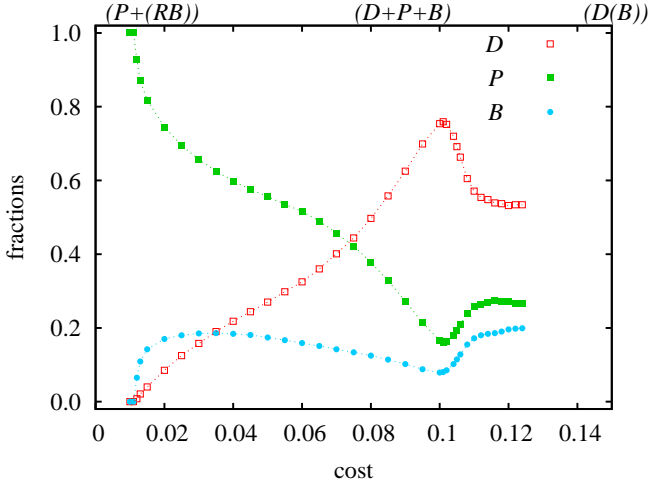,width=8.57cm}}
\caption{Cross-section of the phase diagram depicted in Fig.~\ref{phd_r2_5}, as obtained for $\beta=0.55$. Depicted are stationary fractions of the four competing strategies in dependence on $\gamma$. Stable solutions are denoted along the top axis. Unlike around $\beta=0.37$, here the $(D+P+B) \to D(B)$ phase transition is discontinuous because the amplitude of oscillations diverges independently of the system size (see Fig.~\ref{corrscaling} for details) as $\gamma$ increases. Figure reproduced with permission from \cite{szolnoki_prx13}.}
\label{r2_5_F0_550}
\end{figure}

\begin{figure}
\centerline{\epsfig{file=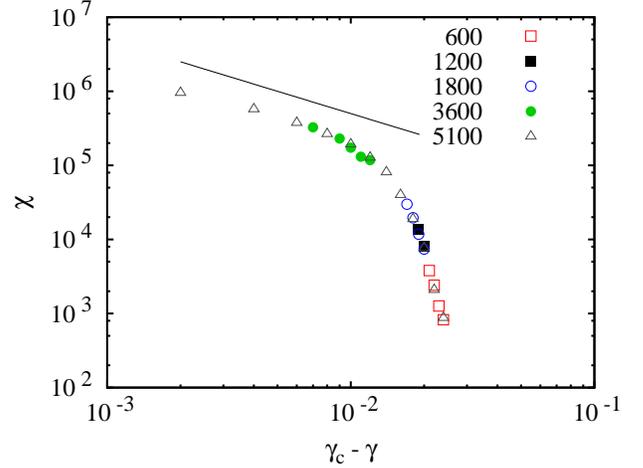,width=8.35cm}}
\caption{Fluctuations of the amplitude of oscillations $\chi$ within the $D+P+B$ cyclic dominance phase in dependence on the vicinity to the critical value of the cost $\gamma_c=0.1242(6)$ for $\beta=0.55$ and different system sizes, as indicated in the legend. The slope of the power-law exponent (solid line) is $1$, indicating divergent fluctuation as $\gamma$ approaches the critical value. Figure reproduced with permission from \cite{szolnoki_prx13}.}
\label{corrscaling}
\end{figure}

The situation around $\beta=0.55$, however, is much more peculiar and interesting. As results presented in Fig.~\ref{r2_5_F0_550} demonstrate, here the average values of all three strategies remain finite. Hence, the termination of the $D+P+B$ phase must have a different origin than at $\beta=0.37$. In fact, for $\beta=0.55$ it is the amplitude of oscillations that increases with increasing values of $\gamma$. And it is the increase in the amplitude that ultimately results in a uniform absorbing phase regardless of the system size. At this point it is crucial to emphasize that the increase of the amplitude of oscillation is not a finite-size effect. Although in spatial systems with cyclic dominance it is typical to observe oscillations with increasingly smaller amplitude as the system size is increased, this does not hold in the present case. To demonstrate this, we measure the fluctuations in the stationary state as defined by Eq.~(\ref{divergeamp}). As Fig.~\ref{corrscaling} shows, the scaled quantity $\chi$ is size-independent, thus indicating a divergent fluctuation as $\gamma$ approaches the critical value. The three-strategy $D+P+B$ phase is therefore unable to exist beyond this value despite the fact that the average fractions of all three strategies are far from zero. Instead, the phase terminates via a discontinuous phase transition towards the $D(B)$ phase, as depicted in Fig.~\ref{phd_r2_5}. Notably, within the $D(B)$ phase either the pure $D$ or the pure $B$ phase can be the final state, depending on which strategy dies out first.

In addition to the above results, the public goods game with correlated positive and negative reciprocity also awaits with indirect territorial competition and the spontaneous emergence of cyclic dominance for certain parameter values \cite{szolnoki_prx13}, much in the same way as already reviewed in Sections~\ref{peerpunresults} and \ref{peerewresults}. Yet despite the high complexity of solutions, the correlated strategy, which conforms to the strong reciprocity hypothesis, can survive only in very narrow and unrealistic parameter regions, thus lending support to empirical research on this subject. Elementary strategies, either in pure or mixed phases, are much more common and likely to prevail. These results highlight the importance of patterns and structure in human cooperation, which should also be taken into account in future experiments.

\section{Institutionalized strategies}
\label{poolresults}
In this section, we turn to institutionalized strategies aimed at promoting human cooperation that, unlike peer-based strategies, rely on contributions to a common pool for the application of punishment and rewarding. Recent economic experiments show that humans prefer pool punishment for maintaining public goods \cite{traulsen_prsb12}, and evidence also exist in favor of competitive advantages of sanctioning institutions \cite{gurerk_s06}. Indeed, the debate how to achieve human cooperation under adverse conditions with a focus on whether this could be best done by means of individual (peer-based) incentives or institutions is a lively one since the advent of this line of research \cite{axelrod_wp85, ostrom_90}. Accordingly, the subject has received its fair share of attention in the scientific literature \cite{bowles_jtb03, henrich_s06, sigmund_n10, isakov_dga12, cressman_jtb12, vasconcelos_ncc13, zhang_by_expecon14, sasaki_dga14, gao_l_srep15}, while of course the involvement of statistical physics in this arena is a comparatively recent one \cite{szolnoki_pre11, szolnoki_pre11b, perc_srep12, szolnoki_prsb15}.

\subsection{Spatiotemporal complexity due to pool punishment}
\label{poolpunresults}
It is likely no coincidence that the enforcement of law and justice in modern human societies relies predominantly on institutionalized punishment \cite{traulsen_prsb12}. The spatial public goods game with pool punishment (see Section~\ref{poolpunmodel} for the definition of the model), where defectors, cooperators, and pool-punishers compete, thus addresses precisely the kind of setup that is commonplace in human societies. Research in well-mixed populations concluded that pool-punishers can prevail over peer-punishers only if the second-order freeriders are punished as well \cite{sigmund_n10}. Conversely, in structured populations self-organizing spatiotemporal structures can maintain pool-punishment viable without such an assumption. In fact, the phase diagrams indicate surprisingly rich and spatiotemporally complex behavior depending on the punishment fine and cost.

\begin{figure}
\centerline{\epsfig{file=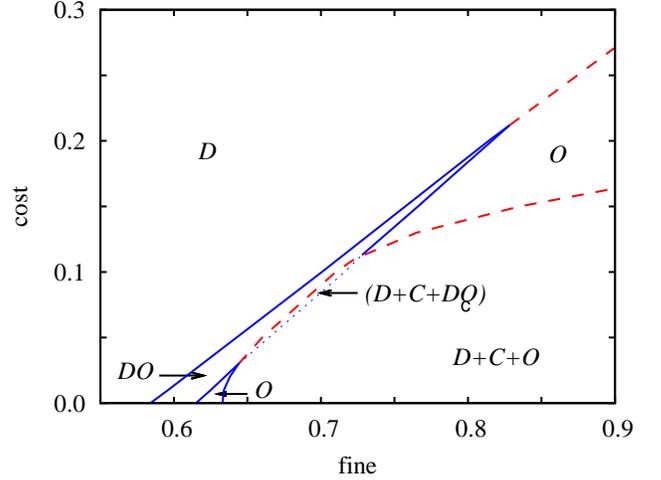,width=8.4cm}}
\caption{Full fine-cost ($\beta-\gamma$) phase diagram for the public goods game with pool punishment, as obtained for $r=2.0$ and $K=0.5$ with a focus on the small cost area. Solid (dashed) lines indicate continuous (discontinuous) phase transitions. The dotted line represents the analytic continuation of the phase boundary separating the pure $D$ and $O$ phases in the absence of cooperators ($C$). The $D+C+O$ phase is governed by cyclic dominance, just like the $(D+C+DO)_c$ phase, however, in the later one strategy in the loop is in fact an alliance of two strategies, namely strategies $D$ and $O$. For further details we refer to the main text. Figure reproduced with permission from \cite{szolnoki_pre11}.}
\label{poolphase}
\end{figure}

First, we illustrate the phase transitions determined by means of Monte Carlo simulations as a function of fine for a low value of cost. Figure \ref{poolphase} shows consecutive transitions from the pure $D$ phase on the left to the final $D+C+O$ phase on the right, in which all the three strategies coexist due to cyclic dominance. Incidentally, the snapshots presented in Fig.~\ref{preparedinitfig} correspond precisely to the emergence of this $D+C+O$ phase, only that the upper row demonstrates  a failure to do so from random initial conditions, while the bottom row show the evolution towards the most stable stationary state from prepared initial conditions. As we have already mentioned, the most stable $D+C+O$ phase can also emerge and spread from a random initial state, but only if the system size is large enough, in this case exceeding $L=1500$ for these $(r, \gamma, \beta)$ parameter values.

Returning to the phase diagram presented in Fig.~\ref{poolphase}, three continuous phase transitions can be observed as we increase the fine $\beta$ at a low cost, for example $\gamma=0.01$. First, the homogeneous defector state ($D$) transforms into the coexistence of defectors and pool-punishers ($DO$). In this phase, pool-punishers form compact clusters to survive in the sea of defectors. This mechanism is identical to the previously identified network reciprocity that enables pure cooperators to coexist with defectors \cite{nowak_n92b}. Cooperators who refuse to bear the cost of punishment, however, are unable to survive due to the low value of $r$. Within the $DO$ phase the frequency of pool-punishers increases continuously until the homogeneous $O$ phase is reached. A quantitative analysis of both these continuous phase transitions supports the directed percolation universality class conjecture \cite{janssen_zpb81} (see Section~\ref{phasetranmethod} for details).

Surprisingly, further increasing $\beta$ leads to an additional phase transition from the $O$ phase into the $D+C+O$ phase, where the self-organizing pattern is maintained by cyclic dominance described above. Within the $D+C+O$ phase $\rho_O$ decreases monotonously as the fine $\beta$ increases, which is in agreement with the anomalous behavior referred frequently as the ``survival of the weakest'' \cite{tainaka_pla93, frean_prsb01}. In the present case, the increase of fine reduces the income of the punished defectors, which allows pure cooperators to survive. The latter strategy behaves as the ``predator'' of pool-punishers, resulting in the decay of $\rho_O$ despite of the increasing fine. The same cyclic dominance mediated complex interaction is able to increase $\rho_O$ when $\gamma$ is increased (in this case the less effective punishment does not allow $C$ players, who are the ``prey'' of $D$, to survive). Similar effects were already reported in several three-strategy models, including the simpler spatial rock-paper-scissors game, and the main features were justified by mean-field approximations and pair-approximations (see \cite{szolnoki_jrsif14} for a review).

We note further that, at such a low punishment cost, the cooperators can invade the sites of pool-punishers, albeit very slowly and only within the territories they have in common. Within these two-strategy territories in the $\gamma \to 0$ limit the strategy evolution reproduces the behavior of the voter model with equivalent strategies exhibiting rough interfaces and extremely slow coarsening \cite{dornic_prl01}. For low but finite values of $\gamma$ the two-strategy system evolves slowly towards the homogeneous $C$ state while the interfaces remain highly irregular. In contrast, the interfaces separating the domains of defectors from cooperators or defectors from pool-punisher are less irregular, thus signaling the more obvious dominance between these two strategy pairs.

\begin{figure}
\centerline{\epsfig{file=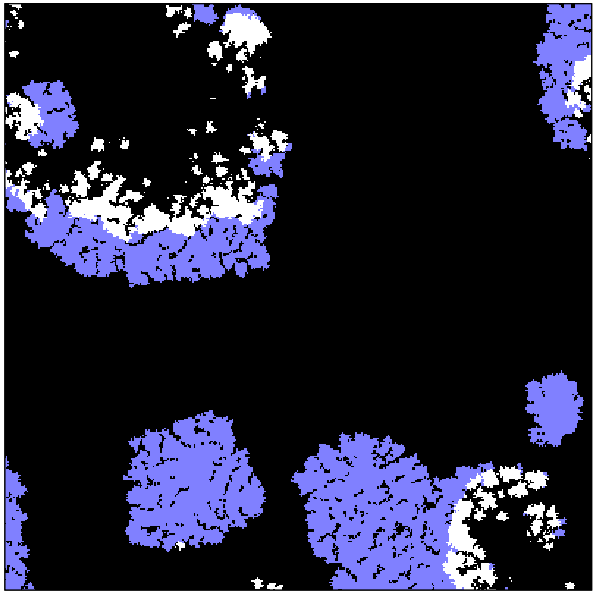,width=8cm}}
\caption{Typical distribution of strategies within the $(D+C+DO)_c$ phase on a $400 \times 400$ portion of a larger square lattice ($L=2000$) for $r=3.5$, $\beta=0.2$ and $\gamma=0.5$ in the public goods game with pool punishment. Notice that one ``strategy'' in the closed loop of dominance is actually an alliance of two strategies $(DO)$. Black, white, and blue denotes players with defector, cooperator, and pool-punisher strategy, respectively. Figure reproduced with permission from \cite{szolnoki_pre11}.}
\label{poolsnap}
\end{figure}

Another novel feature is the appearance of an additional three-strategy phase within a narrow range of the parameter space, namely the $(D+C+DO)_c$ phase. The corresponding snapshot in Fig.~\ref{poolsnap} illustrates clearly that here the cyclic invasions occur between the $D$, the $C$, and the $DO$ phase. This spatiotemporal structure can be reproduced very rarely because of the fast extinction of cooperators if the system is started from a random initial state even for $L>5000$. In such a case the system evolves into the $DO$ phase that is, however, unstable against the invasion of a cooperator block with a sufficiently large size, e.g., $10 \times 10$. It is also worth mentioning that the resultant $(D+C+DO)_c$ phase will appear only after a long transient process. While the cyclic dominance of alliances has also been observed in spatial ecological models \cite{kendall_e99, laird_e08, laird_jtb15}, the public goods game with pool punishers, however, offers an interesting new example when one strategy fights continuously against a group of strategies, resulting in a special and exotic but stable stationary solution.

In terms of the properties of the phase transitions towards the $(D+C+DO)_c$ phase, Monte Carlo simulations have revealed that the transition from $DO$ to $(D+C+DO)_c$ is a weakly first-order one, but can become continuous at higher $r$ values. The discontinuous nature of the phase transitions is likely due to the different time scales characterizing the average formation and lifetimes of the competing phases that depended on $\beta$ and $\gamma$ as well as on the multiplication factor $r$.

As can be inferred from the phase diagram and the subsequent discussion, pool punishment can be stable without any additional assumptions in the spatial public goods game. However, the resulting spatiotemporal dynamics of the model is very complex, and indeed impossible to understand without methods of statistical physics. Closely related research has also considered the competition of individual and institutional punishment in the spatial public goods games \cite{szolnoki_pre11b}, with the conclusion being that peer punishers prevail and control the system behavior in large segments of the parameter space, while pool punishers can survive only in the limit of weak peer punishment, when a rich variety of solutions is observed. Paradoxically, it has also been observed that the two types of punishment may extinguish each other's positive impact on human cooperation, resulting in the triumph of defectors.

To conclude, we note that the sustainability of pool punishment is also important, in that in a population with freeriders, punishers must be strong in numbers to keep the punishment pool from emptying. Failure to do so renders the concept of institutionalized sanctioning unsustainable. It was shown that pool-punishment in structured populations is sustainable, but only if second-order freeriders are sanctioned as well, and to such a degree that they cannot prevail. In this case, a discontinuous phase transition leads to an outbreak of sustainability when punishers subvert second-order freeriders in the competition against defectors \cite{perc_srep12}.

\subsection{The non-existent institutionalized rewarding}
\label{poolrewresults}

The somewhat provocative title of this subsection is not meant exactly literally, but as a reflection of the fact that pool rewarding is much less common in human societies then pool punishment, and that empirical research on this subject is lacking. Perhaps accordingly, a dedicated effort to study pool rewarding has not yet been made in the realm of statistical physics, at least not in the same sense as the peer punishment (see Section~\ref{peerpunresults}) and peer rewarding (see Section~\ref{peerewresults}) have been studied separately in the spatial public goods game. Nevertheless, certain studies have considered pool rewarding before, and we briefly review some of the obtained results next.

In particular, in \cite{chen_xj_fbs14} optimal distribution of incentives for public cooperation in heterogeneous interaction environments have been studied, with a focus on institutional reciprocity and the the effectiveness of pool rewarding. In the simplest case, it can be assumed that, depending on their strategies, all players receive equal incentives from the common pool. The question arises, however, what is the optimal distribution of institutional incentives? In the realm of pool rewarding, how should we best reward individuals for cooperation to thrive? The research has revealed that, if the synergetic effects of group interactions are weak, the level of cooperation in the population can be maximized simply by adopting the simplest ``equal rewards to all'' principle. On the other hand, if the multiplication factor is large, it was shown that it is then best to reward high-degree nodes more than low-degree nodes. Interestingly, for institutional punishment the same optimization problem turned out to be more complex, and its solution depends on whether absolute or degree-normalized payoffs are considered. Extensive Monte Carlo simulations and mean-field calculations revealed that degree-normalized payoffs require high-degree nodes be punished more lenient than low-degree nodes. Conversely, if absolute payoffs are considered, then high-degree nodes should be punished stronger than low-degree nodes. In this place, it is worth noting that payoff normalization in the realm of evolutionary games is of particular importance in heterogeneous interaction networks \cite{masuda_prsb07, szolnoki_pa08, santos_n08}, such as for example on scale-free networks \cite{barabasi_s99}, where the absolute payoffs can be normalized with the degree of each player to arrive at normalized payoffs.

As noted in Section~\ref{poolrewmodel}, pool rewarding exists mainly in certain subcultures of the population, where it is also often mixed with the antisocial variant of the same phenomenon. Antisocial pool rewarding has therefore been considered together with prosocial pool rewarding in \cite{szolnoki_prsb15}, but we review details of this research when focusing on evolutionary outcomes with antisocial strategies in Section~\ref{antisocialresults}, and in particular in Section~\ref{antisocialrewresults}, when we review specifically the lack of adverse effects with antisocial pool rewarding.

\begin{figure*}
\centerline{\epsfig{file=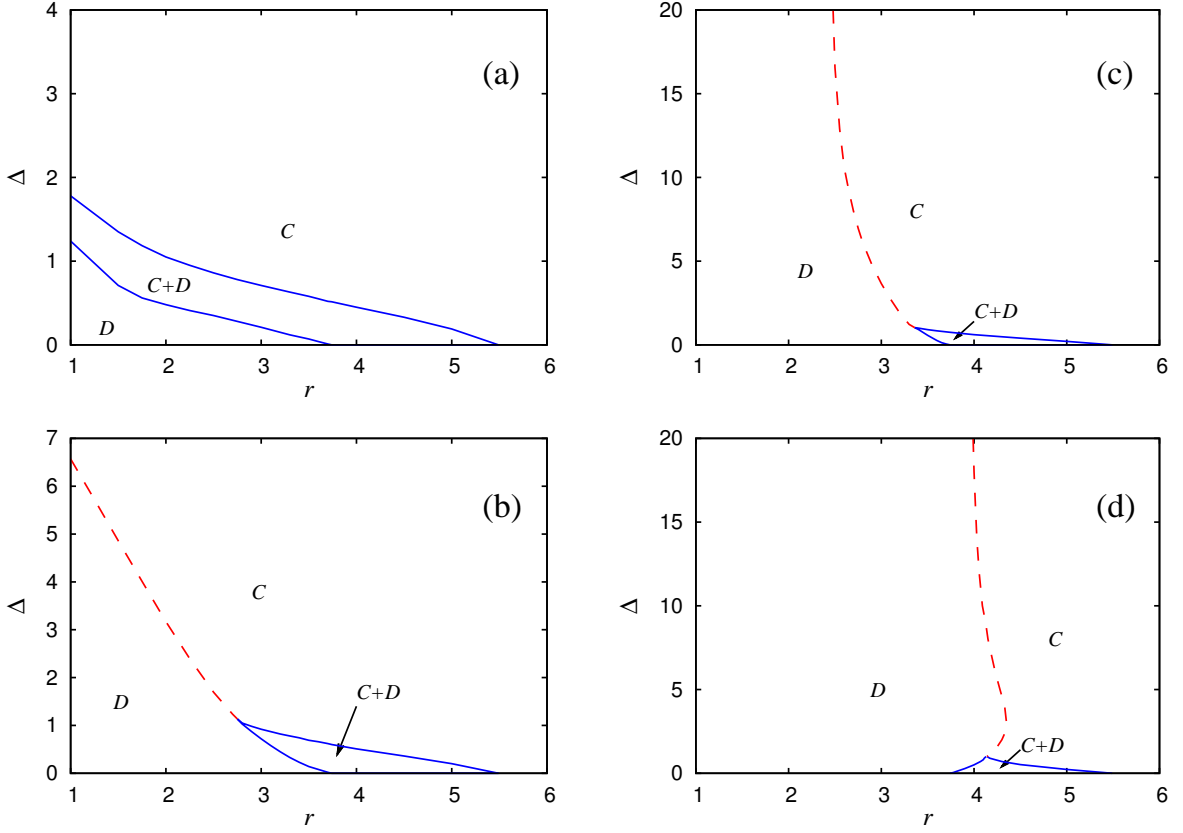,width=16.3cm}}
\caption{Phase diagrams for the spatial public goods game with adaptive punishment. Depicted are cooperators $C$ and defectors $D$ that remain on the square lattice after sufficiently long relaxation times in dependence on the synergy factor $r$ and the incremental step used for adapting the punishing activity $\Delta$. Continuous phase transitions are depicted by solid blue lines, while discontinuous phase transitions are depicted by dashed red lines. The displayed phase diagrams are for (a) $\alpha=0.1$, (b) $\alpha=0.5$, (c) $\alpha=1$, and (d) $\alpha=2$. If the punishment is not costly (panels a and b), the social dilemma can be resolved completely, and accordingly, full cooperator dominance is always possible. That is to say, there exists a sufficiently large value of $\Delta$ irrespective of $r$, such that defectors are unable to spread and are eventually completely eliminated from the population. In this case, cooperators emerge by means of a continuous phase transition from the mixed $C+D$ phase, while for somewhat larger values of $\alpha$ (panel b) the outbreak of cooperation is possible also by means of a discontinuous phase transition. As the sanctioning becomes costly (panels c and d), however, the limits of adaptive punishment become clearly inferable. With increasing values of $\alpha$, both the survivability (mixed $C+D$ phase) as well as the potential dominance (pure $C$ phase) of cooperators shift towards larger values of $r$, whereby as by smaller values of $\alpha$, both continuous as well as discontinuous phase transitions can be observed. Figure reproduced with permission from \cite{perc_njp12}.}
\label{adaptivepunphase}
\end{figure*}

\section{Self-organization of incentives for cooperation}
\label{adaptiveresults}
The assumption thus far, in Sections~\ref{peeresults} and \ref{poolresults}, has been that when somebody rewards or punishes, this is always done equally strongly. Evidently, the reality is different, in that sometimes people are more keen on punishing wrongdoers, while at other times less so. The same considerations apply to rewarding. Here the concept of self-organization provides an apt addition to the mathematical models, where thus the adaptive nature of rewarding and punishing can be taken into account \cite{perc_njp12, szolnoki_njp12}. Akin to the concept of self-organization is also the concept of sharing the cost of such additional prosocial actions, which has also turned out to work well in resolving high-profile obstacles towards human cooperation, such as the problem of costly punishment \cite{chen_xj_njp14}. In what follows, we review research done in the realm of these considerations.

\subsection{Enhanced network reciprocity due to adaptive punishment}
\label{adaptivepunresults}

Adaptive peer punishment on a square lattice has been studied in \cite{perc_njp12}. It has been shown that allowing players to adapt their sanctioning efforts in dependence on the success of
neighboring defectors can explain both, the spontaneous emergence of punishment, as well as its ability to deter defectors and those unwilling to punish them. And perhaps most importantly, this is achieved with globally negligible investments, such that the cost of punishment in the population as a whole does not threaten to nullify the benefits stemming from increased levels of human cooperation. The process of self-organization thus significantly elevates the effectiveness of punishment, and it reveals new mechanisms by means of which this fascinating and widespread social behavior could have evolved.

\begin{figure*}
\centerline{\epsfig{file=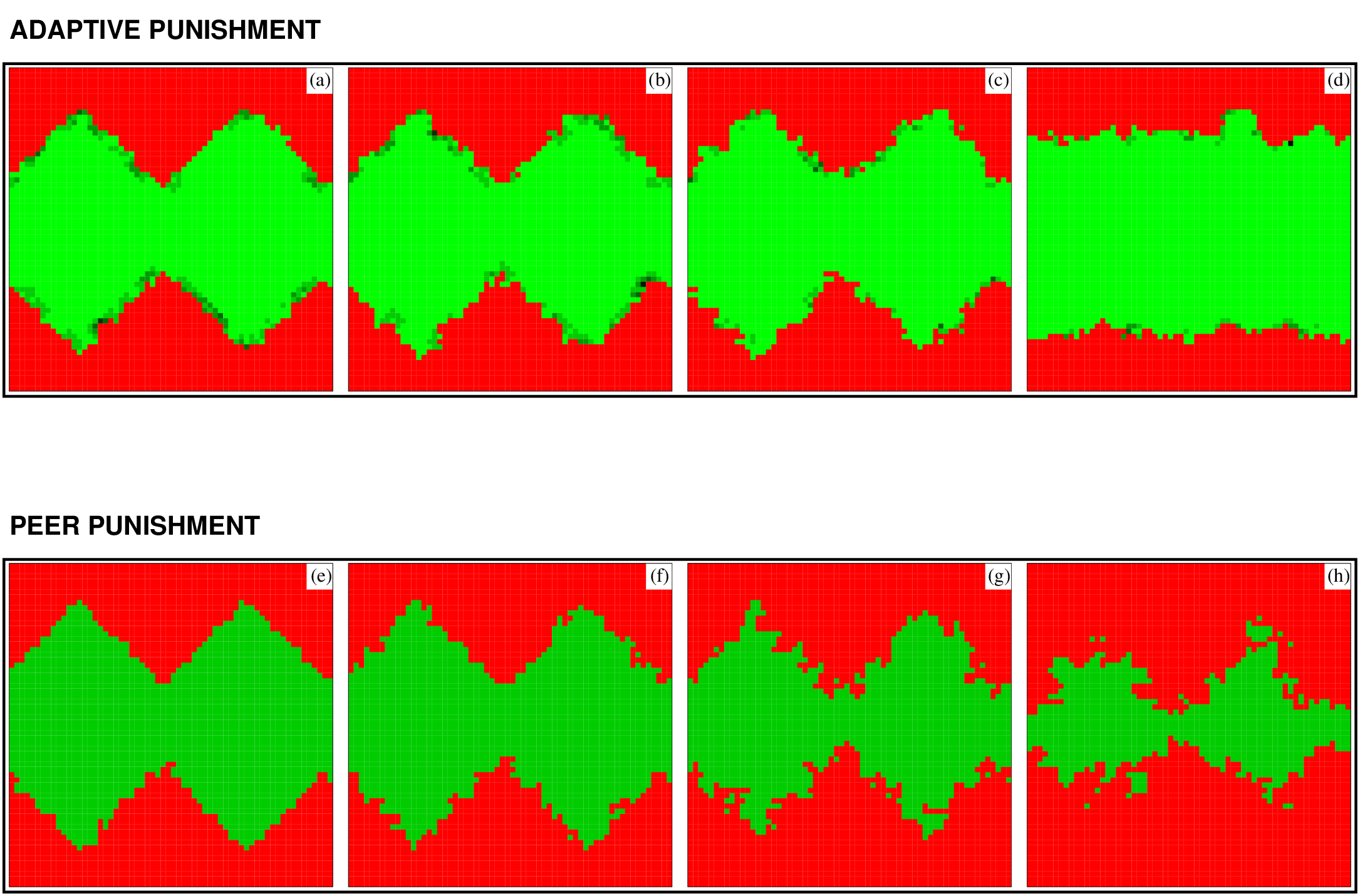,width=16.5cm}}
\caption{The recovery and preservation of smooth interfaces that separate cooperators and defectors enhances network reciprocity. Depicted are characteristic snapshots of the square lattice over time, where cooperators are depicted green and defectors are depicted red. Darker shades of green indicate a higher punishing activity, while non-punishing cooperators are depicted bright green. A prepared initial state, corresponding to a rough interface, is used to reveal the difference in the impact of the two punishing models. Panels (a), (b), (c) and (d) show the evolution of the adaptive punishment model at $1$, $3$, $10$ and $500$ Monte Carlo steps, respectively. Panels (e), (f), (g) and (h), on the other hand, show the evolution of the traditional peer punishment model at $2$, $5$, $10$ and $20$ Monte Carlo steps. Note that all the punishing cooperators in panels (e), (f), (g) and (h) are depicted with a slightly darker shade of green, as representative for players having punishing activity $\pi_x=1$. The parameters used for both models were $r=2.8$, $\alpha=1$, $\Delta=7.4$, and a small $L=62$ linear system size for clarity. The final state in the adaptive punishment model is a pure $C$ phase, while the peer punishment model yields a pure $D$ phase (both not shown). Figure reproduced with permission from \cite{perc_njp12}.}
\label{adaptivepunsnap}
\end{figure*}

To corroborate above statements, we first present in Fig.~\ref{adaptivepunphase} full $r-\Delta$ phase diagrams for different values of $\alpha$ (see Section~\ref{adaptivepunmodel} for the definition of the model). Depending on the parameter values, cooperators ($C$) and defectors ($D$) can dominate completely, although a coexistence of the two strategies ($C+D$) is possible as well. In panel (a), when the sanctioning is cheap (note that $\alpha=0.1$ implies that punishment costs are only $1/10$ of the fines imposed on defectors), already small values of $\Delta$ suffice to restore a mixed $C+D$ phase or even a pure $C$ phase at low synergy factors where otherwise defectors would reign supreme. Continuous phase transitions, depicted by solid blue lines, are the only means through which the stability of the two strategies changes. As the cost of punishment is raised, as in panel (b) to $\alpha=0.5$, introduces a qualitative change in the evolutionary dynamics. Below $r=2.7$ the mixed $C+D$ phase is no longer possible. Instead, discontinuous phase transitions, depicted by the dashed red line, lead to the complete dominance of cooperators for sufficiently high values of $\Delta$. Expectedly, the lower the synergy factor $r$, the higher the value of $\Delta$ required to reach the pure $C$ phase. Still, even at $r=1$ cooperators are able to dominate completely if $\Delta>6.5$. At the border between inexpensive and costly punishment, in panel (c), where $\alpha=1$ and thus the fine and cost of punishment are equal, the evolutionary dynamics does not change but we first observe a hard limit in $r$, below which cooperation can no longer be sustained, irrespective of how large $\Delta$ is. For $\alpha=1$ we find the limiting synergy factor to be $r=2.48$, which, however, is still well below $r>3.74$ where cooperators are able to survive in the absence of punishment. Increasing the cost of sanctioning further to $\alpha=2$ in panel (d) increases the minimal value of $r$ where cooperators are able to survive, and the pure $C$ phase is unattainable if $r<3.99$.

With the analysis of emerging spatial patterns, it is possible to show that adaptive punishment promotes cooperation either through the enhancement of network reciprocity, or the prevention of the emergence of cyclic dominance, or through the provision of competitive advantages to those that sanction antisocial behavior. For the last two we refer to the original research \cite{perc_njp12}, while the enhancement of network reciprocity, also in comparison to traditional peer punishment (see Section~\ref{peerpunresults}), is demonstrated in Fig.~\ref{adaptivepunsnap}. In agreement with the definition of the model incorporating adaptive punishment (see Section~\ref{adaptivepunmodel}), it is important to note that cooperators with non-zero punishing activity ($\pi_x>0$) can exist only along the borders separating the $C$ and $D$ domains. The pure $C$ phase, as well as interiors of large cooperative domains, lack cooperators having $\pi_x>0$ because there is a constant drift towards non-punishment if there are no $D \to C$ invasions occurring in the neighborhood.

We therefore focus on the evolutionary dynamics along the interfaces separating the $C$ and $D$ domains. We note that previous studies emphasized that smooth interfaces between the competing strategies are beneficial for cooperators because it allows the network reciprocity to take full effect. On the contrary, rough interfaces provide ample opportunities for defectors to invade and spread, even at relatively high synergy factors. To demonstrate the positive effect of adaptive punishment, it is instructive to start the simulation from a prepared initial state corresponding to a rough interface, as depicted in Fig.~\ref{adaptivepunsnap}a (adaptive punishment) and Fig.~\ref{adaptivepunsnap}e (peer punishment). By following the snapshots in the lower row (panels e, f, g and h) from left to right, we can observe that peer punishment efforts fail to restore the broken phalanx of cooperators. Due to the additional roughening of the interfaces defectors can invade the cooperative domain very effectively, eventually leading to a pure $D$ phase (not shown). For clarity in this case traditional cooperators were taken out of the model, with only peer punishers and defectors left to compete. Despite of this lenient predisposition that eliminates potential negative consequences of second-order freeriding, however, steady punishment still fails for the considered parameter values.

In contrast, the impact of adaptive punishment is significantly different, as can be observed by following the snapshots in the upper row of Fig.~\ref{adaptivepunsnap} (panels a, b, c and d) from left to right. Here the localized and temporarily very active punishers, which emerge spontaneously as a response to the $D \to C$ invasions, succeed in restoring a smooth (straight) interface between cooperators and defectors. This in turn disables defectors to invade, thus effectively enhancing network reciprocity. The demonstrated recovery and preservation of smoothness along the interfaces is a key advantage that is conferred by adaptive punishment. Importantly, once the regularity of the interfaces is re-established the enhanced effectiveness of network reciprocity spontaneously introduces a decrease in the punishing activity, before eventually the latter altogether seizes once the pure $C $phase is reached. Adaptive punishment thus allows for a spontaneous but prompt and determined response to a threatening invasion of defectors, which is unattainable with previously reviewed models.

\begin{figure*}
\centerline{\epsfig{file=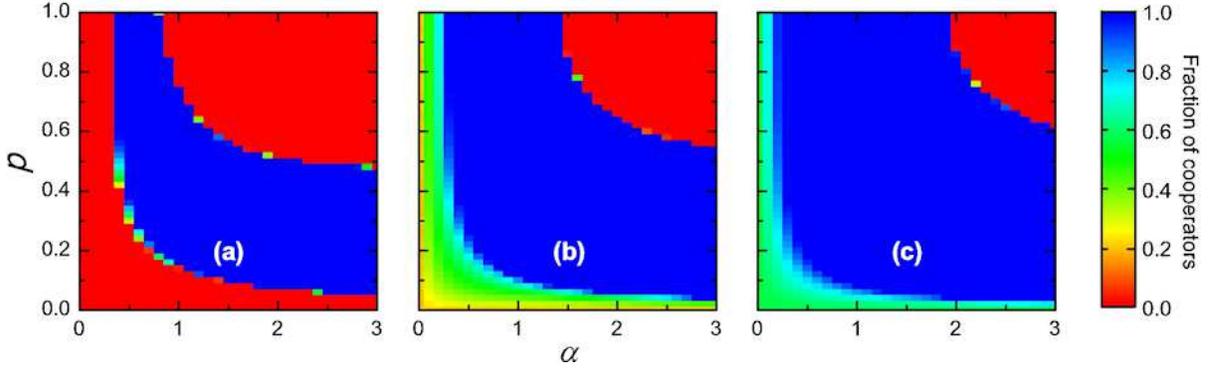,width=16cm}}
\caption{Probabilistic sanctioning in the spatial public goods game promotes the evolution of public cooperation, yet the optimal outcome requires carefully adjusted severity and frequency of punishment. Color maps encode the fraction of cooperators in dependence on the punishment fine $\alpha$ and the probability to punish $p$, as obtained for multiplication factors $r=3.6$ (a), $r=3.9$ (b), and $r=4.2$ (c). Figure reproduced with permission from \cite{chen_xj_njp14}.}
\label{sharepunphase}
\end{figure*}

We conclude this subsection by mentioning that adaptive punishment has also been considered outside the realm of statistical physics research in \cite{boyd_s10}, where it was shown that coordinated punishment of defectors sustains cooperation and can proliferate when rare. The model eliminates key problems associated with punishment, such as the difficulties of explaining the evolutionary emergence of punishment because rare unconditional punishers bear substantial costs and are hence eliminated, or that the sum of costs to punishers and their targets often exceeds the benefits of the increased cooperation that results from the punishment of freeriders. Essentially, the conclusions are thus much the same as reviewed above, which can be seen as an endorsement of the employed methodology.

\subsection{Probabilistic sharing solves the problem of costly punishment}
\label{adaptiveshareresults}
Although conceptually different from adaptive punishment reviewed in the previous subsection, sharing the effort of punishment in a probabilistic manner can also be considered an adaptive from of punishment. The model is a simple alteration of the peer punishment model reviewed in Section~\ref{peerpunmodel}, in that here a probability $p$ is introduced with which cooperators within the group are selected randomly and designated as peer punishers \cite{chen_xj_njp14}. Inspired by the fact that humans have strong but also emotional tendencies for fair play, probabilistic sanctioning was simply the simplest way of distributing the duty. Based on this model, it has been shown that sharing the responsibility to sanction defectors rather than relying on certain individuals to do so permanently can solve the problem of costly punishment, i.e., that punishment can be successful even if the cost of sanctioning is significantly larger than the imposed fines on defectors.

Color maps presented in Fig.~\ref{sharepunphase} depict the stationary fraction of cooperators in dependence on the punishment fine $\alpha$ and the probability to punish $p$ for three intermediate values of the multiplication factor $r$. Going from panel (a) to panel (c), it can be observed that cooperative behavior becomes more and more common, which is expected given that the benefits of collaborative efforts increase through larger values of $r$. The impact of $\alpha$ and $p$ is more subtle. As the values of the two parameters increase along the diagonal in the $\alpha-p$ plane, the fraction of cooperators first increases, reaches a maximum, but then again decreases. Increasing either of the two parameters while the other is kept constant returns the same observation. Both $\alpha$ and $p$ thus have a non-monotonous impact on the cooperation level. At smaller values of $r$, as in panel (a), this distinctive feature is more pronounced, but it remains present at higher values of $r$ as well, as in panel (b) and (c). Probabilistic sanctioning thus promotes human cooperation, yet it requires carefully measured efforts both in terms of severity and frequency of punishment.

\begin{figure*}
\centerline{\epsfig{file=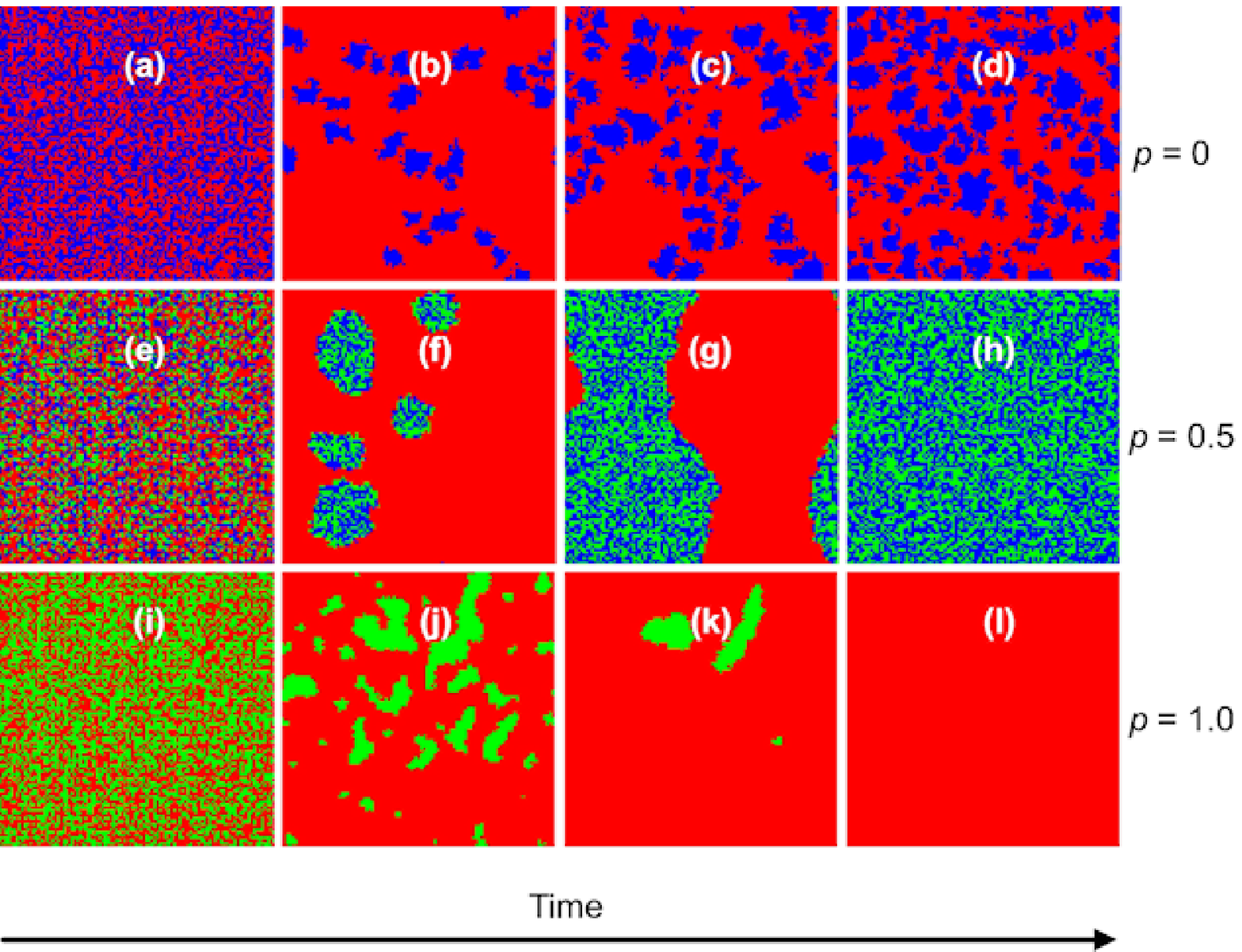,width=12cm}\epsfig{file=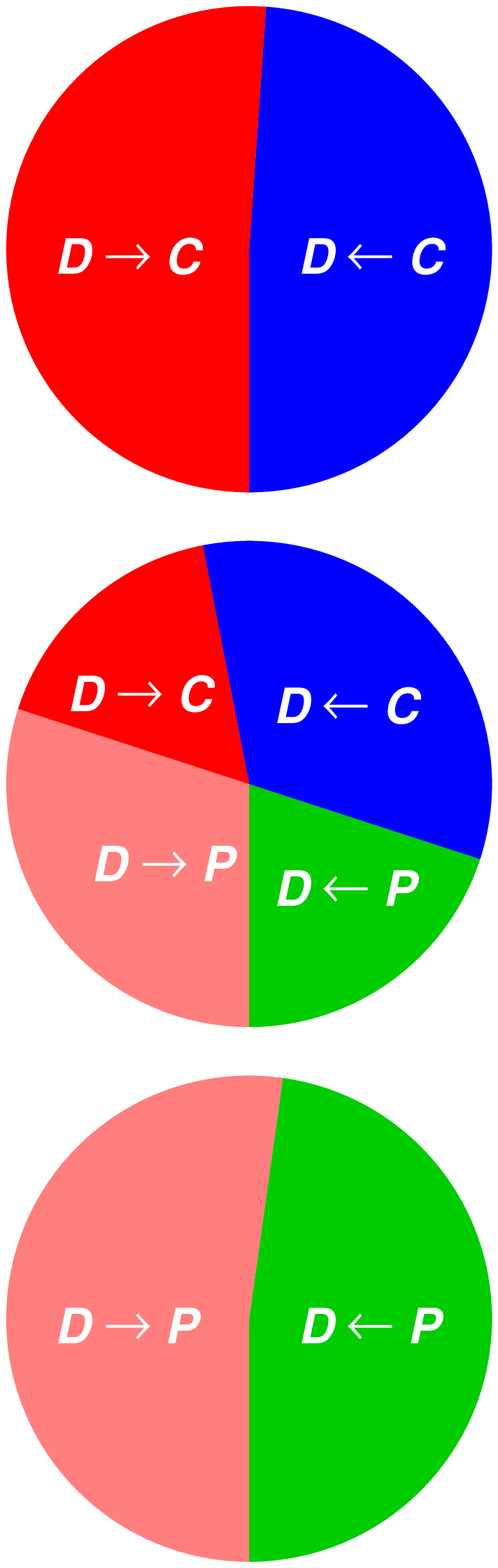,width=4cm}}
\caption{Spatial pattern formation reveals evolutionary advantages of probabilistic sanctioning. In the absence of punishers (panels a to d) cooperators alone struggle to uphold compact cooperative clusters. If everybody punishes the costs of sanctioning are prohibitive to success and defectors win (panels i to l). If the responsibility to sanction is shared 50:50 randomly, cooperative clusters remain compact and smooth, and at the same time their fitness is superior to that of defectors (panels e to h). The direction of invasion therefore reverses and cooperators win. Cooperators who are willing to punish defectors in at least three out of the five groups are depicted green, while other cooperators are depicted blue. Defectors are depicted red. Pie diagrams on the right show the corresponding ratio of elementary invasions between different strategy pairs, confirming that probabilistic sanctioning tips the balance in favor of cooperation. Different shades of red have been used to distinguish between $D \to C$ and $D \to P$ invasions. In all three cases the evolution starts from a random initial state using $r=4$ and $\alpha=2$. Figure reproduced with permission from \cite{chen_xj_njp14}.}
\label{sharepunsnaps}
\end{figure*}

As often before, an understanding of the results presented in Fig.~\ref{sharepunphase} can be obtained with the study of spatial patterns that emerge under the influence of probabilistic sanctioning. In Fig.~\ref{sharepunsnaps}, characteristic snapshots of the square lattice for three different values of $p$ are presented. It turns out that, when plotting the spatial distributions of strategies, it is often helpful to use different colors to distinguish cooperators based on their propensity to punish. Cooperators that are randomly selected as punishers in at least three of the five groups in which they are involved are depicted green, while other cooperators are depicted blue. Defectors are depicted red. If punishment is not an option ($p=0$), cooperators have to rely solely on network reciprocity to survive in the presence of defectors. As panels (a) to (d) illustrate, cooperators form small yet compact clusters that protect them from the invasions of defectors. This is the hallmark of network reciprocity, discovered first by Nowak and May \cite{nowak_n92b}. It is important to note that in the absence of punishment the interfaces that separate cooperators and defectors are not smooth. This creates ample opportunities for defectors to invade successfully, but it also quickly leaves them surrounded by players of the same kind. Since locally there is nobody left to exploit the invasion is stopped, but it also creates new irregularities along the interface which will invite further invasions in the future. The dynamical equilibrium of these elementary processes yields a stable coexistence of cooperators and defectors. At the other extreme, if all cooperators are always ready to punish ($p=1$), the morphology of the spatial patterns is slightly different. As panels (j) and (k) in Fig.~\ref{sharepunsnaps} illustrate, due to the consistent application of punishment the interfaces are somewhat smoother. Individual defectors deep in the bulk of punishers struggle to invade because they are immediately sanctioned. At the same time, the cost of sanctioning is shared by many punishers, which conveys them a local evolutionary advantage. However, at the front where many defectors meet with punishers the cost of sanctioning become prohibitive, and ultimately defectors easily prevail, as shown in panel (l). If the application of sanctioning is probabilistic ($p=0.5$), the direction of invasion is reversed. As illustrated in panels (e) to (h), defectors are eventually completely eliminated from the population. This is because probabilistic sanctioning preserves the smoothness of cooperative interfaces, while at the same time the mixture of pure cooperators and punishers can prevail in the direct competition against defectors. Paradoxically, the option to resort to second-order freeriding provides the necessary relief from the punishment costs, which in turn maintains a healthy fitness of the cooperative domains. The key to success is that the costs of sanctioning are shared.

It is also instructive to monitored the elementary invasion processes between the competing domains of strategies, the results of which are summarized as pie diagrams that depict the ratios of different invasion steps at corresponding values of $p$ at the right of Fig~\ref{sharepunsnaps}. The pie diagrams confirm that the frequency of defector invasions for $p=0$ and $p=1$ is higher than the frequency of cooperator invasions, which ultimately results in states where defection is widespread, as in panels (d) and (l). For $p=0.5$, on the other hand, the combined frequency of $C \to D$ and $P \to D$ invasions is higher than the combined reverse, and as a result the cooperators collectively rise to complete dominance. A careful comparison reveals further that the majority of invasion steps that reduce the number of defectors is due to cooperators that do not punish. In other words, second-order freeriders become stronger against defectors due to the probabilistic presence of punishers. The pie diagrams also highlight that $C$ can beat $D$ only in the presence of $P$, thus indicating that a multi-point interaction is necessary to observe the reported counterintuitive phenomenon.

We conclude this subsection by noting that above observations can be summarized as ``two weaker strategies are able to form a stronger one'', which is reminiscent of the famous Parrondo's paradox \cite{harmer_n99,parrondo_prl00}, where two losing games, if combined, can become a winning game.

\subsection{Evolutionary advantages of adaptive rewarding}
\label{adaptiverewresults}

\begin{figure}
\centerline{\epsfig{file=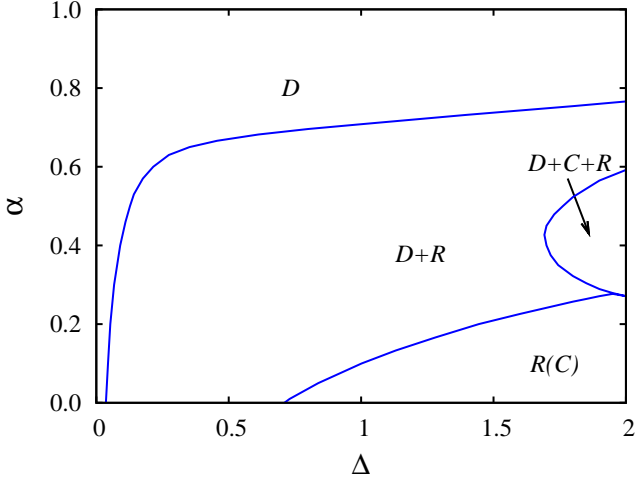,width=8.5cm}}
\caption{Full $\Delta-\alpha$ phase diagram for the public goods game with adaptive rewarding, as obtained for $r=3.5$ and $K=0.5$. Blue solid lines depict continuous, second-order phase transitions, and symbols mark the surviving strategies in the stationary state. Since the multiplication factor $r$ is too small, cooperators can not survive alone in the presence of defectors. Accordingly, the $D+C$ phase is missing. Instead, as $\Delta$ increases, and if $\alpha$ is sufficiently small, the pure $D$ phase gives way to the two-strategy $D+R$ phase, which may further transform into the three-strategy $D+C+R$ phase. At $r=2$, for example (not shown), the three-strategy phase is no longer attainable on the considered $\Delta-\alpha$ plane. For small rewarding costs the defector-free $R(C)$ phase is obtained, although its area shrinks continuously as $r$ increases. Notably, in the absence of defectors strategies $R$ and $C$ become equivalent. The evolutionary process therefore proceeds via slow logarithmic coarsening, as in the voter model \cite{dornic_prl01}, but since at the time of extinction of defectors the majority of players are rewarding cooperators, the system finally arrives at the $R$ phase with a significantly higher probability than to the $C$ phase. Figure reproduced with permission from \cite{szolnoki_njp12}.}
\label{adaptiverewphase}
\end{figure}

In the spirit of properly addressing the long-standing ``stick versus carrot'' dilemma \cite{andreoni_aer03, hilbe_prsb10}, the adaptive punishment model reviewed in Section~\ref{adaptivepunresults} deserve a counterpart with rewarding. Adaptive peer rewarding on a square lattice has therefore been studied in \cite{szolnoki_njp12}, and we have reviewed the model in Section~\ref{adaptiverewmodel}. Statistical physics research in the realm of this model has revealed that allowing for the act of rewarding to self-organize in dependence on the success of cooperation creates several evolutionary advantages that instill new ways through which collaborative efforts are promoted. Ranging from indirect territorial competition, which we have reviewed in detail in Section~\ref{peerpunresults} in the realm of peer punishment, to the spontaneous emergence of cyclic dominance, which we have reviewed in detail in Section~\ref{peerewresults} in the realm of peer rewarding, phase diagrams and the underlying spatial patterns for adaptive rewarding in the spatial public goods game reveal fascinatingly reach social dynamics that explains why this costly behavior has evolved and persevered. At the same time, comparisons with adaptive punishment have uncovered an Achilles heel of adaptive rewarding that is due to over-aggression, which in turn hinders optimal utilization of network reciprocity.

Figure~\ref{adaptiverewphase} features a representative phase diagrams for the public goods game with adaptive punishment, as obtained for $r=3.5$. Interestingly, discontinuous phase transitions are absent, although they do occur in the model for higher values of $r$ (see Fig. 2 in \cite{szolnoki_njp12}). It can be observed that, if the cost of rewarding $\alpha$ ($\alpha$ is actually the ratio between the cost of rewarding and the reward that is allotted to cooperators) is substantial, defectors are the only ones to survive. Naturally, the lower the value of $r$ the lower the value of $\alpha$ that still warrants defector dominance. The pure $D$ phase becomes the two-strategy $D+R$ phase by means of a continuous phase transition, if only the value of $\Delta$, which is the incremental step used for the rewarding activity, is not too small and the value of $\alpha$ is not too large. Continuing further towards more efficient rewarding may lead to the defector-free $R(C)$ state, which has the same properties as the voter model \cite{dornic_prl01}, where the coarsening is logarithmically slow.

\begin{figure}
\centerline{\epsfig{file=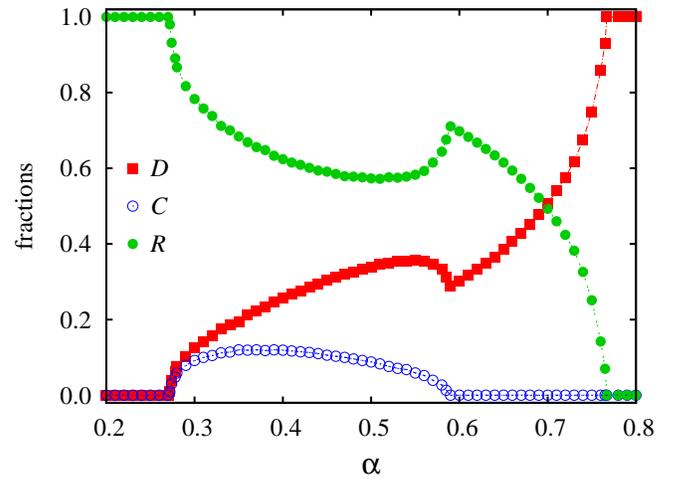,width=8.5cm}}
\caption{A cross section of the phase diagram presented in Fig.~\ref{adaptiverewphase}, as obtained for $\Delta=2.0$. As $\alpha$ increases the rewarding cooperators first give way to a three-strategy $D+C+R$ phase, but further on persevere longer than second-order freeriders. At smaller values of $r$ the latter require a delicate balance of conditions to survive, and can do so only along the $D+R$ interfaces. Figure reproduced with permission from \cite{szolnoki_njp12}.}
\label{adaptiverewcross}
\end{figure}

The second-order freeriders, i.e., traditional cooperators who do not reward, on the other hand, can survive only in the three-strategy $D+C+R$ phase, but its existence is limited to high values of $\Delta$, intermediate values of $\alpha$, and still sufficiently high values of $r$, as can be observed in Fig.~\ref{adaptiverewphase}. Importantly, in this phase cooperators cannot survive alone if surrounded solely by defectors. In fact, they can survive only where defectors and rewarding cooperators meet, i.e., along the $D+R$ interfaces. Precisely how this phase emerges is inferable from the cross-section of the phase diagram presented in Fig.~\ref{adaptiverewcross}, which reveals that as $\alpha$ exceeds a critical value the efficiency of $R$ weakens to the point where defectors are able to survive. The stable presence of a small fraction of cooperators, surviving at the $D+R$ interfaces, accompanies this transition. Interestingly, as $\alpha$ is further increased the first to extinct are not rewarding cooperators but traditional cooperators, who fail to harvest the benefits of decreased rewarding efficiency. This indicates that, especially at small synergy factors, only a fine balance of all the other parameters enables the survival of second-order freeriding.

\begin{figure*}
\centerline{\epsfig{file=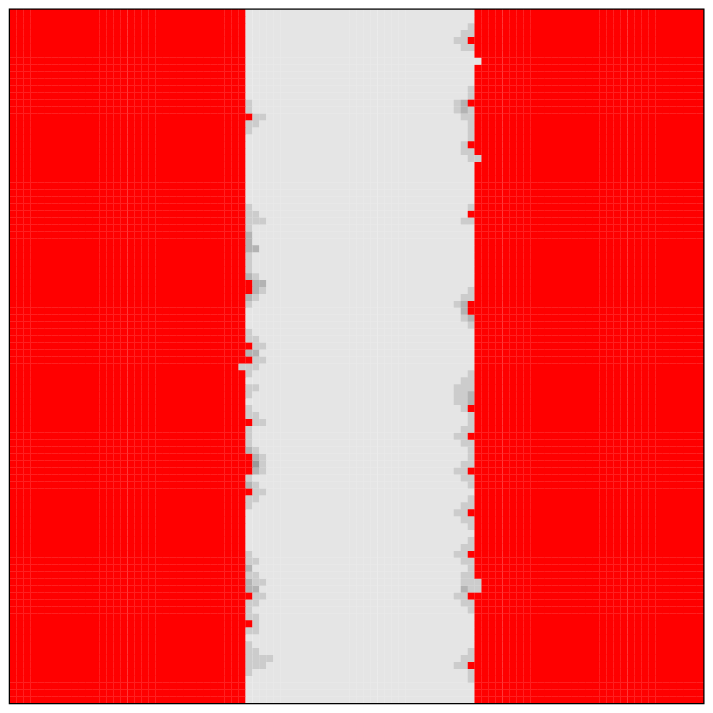,width=2.7cm}\epsfig{file=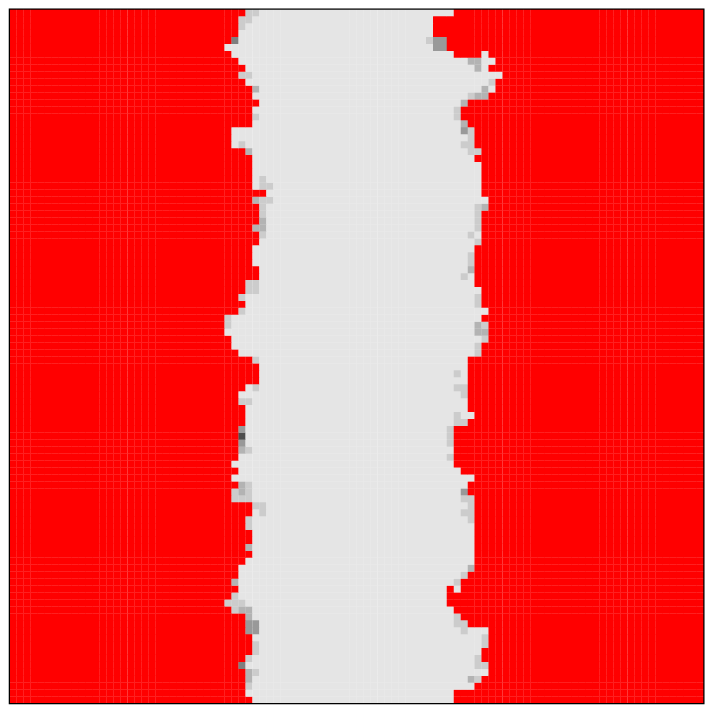,width=2.7cm}\epsfig{file=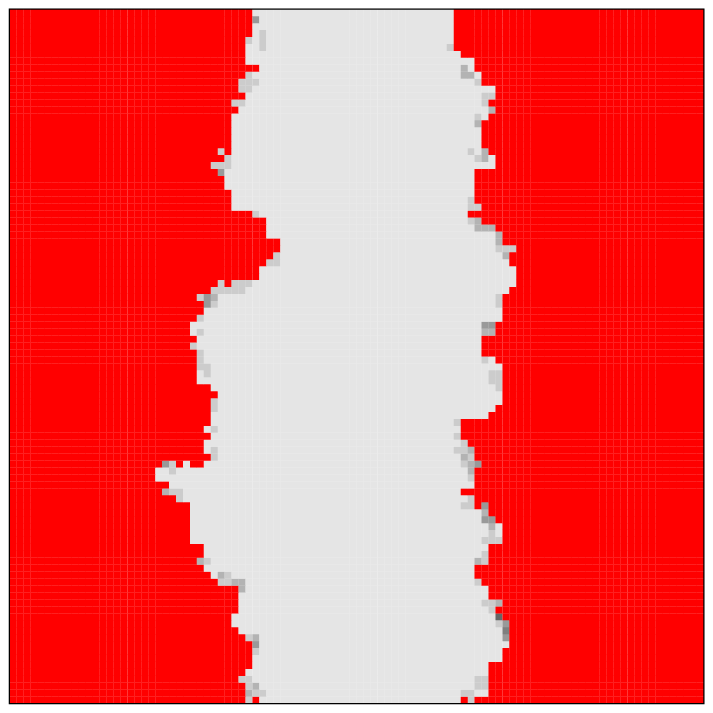,width=2.7cm}\epsfig{file=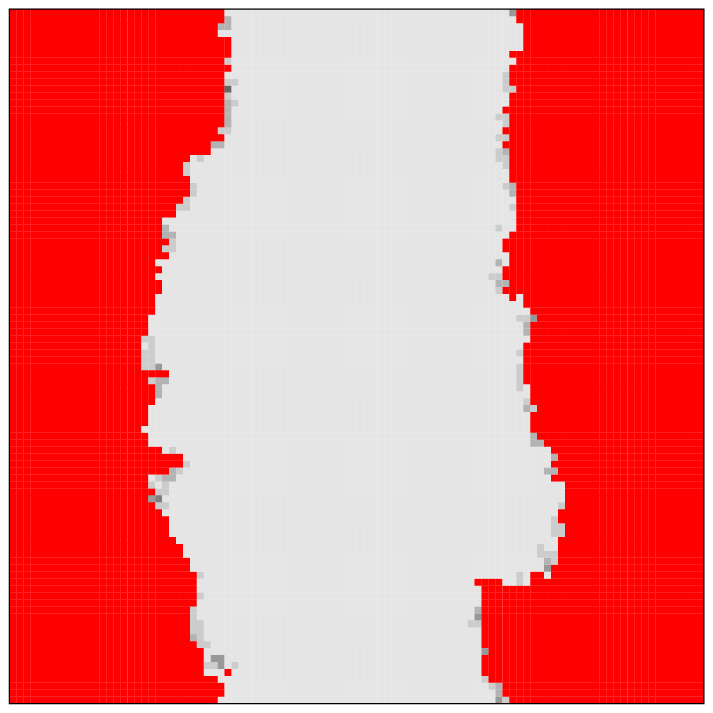,width=2.7cm}\epsfig{file=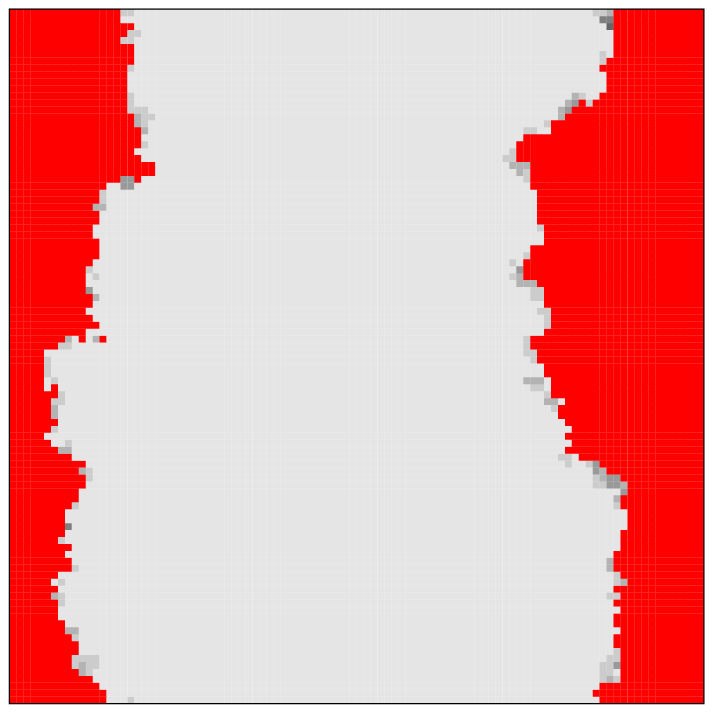,width=2.7cm}\epsfig{file=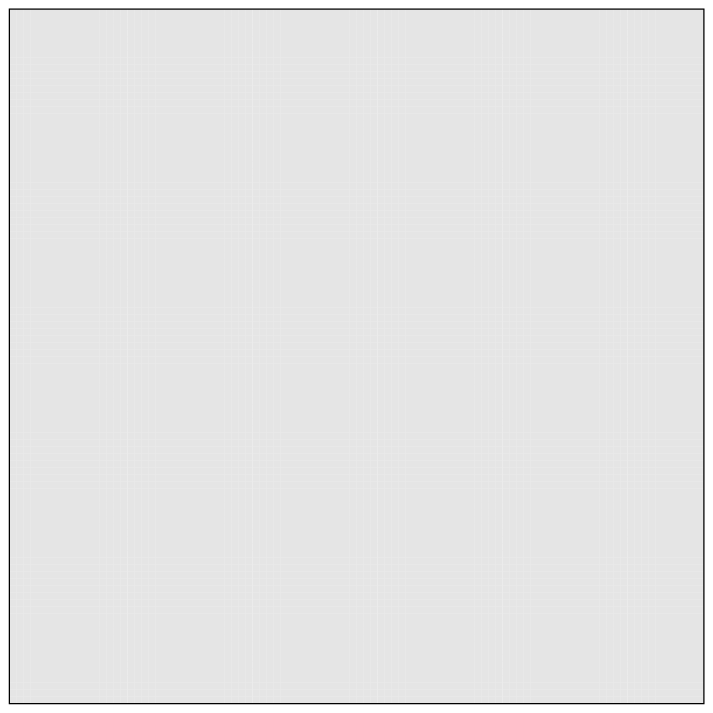,width=2.7cm}}
\centerline{\epsfig{file=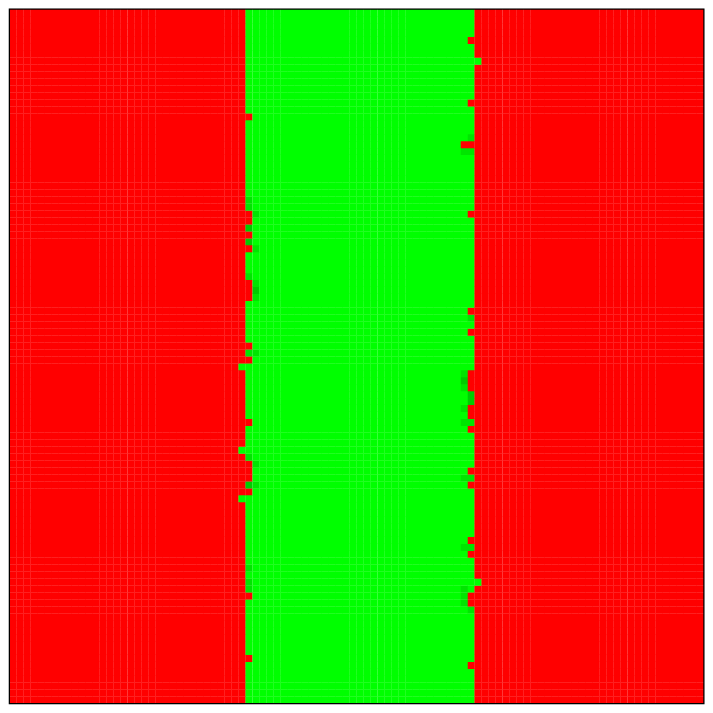,width=2.7cm}\epsfig{file=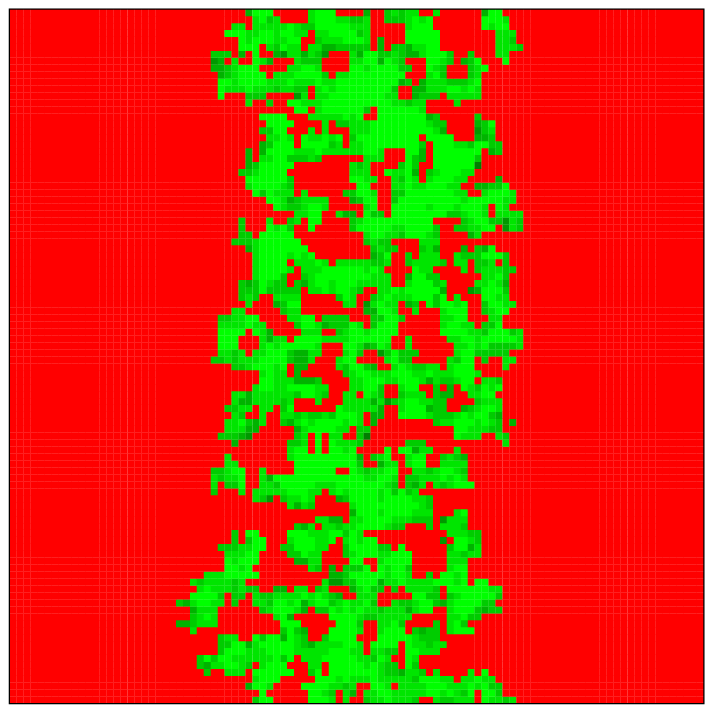,width=2.7cm}\epsfig{file=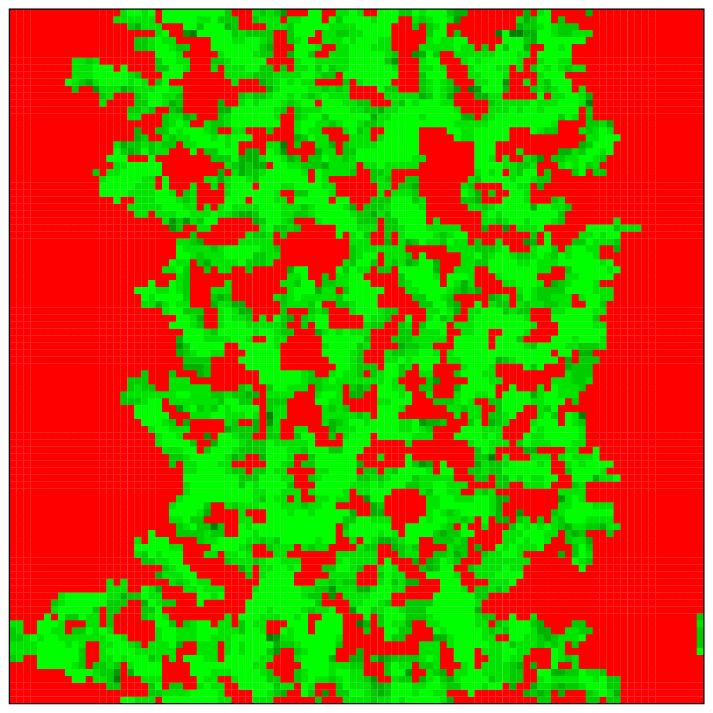,width=2.7cm}\epsfig{file=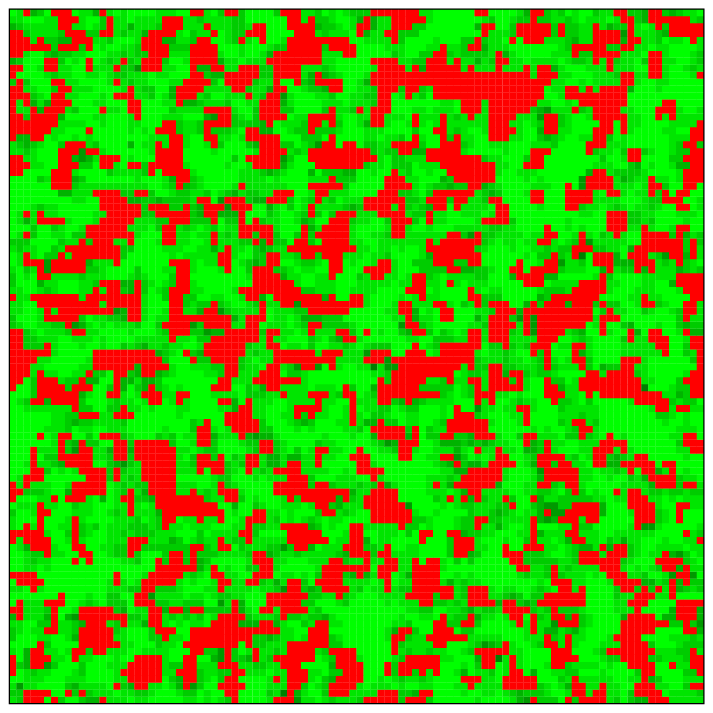,width=2.7cm}\epsfig{file=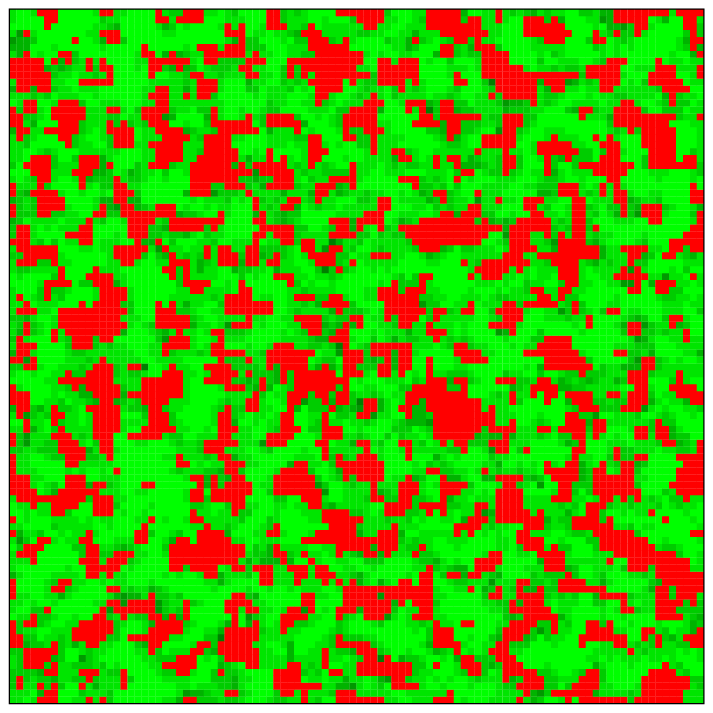,width=2.7cm}\epsfig{file=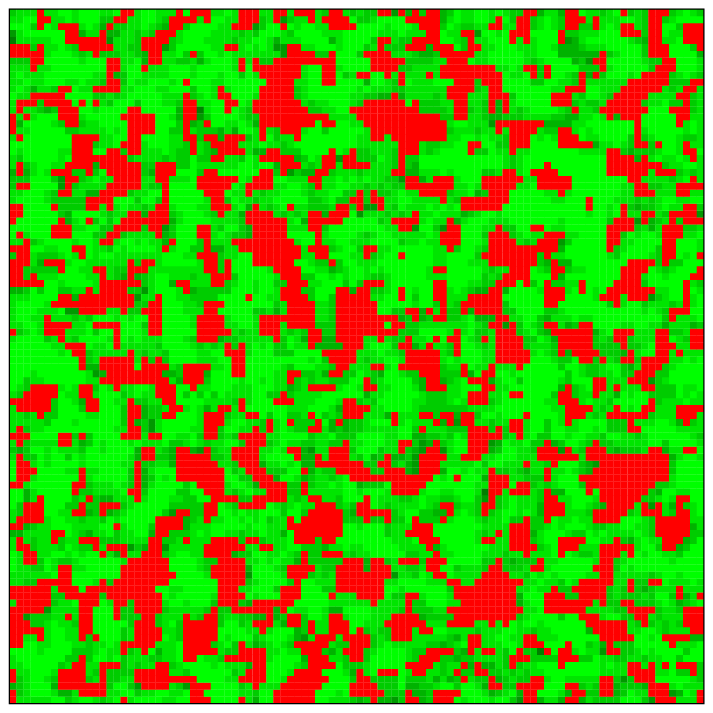,width=2.7cm}}
\caption{Comparison of the evolution of interfaces separating adaptive punishers and defectors (top row), and adaptive rewarding cooperators and defectors (bottom row). It can be observed that, while rewarding cooperators (green) advance faster into the territory of defectors (red), the punishing cooperators (gray) are relentlessly bent on keeping their phase compact. Although the punishers therefore advance slower, they ultimately succeed in completely eliminating the defectors. Rewarding cooperators, on the other hand, have to make do with their coexistence. Note that darker shades of gray (green) denote players with higher punishing (rewarding) activity. The parameter values are the same for both cases, namely $r=2$, $\Delta=2$ and $\alpha=0.4$, while the snapshots were taken at $1$, $70$, $300$, $1000$, $3000$ and $6000$ full Mote Carlo Steps. Figure reproduced with permission from \cite{szolnoki_njp12}.}
\label{adaptiverewsnaps}
\end{figure*}

We conclude this subsection by reviewing results where the efficiency of adaptive rewarding is directly compared with the efficiency of adaptive punishment, thus addressing the ``stick versus carrot'' dilemma. By determining the minimally required value of $\Delta$ that warrants the complete elimination of defectors with either adaptive rewarding or adaptive punishment, it was shown that adaptive punishment, which was studied separately in \cite{perc_njp12} and reviewed in Section~\ref{adaptivepunresults}, is more effective than adaptive rewarding in warranting defector-free states (see Fig.~7 in \cite{szolnoki_njp12}). An intuitive explanation as to why this is the case is presented in Fig.~\ref{adaptiverewsnaps}, where we can follow the evolution of interfaces separating defectors and adaptively punishing cooperators (top row) as well as defectors and adaptively rewarding cooperators (bottom row) under identical conditions. It can be observed that while rewarding cooperators are more successful in penetrating the area of defectors, the punishing cooperators advance less fast but maintain a compact phase. For example, in the third snapshot from the left, some rewarding cooperators have already reached the border of the lattice while punishing cooperators have yet to advance notably. However, rewarding cooperators have to pay a price for their over-aggressive invasion, namely an irregular interface that facilitates the coexistence with defectors. Paradoxically, the less aggressive effect of punishment, which focuses on repairing the cracks in the phalanx rather than on advancing into the territory of defectors at any cost, turns out to be more effective at the end. Punishing cooperators rise to complete dominance with the aid of a near flawless support of network reciprocity \cite{nowak_n92b} (see Section~\ref{adaptivepunresults} for more details). Rewarding cooperators, on the other hand, sacrifice the latter for a faster advancement, but therefore fail to create the desired defector-free state. The Achilles heel of adaptive rewarding is thus an excessively aggressive invasion of defectors that neglects the benefits of network reciprocity. This, in turn, may explain why, despite of its success, rewarding is not as firmly weaved into our societal organization as punishment.

\section{Evolutionary outcomes with antisocial strategies}
\label{antisocialresults}
Another important but thus far overlooked aspect of punishment and rewarding is that people do not always use this actions for the right reasons. Sometimes people are rewarded for antisocial actions, and sometimes punishment is applied to people despite their prosocial activities. In fact, research across different human societies has shown that antisocial punishment is significantly more common than one might have expected \cite{herrmann_s08}. There also exist evidence that human punishment is motivated by inequity aversion rather than by the desire for reciprocity \cite{raihani_bl12}, which also leaves ample room for reasoning as to why someone might engage in misuse of punishment. In what follow, we review research on antisocial strategies in the public goods game separately for punishment and rewarding.

\subsection{Antisocial punishment and the crumbling of cooperation enforcement}
\label{antisocialpunresults}

\begin{figure}
\centerline{\epsfig{file=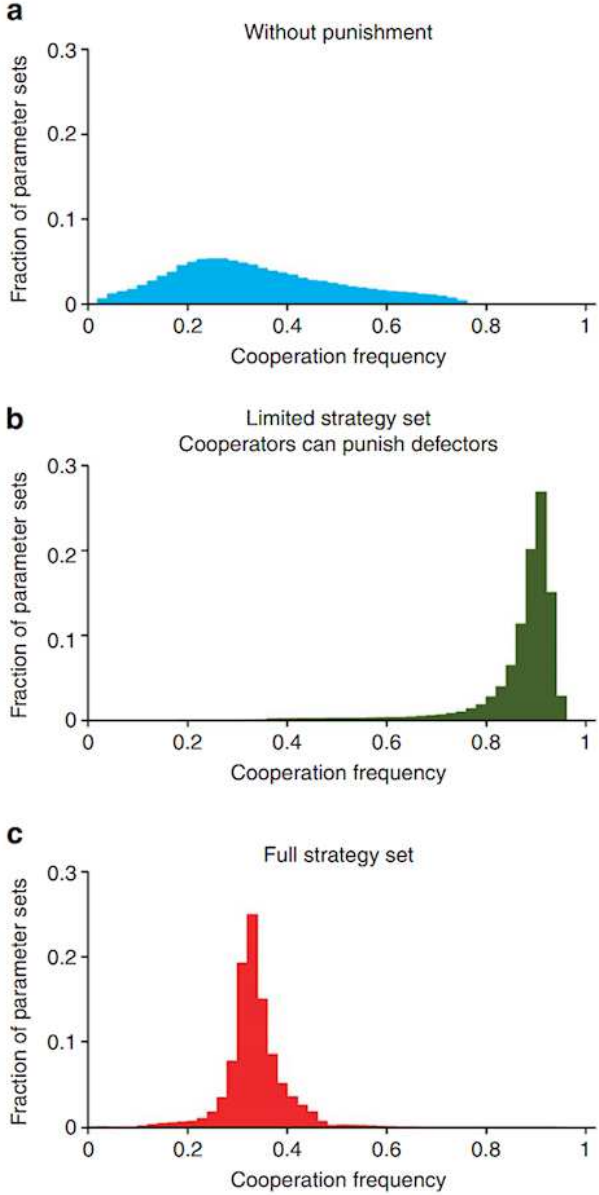,width=8.1cm}}
\caption{Antisocial punishment and the evolution of cooperation in the optional public goods game. For details of the model we refer to \cite{rand_nc11}. It can be observed that the inclusion of antisocial punishment to the strategy set (panel c) returns the level of cooperation in the population to where it was without punishment (panel a). Only if solely the traditional punishment of defectors is allowed, then the likelihood of a more cooperative outcome is higher (panel b). The statistics has been obtained by randomly drawing game parameters from representative intervals and simulating $10^5$ independent outcomes. Figure reproduced with permission from \cite{rand_nc11}.}
\label{antisocpun}
\end{figure}

Unfortunately, a dedicated effort to study antisocial punishment has not yet been made in the realm of statistical physics, at least not in the same sense as peer punishment (see Section~\ref{peerpunresults}), pool punishment (see Section~\ref{poolpunresults}), and adaptive punishment (see Section~\ref{adaptivepunresults}) have been studied separately in the spatial public goods game with a focus on phase transitions, pattern formation, and self-organization. Nevertheless, in \cite{rand_jtb10, rand_nc11} authors have considered antisocial punishment in the realm of human cooperation, and we here briefly review their main results.

Results presented in Fig.~\ref{antisocpun} show convincingly that allowing antisocial punishment significantly reduces the effectiveness of punishment to promote cooperation in the optional public goods game. In fact, the likelihood of a cooperative outcome with antisocial punishment present is about the same as in the absence of any punishment. Furthermore, behavioral experiments conducted to verify these theoretical predictions also deliver much the same results \cite{rand_nc11}, as do results obtained in the realm of score-dependent viability dynamics \cite{nakamaru_eer05} and the corresponding spatially structured lattice model reported in \cite{rand_jtb10}. In short, antisocial punishment is clearly responsible for the crumbling of cooperation enforcement in the optional public goods game with punishment. Given the relative abundance of such antisocial behavior in human societies \cite{herrmann_s08}, one rightfully questions the whole concept of punishment, and whether positive interactions might be the better way forward \cite{dreber_n08, rand_s09}. Indeed, while the majority of previous studies addressing the ``stick versus carrot'' dilemma concluded that punishment is more effective than reward in sustaining cooperation \cite{sigmund_tee07}, evidence suggesting that rewards may be as effective as punishment and lead to higher total earnings without potential damage to reputation \cite{milinski_n02, dos-santos_m_prsb11} or fear from retaliation \cite{dreber_n08} is certainly there. Research in the realm of antisocial punishment done thus far delivers solid evidence that antisocial punishment renders the concept of sanctioning ineffective, and further support the idea that healthy levels of human cooperation are likelier to be achieved through less destructive means.

\subsection{Lack of adverse effects with antisocial rewarding}
\label{antisocialrewresults}

\begin{figure}[b]
\centerline{\epsfig{file=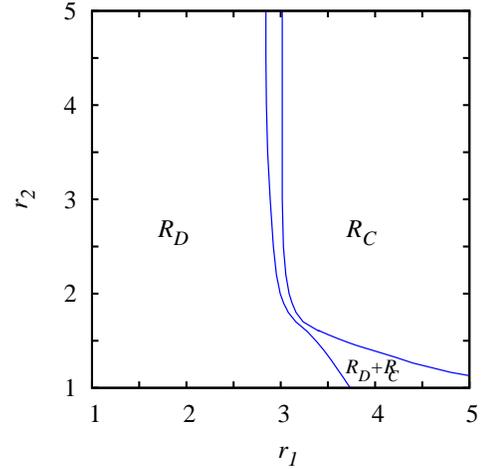,width=6.5cm}}
\caption{Phase diagram of the spatial public goods game with antisocial and prosocial pool rewarding, demonstrating that antisocial rewarding does not hinder prosocial rewarding to promote cooperation. Depicted are strategies that remain on the square lattice after sufficiently long relaxation times as a function of the multiplication factor for the public goods pool $r_1$ and the multiplication factor for the antisocial and prosocial rewarding pool $r_2$. Solid blue lines denote continuous phase transitions. Figure reproduced with permission from \cite{szolnoki_prsb15}.}
\label{antisocrewphase}
\end{figure}

\begin{figure*}
\centerline{\epsfig{file=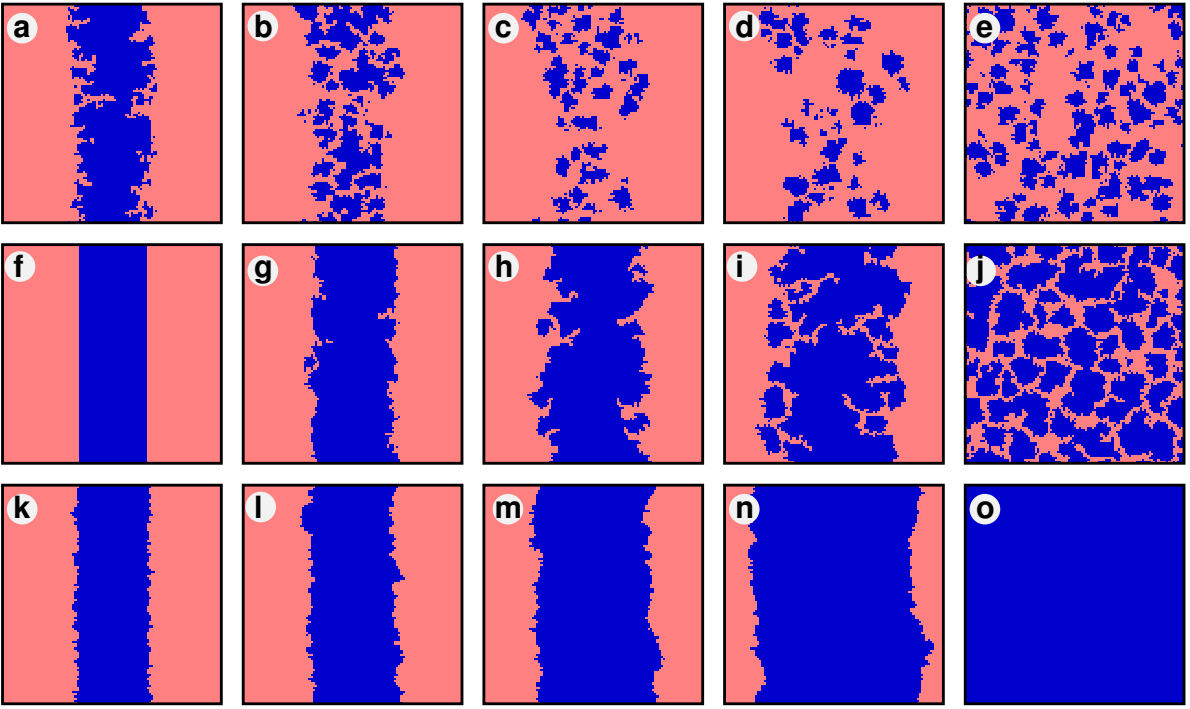,width=16.0cm}}
\caption{Evolution of the spatial distribution of strategies in the spatial public goods game with antisocial and prosocial pool rewarding over time reveals that, even in the presence of equally effective antisocial rewarding, prosocial rewarding promotes the spatial selection for cooperation. Depicted are snapshots of the square lattice over time from left to right, as obtained for $r_2=1$ (top row), $r_2=1.3$ (middle row), and $r_2=2$ (bottom row). For clarity, a prepared initial state has been used for all cases with only a stripe of rewarding cooperators (blue) and rewarding defectors (pale red) initially present in the population, as depicted in panel (f). It can be observed that in the absence of rewarding (top row) the interface separating the two competing strategies is broken easily, and network reciprocity alone can ultimately sustain only small cooperative clusters. However, as the effectiveness of pool rewarding increases (middle and bottom row), the interface is strengthened, which makes the phalanx of cooperators more effective. The latter helps to reveal the benefit of aggregated cooperators in structured populations. In all three cases the synergy factor for the main public goods game is $r_1=3.8$. Figure reproduced with permission from \cite{szolnoki_prsb15}.}
\label{antisocrewsnaps}
\end{figure*}

Unlike antisocial punishment briefly reviewed in the previous subsection, antisocial rewarding has received a dedicated effort in terms of the application of statistical physics methods in \cite{szolnoki_prsb15}, where the focus was on the public goods game with antisocial and prosocial pool rewarding in order to determine the potential negative consequences on the effectiveness of positive incentives to promote cooperation (see Section~\ref{poolrewmodel} for the definition of the model). Contrary to a naive expectation, it was shown that the ability of defectors to distribute rewards to their like does not deter public cooperation as long as cooperators are able to do the same. In fact, even in the presence of antisocial rewarding the spatial selection for cooperation, i.e., network reciprocity, was found to be enhanced.

Figure~\ref{antisocrewphase} shows the phase diagram for the whole $r_1-r_2$ parameter plane, as obtained with Monte Carlo simulations. Starting with the $r_2=1$ line, which implies the absence of pool rewarding, it can be observed that cooperators survive only if the critical value of $r_1$ is $r_{1_c}>3.74$ \cite{szolnoki_pre09c} (see Section~\ref{nullmodel}). The fact that this value is still lower than the group size $G=5$, which would be the threshold in a well-mixed population, is due to network reciprocity. The latter enables cooperators to form compact clusters and so protect themselves against being wiped out by defectors \cite{nowak_n92b}. Taking this as a reference value, we can appreciate at a glance that, even in the presence of antisocial rewarding, prosocial rewarding still promotes the evolution of cooperation. However, neither defectors ($D$) nor cooperators ($C$) who abstain from pool rewarding can survive if $r_2>1$. Indeed, only rewarding defectors ($R_D$) and rewarding cooperators ($R_C$) remain in the stationary state, depending on the value of $r_1$ and $r_2$. This outcome can be understood since players that do engage in pool rewarding collect payoffs that exceed their initial contributions to the rewarding pool.

In terms of the relation between $R_D$ and $R_C$ players, results in Fig.~\ref{antisocrewphase} show that the introduction of strategy-neutral pool rewarding unambiguously supports the cooperative strategy. In particular, as the value of $r_2$ increases and with it also the efficiency of rewarding, the critical value of $r_1$ where $R_C$ players are able to survive decreases steadily. Likewise decreasing is the $r_1$ threshold for complete dominance of the $R_D$ strategy. At specific values of $r_1$, for example at $r_1=3.5$, it is even possible to go from the pure $R_D$ phase to the pure $R_C$ phase solely by increasing the value of $r_2$. Thus indeed, even if the prosocial pool rewarding scheme is accompanied by an equally effective antisocial pool rewarding scheme, in structured populations the evolution of cooperation from a neutral or even from an adverse initial state is still promoted well past the boundaries imposed by network reciprocity alone.

An obvious question is why this is the case? As frequently emphasized throughout this review, an in-depth analysis of the emerging special patters from prepared initial conditions provides the answer, which in this case is rooted in the possible aggregation of cooperators, which can easily emerge spontaneously in a structured population. It is therefore instructive to monitor the evolution of the spatial distribution of strategies over time, as obtained for different values of $r_2$. Results, where for clarity a prepared initial state with only a stripe of rewarding cooperators (blue) and rewarding defectors (pale red) has been used, as illustrated in panel (f), are presented in Fig.~\ref{antisocrewsnaps}.

The top row of Fig.~\ref{antisocrewsnaps} shows the evolution obtained at $r_2=1$, which corresponds to the traditional, reward-free public goods game. It can be observed that the initially straight interface separating the two competing strategies disintegrates practically immediately. There is a very noticeable mixing of the two strategies, which ultimately helps defectors to occupy the larger part of the available space. Here cooperators are able to survive solely due to network reciprocity, but at such a relatively small value of $r_1$ only small cooperative clusters are sustainable. Snapshots depicted in the middle row of Fig.~\ref{antisocrewsnaps} were obtained at $r_2=1.3$, where thus both antisocial and prosocial pool rewarding mechanisms are at work. Here the final state is still a mixed $R_C+R_D$ phase (see also Fig.~\ref{antisocrewphase}), but the fraction of cooperators is already significantly larger than in the absence of rewarding. Larger cooperative clusters are sustainable in the stationary state, which is due to an augmented interfacial stability between competing domains. In addition to traditional network reciprocity, clearly the formation of more compact cooperative clusters is further promoted by the introduction of pool rewarding, and this despite the fact that both antisocial and prosocial rewarding mechanisms are equally strong. If an even higher value of $r_2$ is applied, the interface that separates $R_C$ and $R_D$ players becomes impenetrable for defectors. The two strategies do not mix at all, which maintains the phalanx of cooperators. Accordingly, the latter players simply spread into the region of defectors until they dominate completely. This scenario is demonstrated in the bottom row of Fig.~\ref{antisocrewsnaps}, where the final stationary state is indeed a pure $R_C$ phase.

As demonstrated in the middle and the bottom row of Fig.~\ref{antisocrewsnaps}, the introduction of pool rewarding supports the aggregation of akin players and results in more stable interfaces between competing domains. Since the administration of rewards to either strategy requires a considerable degree of aggregation, cooperators can enjoy the benefits of their prosocial contributions as well as the corresponding rewards. Defectors when aggregated, on the other hand, can enjoy antisocial rewards, but due to their lack of contributions to the public good they ultimately succumb to their inherent inability to secure a sustainable future. Strategies that facilitate the aggregation of akin players, even if they seek to promote antisocial behavior, thus always enhance the long-term benefits of cooperation. This argument also explains why the same positive outcome is not attainable from a random initial state in well-mixed populations, where it was concluded that the possibility of antisocial rewarding utterly shatters any evolutionary benefits to cooperators that might be stemming from prosocial rewards \cite{dos-santos_m_prsb15}. If the interactions among players are well-mixed, then of course neither cooperators nor defectors can aggregate locally, which is a fundamental condition to reveal the long-term benefits of cooperation in a collective enterprize, even if the population contains strategies that seek to actively promote antisocial behavior.

\section{Tolerance and cooperation}
\label{toleranceresults}
As stated in the introduction, tolerance is a rather unique human ability, which manifests as a steadfast endurance of a trying circumstance with a fair and objective attitude. In this sense, it is quite far removed from positive and negative reciprocity, like rewarding and punishment, that we have focused on thus far in this review. Nevertheless, it is important to take tolerance into account when studying human cooperation, which is why, in the next two subsections, we will review statistical physics research dedicated to the benefits of tolerance \cite{szolnoki_pre15} and the impact of diverse tolerance levels \cite{szolnoki_njp16} in the realm of the spatial public goods game.

\subsection{The benefits of tolerance}
\label{singletoleranceresults}

\begin{figure*}
\centerline{\epsfig{file=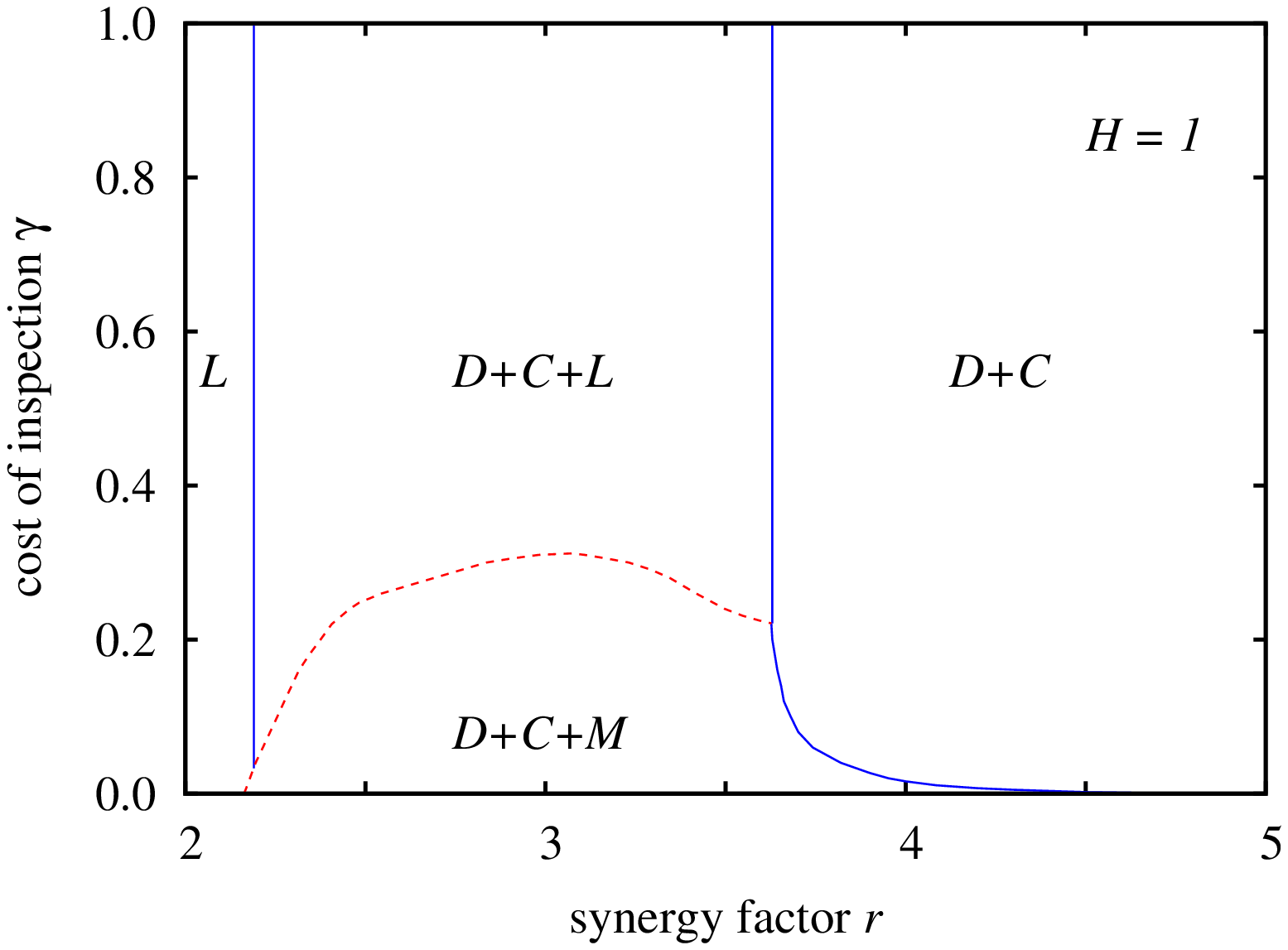,width=8.5cm}\epsfig{file=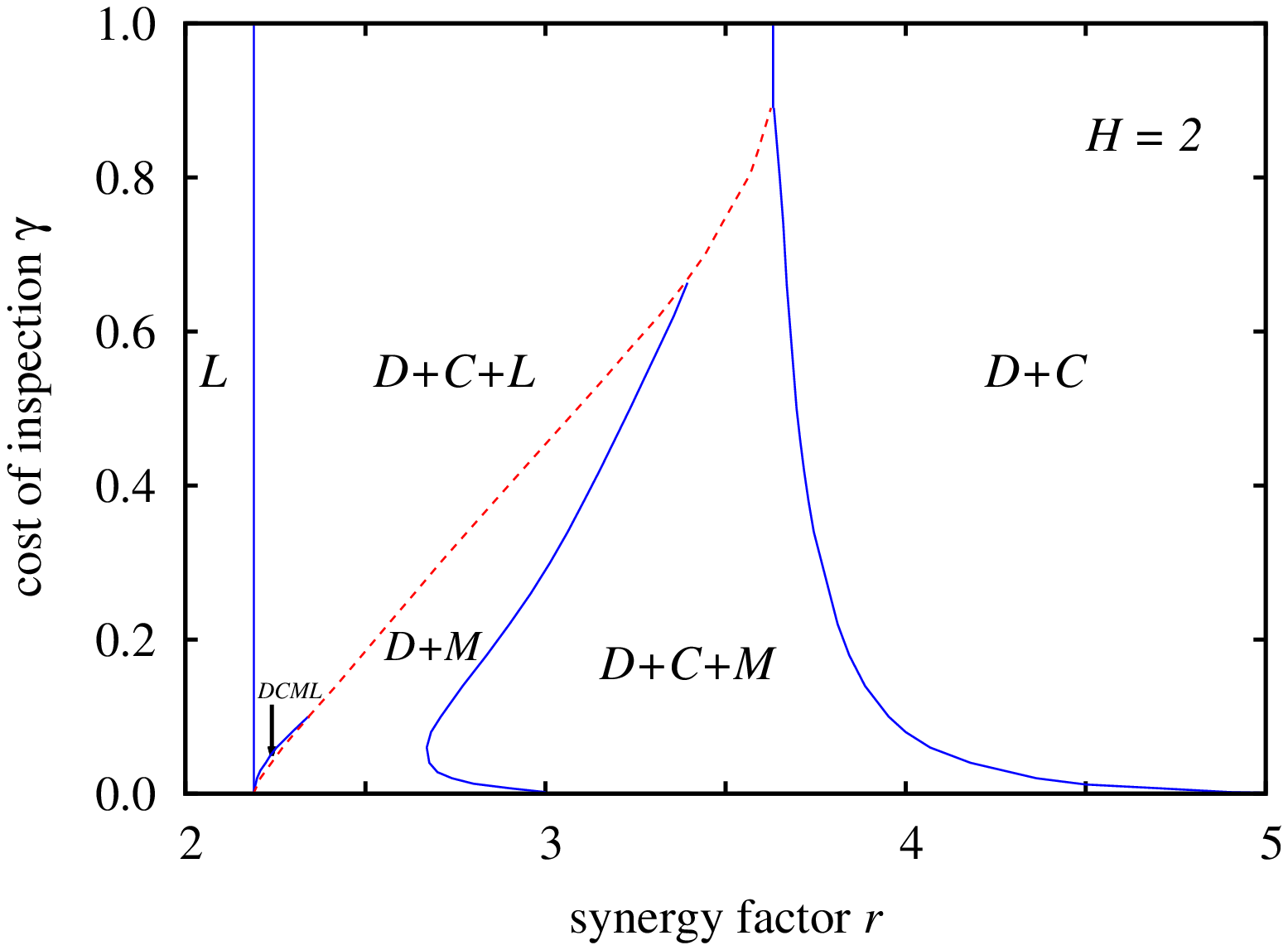,width=8.5cm}}
\centerline{\epsfig{file=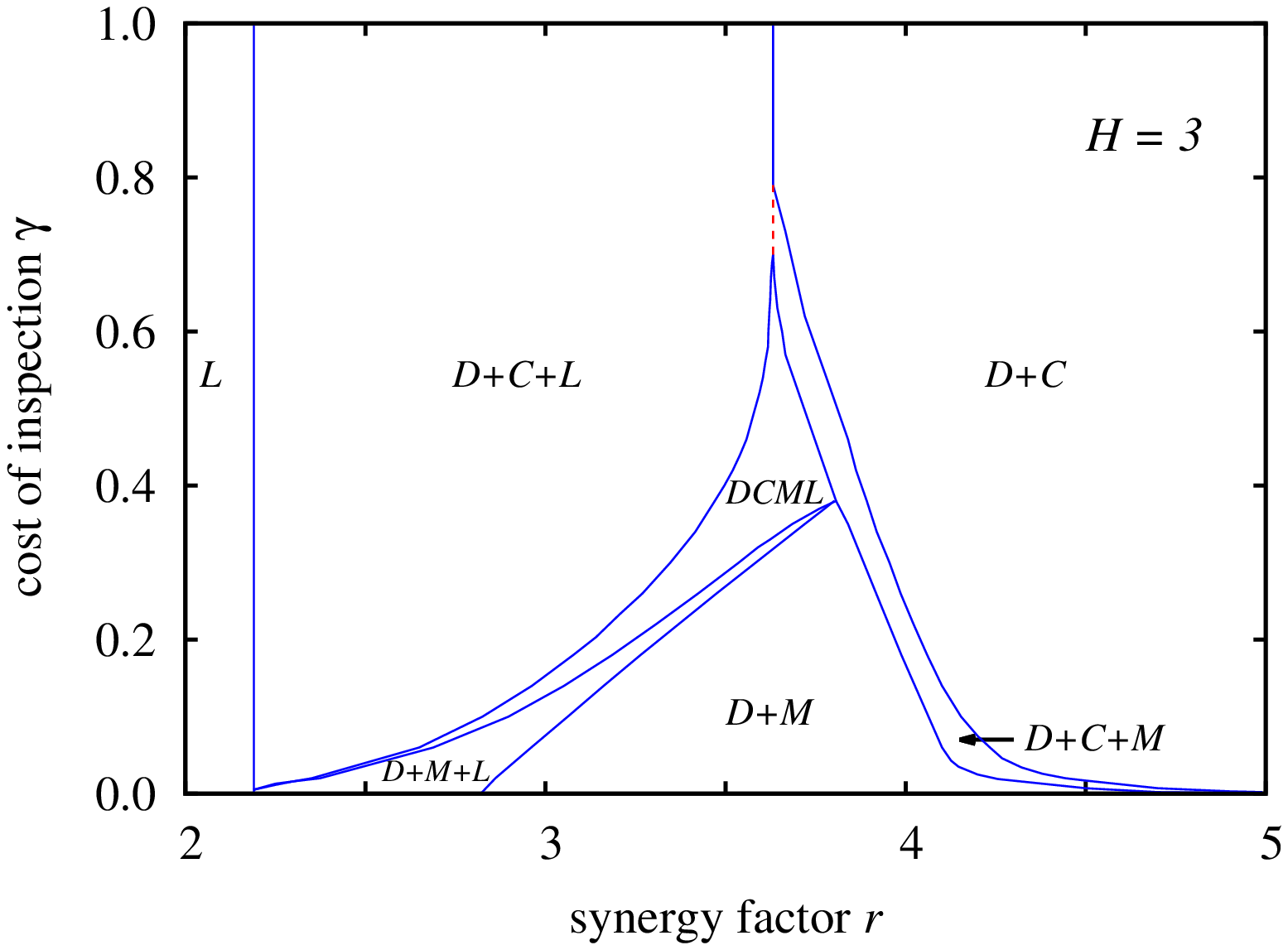,width=8.5cm}\epsfig{file=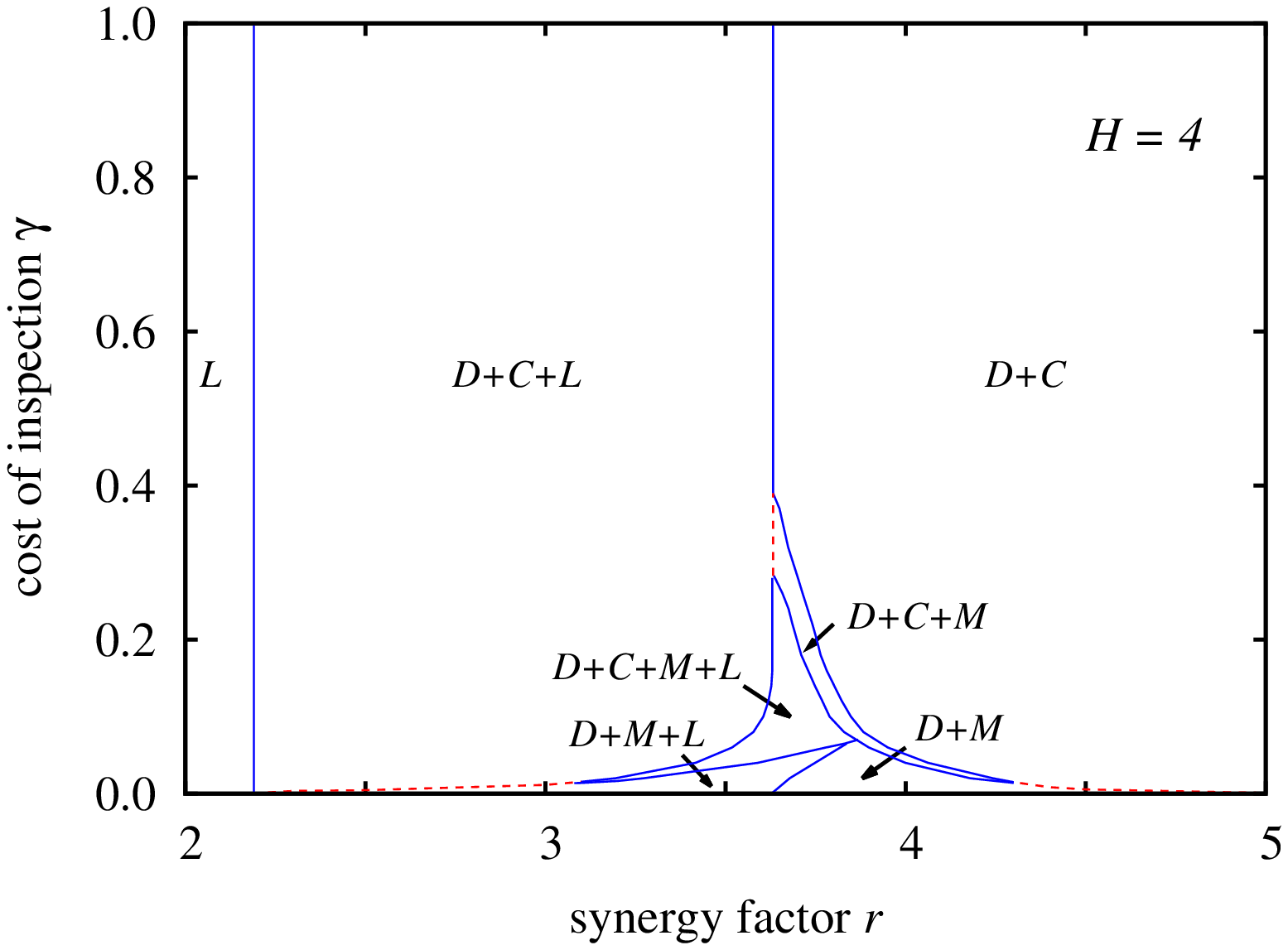,width=8.5cm}}
\caption{Full $r-\gamma$ phase diagrams for the spatial public goods game with tolerant players, as obtained for $H=1$, $2$, $3$ and $4$ on the square lattice. Different phases are denoted by the symbols of the strategies that survive in the final strategy distribution. Solid lines represent continuous, whereas dashed lines indicate discontinuous phase transitions between stable solutions. Figure reproduced with permission from \cite{szolnoki_pre15}.}
\label{tolerancephase}
\end{figure*}

The benefits of tolerance can be studied best with a slightly simplified version of the model reviewed in Section~\ref{tolerancemodel}, where instead of several different types of tolerant players only one type is present in the population, with a tolerance threshold $H$. Research based on this model revealed that some tolerance toward defectors can not only save cooperation if the multiplication factor $r$ is low, but also result in a surprisingly high average payoff in the population \cite{szolnoki_pre15}. Notably, if only loners are added to the public goods game cooperation can be sustained \cite{hauert_s02}, but the average payoff is no larger than the low default income of the loners \cite{hauert_jtb02, semmann_n03}. Phase diagrams and the underlying spatial patterns reveal a high complexity of stationary states, where cyclic dominance and two-strategy alliances characterize evolutionary outcomes.

Figure~\ref{tolerancephase} shows representative phase diagrams in dependence on the cost of knowing the strategy of the other players in the group $\gamma$ and the multiplication factor $r$. In general, it can be observed that, if the inspection cost $\gamma$ is too large, tolerant players ($M$) always die out. As a result, the classical three-strategy $(D,C,L)$ model is recovered \cite{hauert_s02}. In this case, traditional cooperators ($C$) can survive for $r \ge 2.19$ due to cyclic dominance and form the three-strategy $DCL$ phase. If the synergy factor $r$ is high enough, and consequently cooperators and defectors can coexist due to network reciprocity, the cyclic dominance among the three strategies is broken and the loners die out. However, the evolutionary outcomes are significantly different for lower $\gamma$ values and intermediate values of $r$, where the $M$ strategy is able to survive.

The lowest nonzero threshold $H=1$ is a special case because here the tolerant players ($M$) can survive with defectors ($D$) only in the presence of traditional cooperators ($C$). Due to the low threshold, an $M$ player can change from the $C$ to the $L$ state immediately when it recognizes the presence of a defector in the group. The previously mentioned $DCL$ cyclic dominance is established, but instead of this cycle the strategies $M \to D \to C \to M$ form the closed loop of dominance. In other words, $L$ is simply replaced by $M$. The evolutionary stability of this three-strategy phase in relation to other possible phases depends on the average rotation speed within the phase. Namely, the faster the invasion rate, the more stable a cyclic dominance phase \cite{perc_pre07b}. By increasing the value of $r$, it is possible to observe a reentrant phase transition $(D+C+L) \to (D+C+M) \to (D+C+L)$, which is a general behavior when the average invasion rates within a cycle can be adjusted by varying a control parameter \cite{szolnoki_epl15}.

Staying with the phase diagrams in Fig.~\ref{tolerancephase}, if the tolerance threshold $H$ increases, thus allowing $M$ players to tolerate more defectors within the group, then new types of solutions emerge. Namely, tolerant players can coexist with defectors without the presence of a third strategy. As was shown in \cite{szolnoki_pre15}, the $D+M$ phase can be particularly efficient in terms of the average payoff in the population. In addition to the two-strategy $D+M$
phase, there are parameter values where all four competing strategies coexist, and there are some specific cases where tolerant players crowds out cooperators but stay together with loners in the presence of defectors. Here defectors and tolerant players are still capable to form a two-strategy alliance, but loners can invade defectors. As a result, small loner patches emerge temporarily, but they are vulnerable against the invasion of tolerant players, who are capable of utilizing network reciprocity and thus close the $D+M+L$ cycle of dominance. A stable coexistence of all four strategies, namely the $D+C+M+L$ phase, is also observable.

The comparison of phase diagrams obtained for different tolerance thresholds in Fig.~\ref{tolerancephase} highlights that there is an optimal intermediate tolerance level which provides the best condition for tolerant players to survive, even if the inspection cost $\gamma$ is relatively high. It is important to note that tolerant players always bear this cost, in addition to the cost of cooperation when they act as cooperators within a group (when the number of defectors in the group is sufficiently low). At $H=2$, for example, tolerant players pay nearly double the cost of traditional cooperators, yet are still able to outcompete them as well as the loners. Taken together, neither too small nor too high tolerance levels are good. At an intermediate tolerance level, however, tolerant players provide a significant boost to cooperation, assuring moreover that the average payoffs in the population remain high \cite{szolnoki_pre15}.

\begin{figure}
\centerline{\epsfig{file=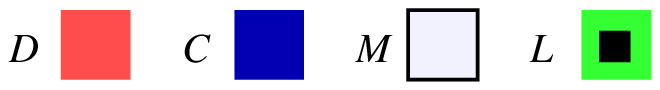,width=3.5cm}}
\centerline{\epsfig{file=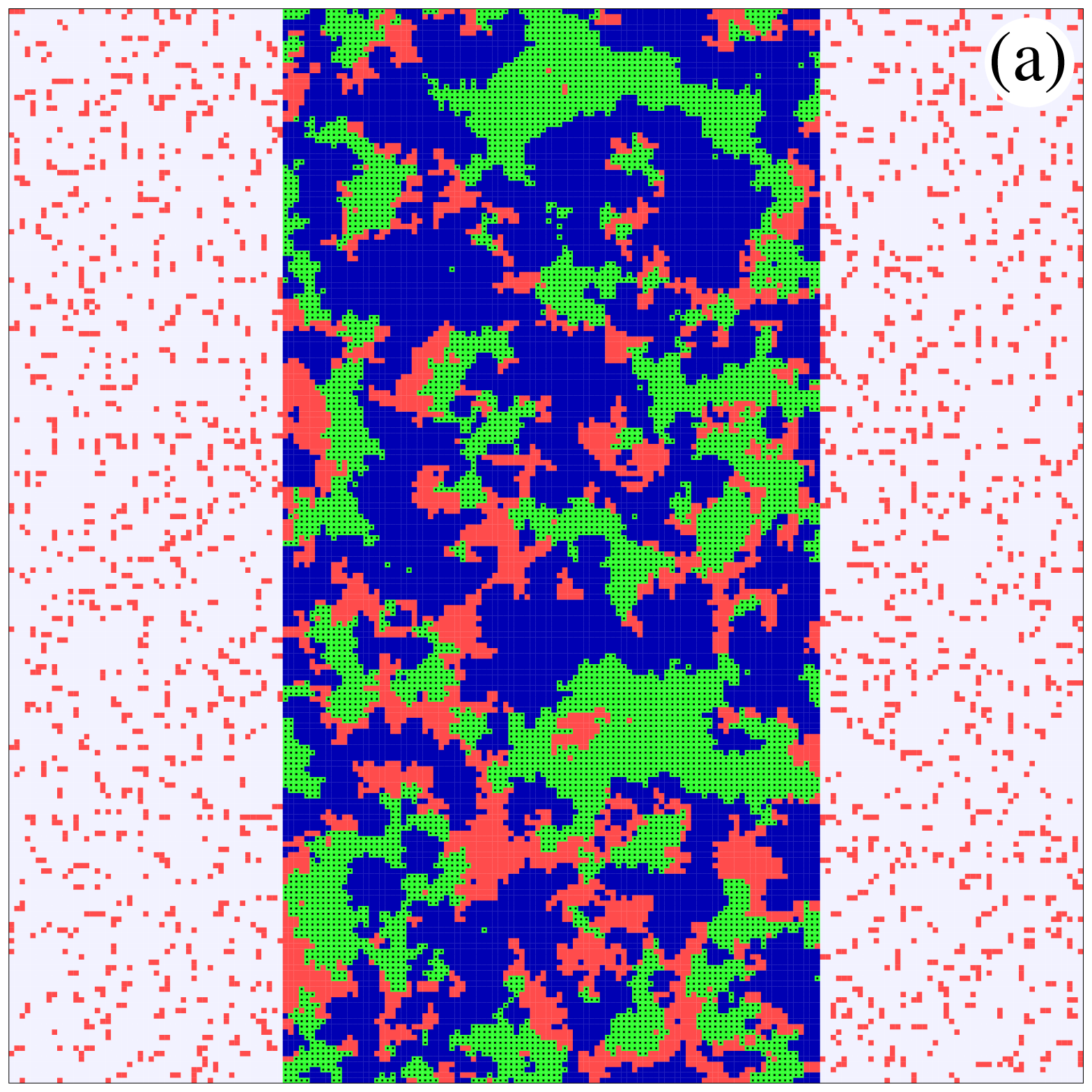,width=2.8cm}\epsfig{file=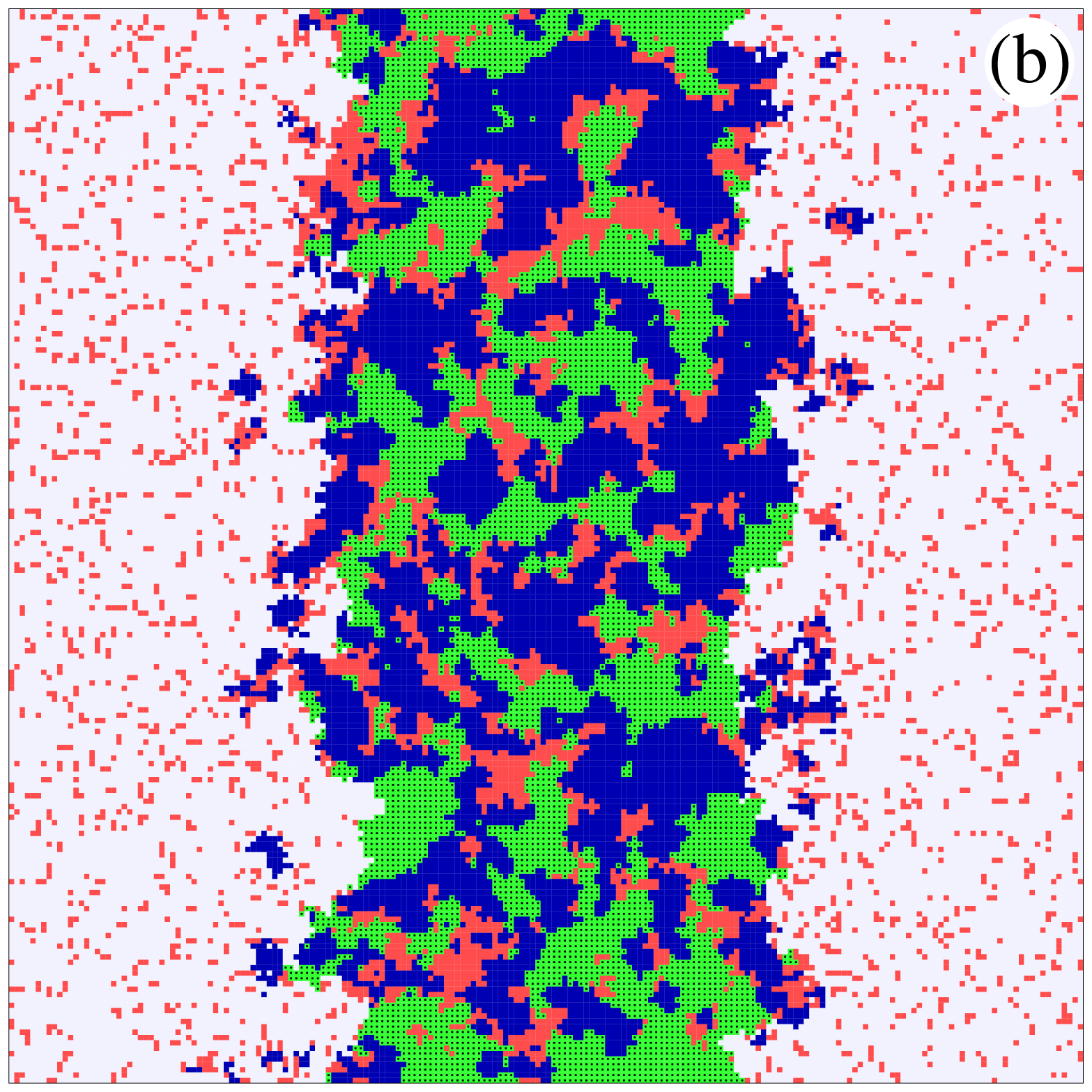,width=2.8cm}\epsfig{file=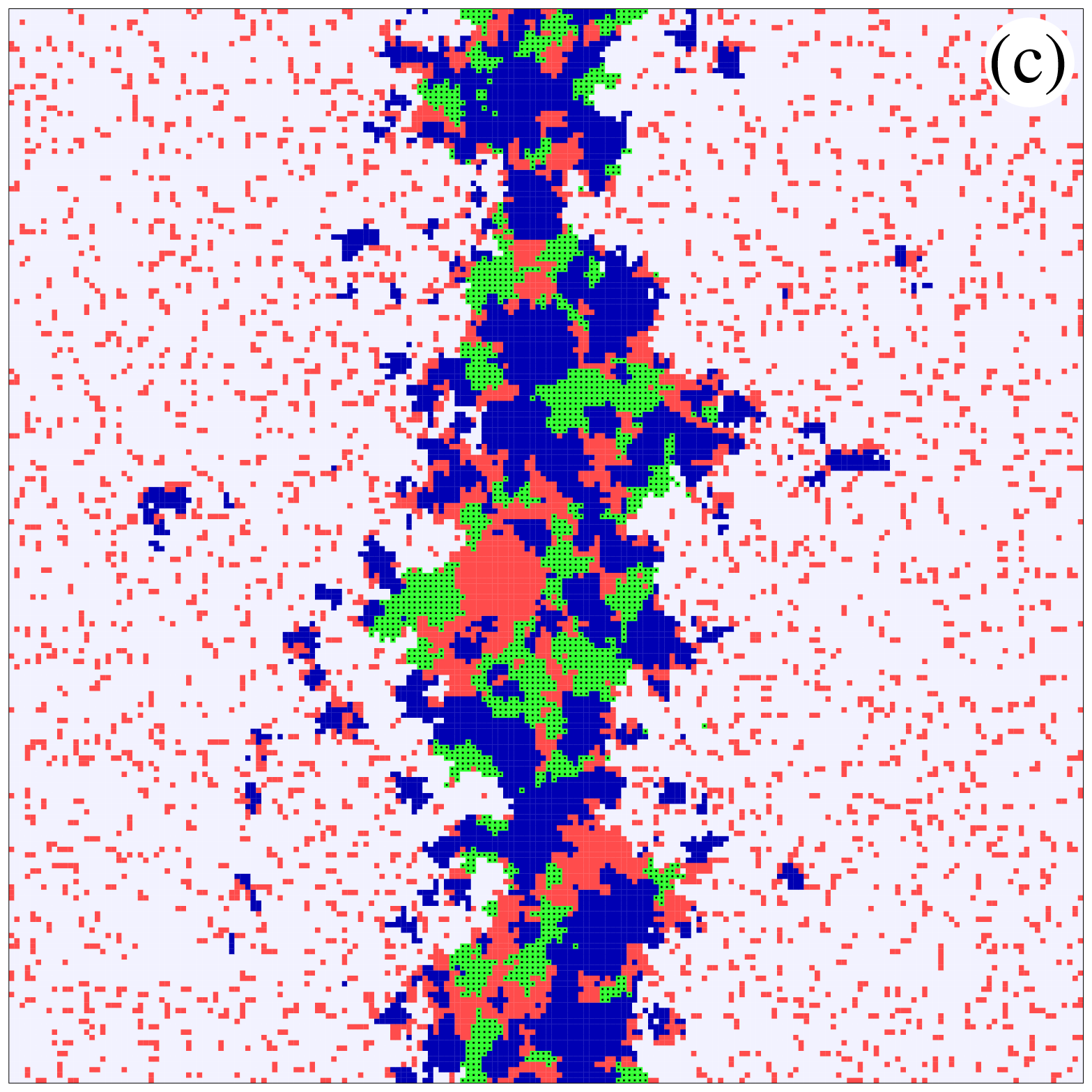,width=2.8cm}}
\centerline{\epsfig{file=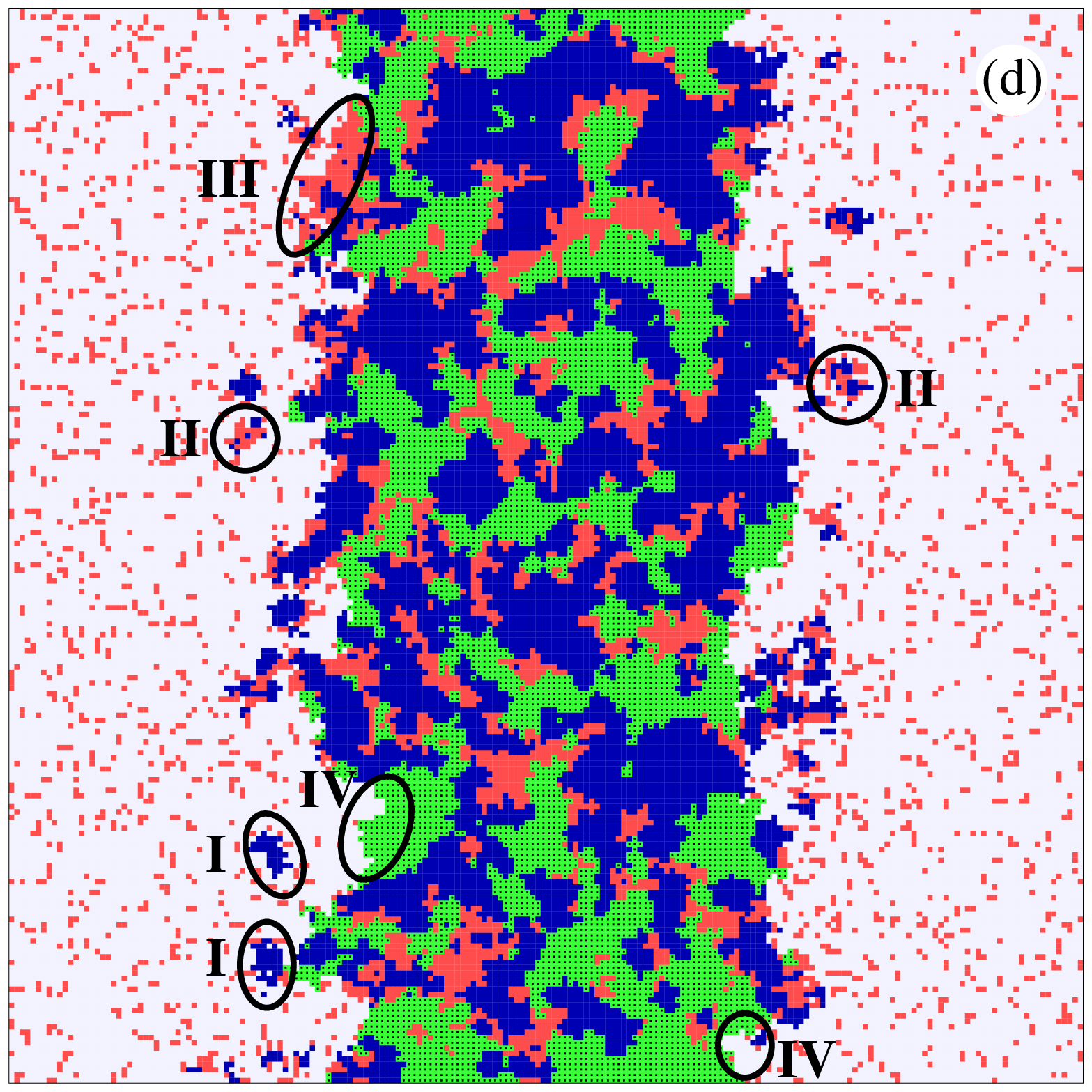,width=8.5cm}}
\caption{The competition of two possible solutions in the spatial public goods game with tolerance at $r=2.7$, $\gamma=0.15$, and $H=2$. Defectors are denoted red, cooperators are denoted dark blue, tolerant players are denoted light blue, and loners are denoted green, as indicated by the legend at the top of the figure. At these parameter values both the $D+C+L$ cyclic dominance phase and the two-strategy $D+M$ phase are stable subsystem solutions (see Section~\ref{subsystemethod} for details). Panel~(a) illustrates a prepared initial state where the $D+C+L$ phase is embraced by the $D+M$ phase. In panel~(b), the interface is opened up for strategy invasions, and hence the two subsystem solutions can start competing for space. Eventually the $D+M$ solution starts to dominate, as illustrated in panel~(c), and finally the two-strategy phase prevails completely (not shown). Panel~(d) shows the enlarged part of panel~(b) to illustrate the microscopic mechanisms that are responsible for the successful invasion of the $D+M$ phase. The snapshots were taken at $0$, $70$ and $210$ full Monte Carlo steps. Figure reproduced with permission from \cite{szolnoki_pre15}.}
\label{tolerancesnaps}
\end{figure}

A deeper understanding as to why tolerance promotes cooperation can be obtained, once again, by studying the evolution of emerging spatial patterns from a prepared initial state, as illustrated in Fig.~\ref{tolerancesnaps}. It should be emphasized that the three-strategy $D+C+L$ phase is always a solution in the low-$r$ region \cite{hauert_s02}. To understand the superiority of the
$D+M$ phase, it is necessary to start the evolution from a special, prepared initial state where the two phases can first evolve calmly in a restricted area. Panel~(a) of Fig.~\ref{tolerancesnaps} illustrates the final result of the two isolated evolutions. After this, the borders open up, and the competition of subsystem solutions starts. The elementary steps of this competition can be identified in panel~(b), which is zoomed out in panel~(d) for clarity. In this snapshot, we can distinguish three different cases of how the three-strategy solution meets with the two-strategy
$D+M$ phase. If a $C$ domain, marked by dark blue, is at the frontier, then cooperators start spreading in the sea of $M$. These invasions are marked by ``I'' in panel~(d). The success of $C$, however, is temporary, because defectors, marked by red, will follow them and gradually invade the invaders. This stage is marked by ``II'' in panel~(d). Subsequently, when $D$ players remain alone with $M$ players, the latter (marked by light blue) regulate the defectors and lower their concentration to a minimal level. The second option of how competing solutions meet in this case is when a $D$ spot from the $D+C+L$ phase meets with the $D+M$ phase. This is marked by ``III'' in panel~(d). Here the previously described regulation process starts immediately, which will decrease the area of the $D+C+L$ phase. Finally, when an $L$ domain (marked by green) is at the interface then it will shrink immediately because $M$ is able to utilize the positive impact of network reciprocity. This process is marked by ``IV'' in panel~(d). Together, these four elementary processes will reduce the middle area occupied by the three-strategy phase. Lastly, a negative feedback sets in, which is related to the shrinkage of the three-strategy area. Namely, as space becomes sparse, the $D+C+L$ phase becomes more vulnerable against external invasions because the amplitude of local oscillations within the phase becomes larger and larger the smaller the patches, which is a well-known phenomenon in the realm of cyclic dominance \cite{szolnoki_jrsif14} (although the amplitude can increase also due to other reasons, as reviewed in Section~\ref{peercorrresults}, Fig.~\ref{corrscaling} in particular). Consequently, when the width of the three-strategy phase in the middle becomes comparable to the typical size of the smallest sustainable $D+C+L$ patches the three-strategy phase can be easily trapped into a homogeneous state because one strategy within the cyclic alliance is destined to die out. This stage is illustrated in panel~(c) of Fig.~\ref{tolerancesnaps}, after which the $D+M$ phase can easily invade the rest of the population.

We conclude this subsection by noting that tolerance has been considered before also in myopically selective interactions in the realm of social dilemmas, showing that it promotes cooperation \cite{chen_xj_pre09}, as well as in the realm of an adaptive environment, where it was found that social tolerance allows cooperation to prevail \cite{chen_xj_pre09b}.

\subsection{Diverse tolerance levels}
\label{diversetoleranceresults}
To determine whether evolutionary advantages might be stemming from diverse levels of tolerance in a population, an upgrade to the model reviewed in the above subsection has been studied in \cite{szolnoki_njp16}, where instead of a single tolerance level in the population, all four possible tolerant strategies could compete at once (see Section~\ref{tolerancemodel} for the definition of the model). Research revealed that the diversity of tolerance can give rise to synergistic effects, wherein players with a different threshold in terms of the tolerated number of defectors in a group compete most effectively against defection and loners. Such synergistic associations can stabilize states of full cooperation where otherwise defection would dominate. A highlight of this model are almost invisible yet stable strategy alliances that are driven by complex pattern formation in a structured population.

\begin{figure*}
\centerline{\epsfig{file=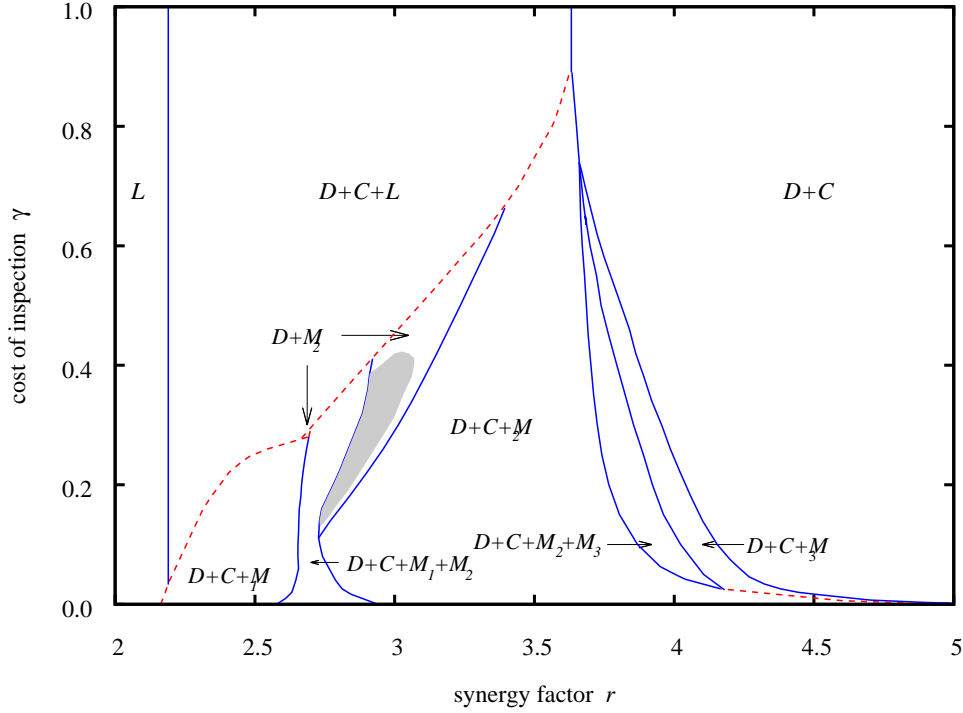,width=13cm}}
\caption{Full phase diagram of the $8$-strategy public goods game with diverse tolerance levels on the $r-\gamma$ parameter plane, as obtained on a square lattice. Dashed lines denote discontinuous, while solid lines mark continuous phase transitions. There exist regions where the coexistence of different tolerance levels is the most stable solution. In the latter case the grey-shaded region in the two-strategy $D+M_2$ phase denotes those parameter values where the population evolves into a full-cooperator, defector-free $M_1+M_2$ phase, but only if the system size is large enough. We attend to these invisible solutions in detail in Fig.~\ref{diversetolerancesize}. Figure reproduced with permission from \cite{szolnoki_njp16}.}
\label{diversetolerancephase}
\end{figure*}

The main phase diagram of the model is presented in Fig.~\ref{diversetolerancephase}. These results reveal several fundamental features that can be associated with the viability of tolerant strategies in the public goods game. It can be observed that the higher the value of $r$, the higher the tolerance can be, and vice versa. This observation resonates with our naive expectation and perception of tolerance in that overly tolerant strategies cannot survive in the presence of other less tolerant strategies. From the viewpoint of the considered evolutionary game this is not surprising, because players adopting the $M_4$ strategy act as loners only if everybody else in the group is a defector. And such sheer unlimited tolerance is simply not competitive with other less tolerant strategies. Also, if the cost of inspection $\gamma$ is too high, or if the value of the synergy factor is either very low or very high, then tolerant players cannot survive even if they exhibit different levels of tolerance. This observation is in agreement with research reviewed in Section~\ref{singletoleranceresults}, where only uniform tolerance levels were considered \cite{szolnoki_pre15}.

More precisely, as results in Fig.~\ref{diversetolerancephase} reveal, at very low values of $r$, similarly as in the simplest three-strategy $DCL$ model, the loners prevail. At slightly larger values of $r$, this single-strategy phase gives way to the three-strategy phase where $D, C$, and $L$ strategies dominate each other cyclically. Interestingly, unlike in the uniform tolerance public goods game \cite{szolnoki_pre15}, here the $D \to C \to L \to D$ closed loop of dominance is the only way for loners to survive at higher $r$ values \cite{hauert_s02}. In all the other cases the loners die out due to the fact that the diversity of tolerant players is able to provide a more competitive response to the exploitation of defectors. Indeed, if the synergy factor is increased further, we find that, through a succession of different phase transitions, the $D+C+L$ phase gives way to a rich variety of two-, three- or even four-strategy phases, in all of which tolerant strategies are present. These solutions are the $D+C+M_1$, the $D+M_2$, the $D+C+M_2$, and the $D+C+M_3$ phase, as well as two four-strategy phases $D+C+M_1+M_2$ and $D+C+M_2+M_3$, which are unique to the public goods game with diverse strategy levels.

\begin{figure*}
\centerline{\epsfig{file=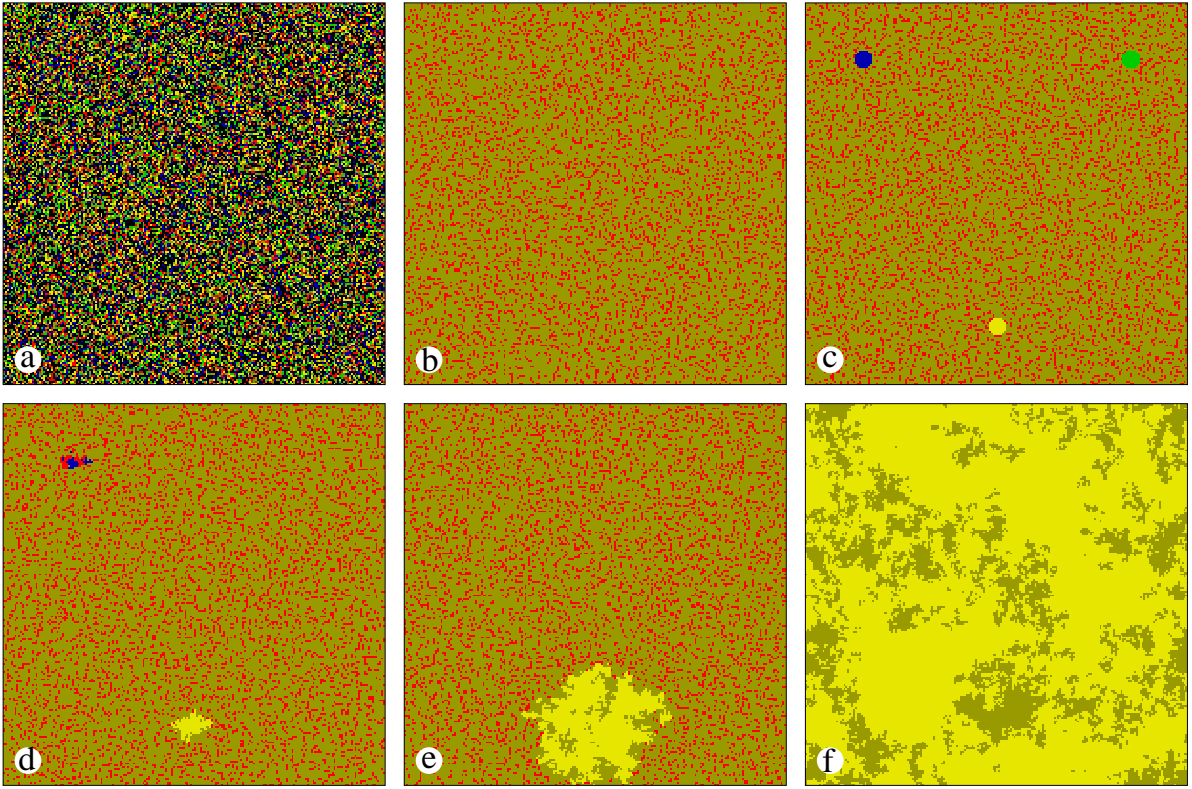,width=16cm}}
\caption{Finite-size effects may lead to misleading evolutionary outcomes in the $8$-strategy public goods game with diverse tolerance levels. At the beginning, the evolution is started from a random initial state where all the eight strategies are present [panel~(a)]. After the relaxation of $1000$ MCS [panel~(b)], only $D$ (red) and $M_2$ (ochre) strategies survive, which also form the stable two-strategy phase in the uniform-tolerance public goods game at the same parameter values ($r=2.97$ and $\gamma=0.35$) (see Section~\ref{singletoleranceresults}). In panel~(c), small compact patches of three other strategies are then manually re-introduced, namely strategy $C$ (blue), strategy $L$ loner (green), and strategy $M_1$ (yellow). Soon afterwards, loners die out (again) very soon within the $D+M_2$ phase, as shown in panel~(d). Later, $C$ players also die out, but tolerant strategies $M_1$ and $M_2$ form a successful alliance, as shown in panel~(e). Indeed, the $M_1+M_2$ subsystem solution turns out to be stronger than the $D+M_2$ phase, and the population terminates in a defector-free state, as shown in panel~(f). The system size used in this example is $L=200$, while stages presented in panels~(d), (e), and (f) were obtained after $30$, $250$, and $1300$ full Monte Carlo steps from panel~(c) onwards. Importantly, as we review in Fig.~\ref{diversetolerancesize}, the defector-free state can evolve naturally from a random initial state if the system size is large enough. Figure reproduced with permission from \cite{szolnoki_njp16}.}
\label{diversetolerancesnaps}
\end{figure*}

According to the phase diagram in Fig.~\ref{diversetolerancephase}, the first transition from the
$D+C+L$ phase to one of the mentioned phases is always discontinuous. The accurate position of these kind of phase transition points, as denoted by the dashed line in the phase diagram, can only be determined by means of a stability analysis of competing subsystem solutions, as described in Section~\ref{subsystemethod}. In this Section, Fig.~\ref{subsystemstability} shows the competition between the three-strategy $D+C+L$ phase and the four-strategy $D+C+M_1+M_2$ phase at both sides of the discontinuous phase transition point, although on both sides the two phases are individually stable, i.e., are proper subsystem solutions. We refer to Section~\ref{subsystemethod} for details, and we emphasize again this to be the only viable way to accurately determine the position of such phase transitions.

In order to understand how the combination of different tolerant strategies is the most effective response to a public goods dilemma, a series of snapshots that is carefully engineered to illustrate exactly what otherwise remains invisible if a small population starts from a random initial state, is shown in Fig.~\ref{diversetolerancesnaps}. In panel~(a), the evolution starts from a small $200 \times 200$ sqaure lattice where initially all eight strategies are distributed uniformly at random. After a relatively short relaxation time only $D$ and $M_2$ players survive in panel~(b) to form what appears to be the dominating two-strategy phase at these parameter values. In fact, this $D+M_2$ phase is the dominant stable solution in the previously reviewed uniform-tolerance public goods game in Section~\ref{singletoleranceresults}. But in the game reviewed here, we have other options because of the diverse tolerance levels that we consider. Therefore, to test the stability of the $D+M_2$ phase properly, small compact patches of other strategies are inserted manually in panel~(c). These are the pure cooperators ($C$), the loners ($L$), and the tolerant strategy ($M_1$), positioned from the left-upper part of the lattice in a clockwise manner. Subsequently, panels~(d) and (e) demonstrate clearly that first the loners die out, followed by the pure cooperators. However, the $M_1$ strategy forms a powerful alliance with the $M_2$ strategy, whose domain is able to grow, as shown in panel~(e). Gradually, all the defectors are crowded out, and in the final state illustrated in panel~(f), only the two different tolerant strategies remain to provide the maximal cooperation level in the defector-free state. Thus, it turns out that the $M_1+M_2$ subsystem solution is ultimately stronger than the
$D+M_2$ subsystem solution. This outcome can be interpreted so that $M_2$ players are efficient at sweeping out $C$ and $L$ players, who, as the two extreme limits of tolerance, prevent other $M$ strategies to function efficiently. At the same time, $M_1$ players are capable to beat strategy $D$. Accordingly, we need the features of both $M_1$ and $M_2$ players to reach the happy end.

\begin{figure}
\centerline{\epsfig{file=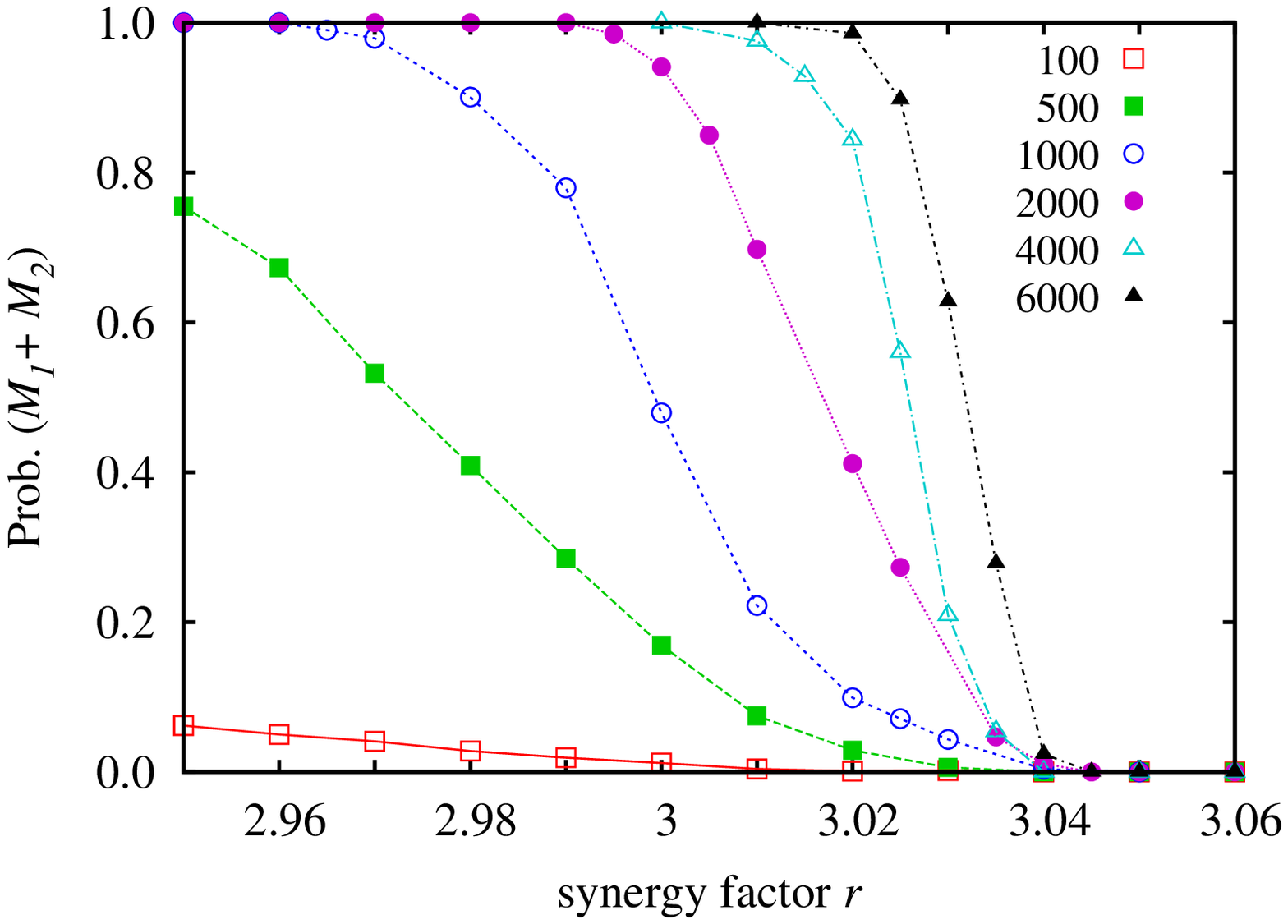,width=8.5cm}}
\centerline{\epsfig{file=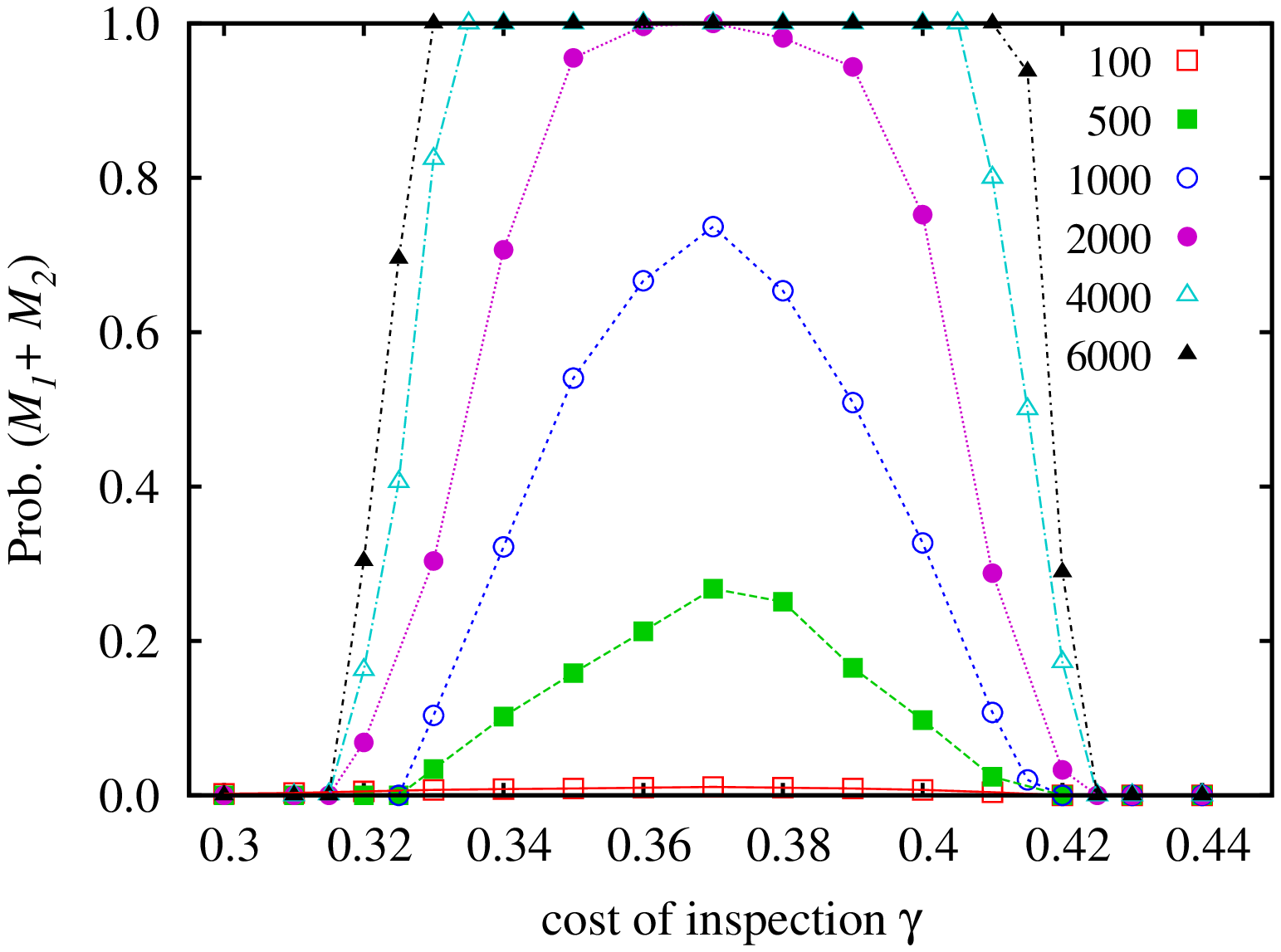,width=8.5cm}}
\caption{The invisible solutions in the public goods game with diverse tolerance levels. Fixation probability for the $M_1+M_2$ phase, as obtained for $\gamma=0.35$ (top) and for $r=3.0$ (bottom) in dependence on the other key parameter, respectively. The linear size of the applied square lattices are indicated in the legend. Remarkably, there exist combinations of parameters $r$ and $\gamma$ at which, even at $500 \times 500$ system size, the $M_1+M_2$ solution remains completely invisible, even though at a much larger $6000 \times 6000$ system size the fixation probability for the exact same solution is $1$. Depicted probabilities are averaged over 500-1000 independent runs. Figure reproduced with permission from \cite{szolnoki_njp16}.}
\label{diversetolerancesize}
\end{figure}

The example shown in Fig.~\ref{diversetolerancesnaps} highlights that the order in which the strategies die out is crucial for the final evolutionary outcome. To give a more quantitatively accurate prediction of whether the system will terminate into the
$D+M_2$ or the $M_1+M_2$ phase, we show in Fig.~\ref{diversetolerancesize} the fixation probability to the latter for different combinations of $r$ and $\gamma$. For these results, the evolution always started from a random initial state (containing all the eight strategies distributed uniformly at random), and it was simply counted how many times the system terminates into the
$M_1+M_2$ phase. The alternative destination of the evolutionary process was to arrive to the
$D+M_2$ phase, which for all the considered parameter combinations used in Fig.~\ref{diversetolerancesize} is the only other stable subsystem solution. To get a reliable statistics, independent runs were repeated $1000$ times for $L=100-1000$ linear system sizes, while for even larger lattices $500$ independent runs were made. The results depicted in Fig.~\ref{diversetolerancesize} show a remarkable finite-size effect. More precisely, there are parameter values of $r$ and $\gamma$ where the strategy $C$ dies out first, thus allowing the
$M_1+M_2$ phase to conquer the whole system. But this solution remains completely invisible if the system size is too small. To conclude, these are the specific combinations of the parameters $r$ and $\gamma$ that are shaded grey in the phase diagram shown in Fig.~\ref{diversetolerancephase} -- these are the invisible solutions for the large majority of the system sizes that are in widespread use in the literature nowadays.

These results ultimately highlight the delicate importance of diversity and tolerance for human cooperation, and they reveal fascinating subtleties of the spatiotemporal dynamics that is due to the competition of subsystem solutions in structured populations. The latter can be studied and understood only by means of statistical physics methods, applied in a rigorous and systematic manner. As we hope this review shows, this is the fundamental nature of research in the realm of statistical physics of human cooperation.

\section{Summary}
\label{summary}
We have provided a systematic review of statistical physics research done to advance our understanding of human cooperation, in particular focusing on the application of Monte Carlo methods and the theory of phase transitions to understand pattern formation, the spatiotemporal dynamics of solutions, and the principles of self-organization that may lead to socially favorable evolutionary outcomes. In the Introduction, we have described human cooperation as the result of our evolutionary struggles for survival, and as the cornerstone of our evolutionary success story. We have also outlined the pressing challenges of our time, which are in large parts due to failing cooperation in our societies, and we have introduced statistical physics as the key to understanding counterintuitive evolutionary outcomes in structured populations.

In Section~\ref{humex}, we have first reviewed human experiments as a critical tool for testing predictions of theoretical models and investigating human cooperation. We have emphasized that existing experiments provide clear evidence that people behave prosocially, that they are willing to enforce prosociality, and that it matters whether the structure of the interaction network is taken into account.

In Section~\ref{mathmodels}, we have introduced the spatial public goods game as the null model for human cooperation, and we have subsequently provided a description of various extensions of this model incorporating punishment, rewards, correlated positive and negative reciprocity, as well as tolerance.

In the continuation, we have then provided a description of the most important Monte Carlo methods for the study of human cooperation in Section~\ref{mcmethods}. In particular, we have described random sequential strategy updating, we have discussed the role and limitations of random initial conditions, we have briefly reviewed the most important concepts of phase transitions, and we have presented in detail the stability analysis of subsystem solutions.

The main results obtained in the realm of statistical physics of humans cooperation have been reviewed from Section~\ref{peeresults} onwards, firstly for peer-based strategies, secondly for institutionalized strategies, thirdly in the realm of self-organization of incentives for cooperation, then for evolutionary outcomes obtained with antisocial strategies, and lastly related to the impact of tolerance on human cooperation. By peer-based strategies in Section~\ref{peeresults}, we have emphasized the importance of indirect territorial competition in peer punishment, the spontaneous emergence of cyclic dominance in peer rewarding, and an exotic first-order phase transition observed with correlated strategies. By institutionalized strategies in Section~\ref{poolresults}, we have highlighted the fascinating spatiotemporal complexity that is due to pool punishment, as well as the virtual non-existence of institutionalized rewarding on its own right. In the realm of self-organization of incentives for cooperation in Section~\ref{adaptiveresults}, we have emphasized the elevated effectiveness of adaptive punishment, the possibility of probabilistic sharing to solve the problem of costly
punishment, and the many evolutionary advantages of adaptive rewarding. By antisocial strategies in Section~\ref{antisocialresults}, we have reviewed the devastating effects of antisocial punishment on cooperation enforcement, and the rather surprising lack of similarly adverse effects with antisocial rewarding. Lastly, in Section~\ref{toleranceresults}, we have reviewed the benefits of tolerance for human cooperation, as stemming from unique and diverse tolerance levels.

\section{Future research and outlook}
\label{future}
Statistical physics of human cooperation has, over the past decade, led to many new insights into which strategies, mechanisms, and external factors promote prosocial outcomes in competitive settings, and it has led to a fundamentally better understanding as to why exactly this is the case \cite{szabo_pr07, perc_bs10, wang_z_epjb15, szabo_pr16}. However, an integrative approach that would aim to merge all the different mathematical models together into a more cohesive, and thus also relevant, theoretical framework is lacking. While this review can be considered a modest step in this direction, from afar, it likely still looks like we are considering little fractions of the problem at a time. This proposition is of course a formidable challenge, but it is also a necessity for the wider recognition of this line of research outside the realm of physics. Some inspiration and guidance to aid the cause can be found in the many works on the subject stemming from anthropology, psychology, and sociology \cite{henrich2006culture, boyd2009culture, boyd_pnas11, kraft_cobs15, poncela2016humans, wang2017onymity}, which to date are still insufficiently, if at all, integrated into theoretical research.

There are also several hypotheses available to explain why cooperation is so widespread and common in human societies that have not yet been introduced to evolutionary games. The ``heart on your sleeve'' hypothesis, for example, asserts that humans are cooperative because they can truthfully signal cooperative intentions. Indeed, recent research indicates that third-party punishment is precisely such a costly signal of trustworthiness among humans \cite{jordan_n16}. Moreover, cultural group selection hypotheses argue that the importance of culture in determining human behavior causes selection among groups to be more important for humans than for other animals. There also exist moralistic reciprocity hypotheses, which argue that greater human cognitive abilities and advanced language allow us to manage larger networks of reciprocity. Such hypotheses could potentially be verified in the realm of evolutionary games in structured populations, and thus become susceptible to research by means of statistical physics methods. More precisely, information sharing and the reliability of shared information across different networks representing different groups or populations could reveal whether honest signalling is indeed crucial. In terms of the integration of cognitive abilities, Bear and Rand \cite{bear_pnas16} have recently introduced a new model for looking at the evolution of intuition versus deliberation as well as at the evolution of cooperation versus defection, which also lends itself to research in structured populations and in presence of coevolutionary rules~\cite{perc_bs10}.

The current theoretical framework also fails to account for the differences among us in terms of our goals and status. Both these factors determine, or at least affect, our behavior. A wealthy individual might look at the potential personal loss in a social dilemma situation differently than a poor individual. Thus, each time we are faced with the choice of either cooperating or defecting, we should also take into account what we risk to loose as individuals, and what we are striving for. In the realm of these considerations, evolutionary multigames \cite{szolnoki_jtb13, wang_z_pre14b, szolnoki_epl14b, amaral2016evolutionary} appear as a promising mathematical foundation for properly taking into account such upgrades to existing models. Alternatives to imitation dynamics, such as myopic best response microscopic dynamics \cite{sysiaho_epjb05, szabo_jtb12, amaral2017role} and its variants \cite{amaral2016stochastic}, also merit more research in the realm of the public goods game, in particular as they are more akin to an innovative analysis of a particular situation and thus useful to model human behavior.

Antisocial strategies, in general, have also not yet been studied systematically in the realm of statistical physics, an exception being antisocial pool rewarding in competition with prosocial pool rewarding \cite{szolnoki_prsb15}, where it was shown that antisocial efforts, rather surprisingly, do not deter public cooperation. Existing research points to a rather devastating effect of antisocial punishment on cooperation enforcement \cite{rand_jtb10, rand_nc11}, but a dedicated effort to check the stability of subsystem solutions in the spatial public goods game with antisocial punishment, either peer- or pool-based, as well as with antisocial rewarding, is still lacking.

Looking forward, physicists have to reach out and work together more closely with social scientists, and with their help merge, refine, and upgrade current models so that they will become even more relevant for human cooperation. An outline of the challenges ahead in this endeavor is provided in Section~\ref{humex}. While we should of course utilize methods of statistical physics and network science to the fullest, we should also do our best to ensure a solid and plausible social embedding for our research. In this, we can connect to existing interdisciplinary applications of statistical physics, for example to better understand the economy \cite{mantegna1999introduction, plerou1999universal, plerou2002random, onnela2003dynamics, matteo2005long, aste2010correlation, battiston2016price}, to mitigate crime \cite{orsogna_plr15}, to promote vaccination \cite{wang_z_pr16}, to predict and prevent epidemics \cite{pastor_rmp15}, and to save lives \cite{helbing_jsp15}, but we also need help from outside of physics if we wish to come up with useful models that will help guide our efforts towards a sustainable and better future. Research reviewed above has the potential to have a deeply positive impact on pressing challenges of our time, many of which are due precisely because of large-scale failures of cooperation. Ultimately, we must learn how to create organizations, governments, and societies that are more cooperative and more egalitarian, and perhaps most importantly, are driven not by policy and law that can often be tricked, but simply by a higher level of collective intelligence stemming from each individual.

\begin{acknowledgments}
This work was supported by the Slovenian Research Agency (Grants J1-7009 and P5-0027), the Templeton World Charity Foundation (Grant TWCF0209), the Defense Advanced Research Projects Agency NGS2 program (Grant D17AC00005), the National Key Research and Development Program (Grant 2016YFB0800602), the National Natural Science Foundation of China (Grant 61471299), and by the Hungarian National Research Fund (Grant K-120785).
\end{acknowledgments}

\end{document}